\documentclass[11pt,a4paper,oneside]{article}

\usepackage{setspace}

\usepackage[top=3cm, bottom=3cm, left=2cm, right=2cm]{geometry}
\usepackage{sectsty}
\allsectionsfont{\centering \normalfont\scshape} 

\usepackage{authblk}
\usepackage{dirtytalk}

\usepackage{amsmath}
\usepackage{graphicx,psfrag,epsf}
\usepackage{enumerate}
\usepackage{natbib}
\usepackage{xcolor}
\usepackage{url} 
\usepackage{amsfonts}
\usepackage{amsthm}
\usepackage{amssymb}

\usepackage{mathrsfs} 
\usepackage{algpseudocode}
\usepackage[ruled,vlined]{algorithm2e}
\usepackage{xcolor}
\usepackage[caption = false, labelformat=simple]{subfig}
\usepackage{bm}
\usepackage{bbm}
\usepackage{mathtools} 
\usepackage{comment}
\usepackage{dirtytalk}
\usepackage[backref=false]{hyperref}

\theoremstyle{plain}

\newtheorem{theorem}{Theorem}[section]

\theoremstyle{definition}
\newtheorem{definition}{Definition}[section]
\newtheorem{remark}{Remark}

\theoremstyle{remark}

\newcommand{\Autoref}[1]{%
  \begingroup%
  \def\chapterautorefname{Chapter}%
  \def\sectionautorefname{Section}%
  \def\subsectionautorefname{Section}%
  \def\subsubsectionautorefname{Section}%
  \def\appendixautorefname{Section}%
  \def\theoremautorefname{Theorem}%
  \def\algorithmautorefname{Algorithm}%
  \autoref{#1}%
  \endgroup%
}

\def\drm{\mathrm{d}}
\def\P{\mathbb{P}}
\def\simind{\stackrel{\text {ind}}{\sim}}
\def\simiid{\stackrel{\text {iid}}{\sim}}

\definecolor{amber}{rgb}{1.0, 0.75, 0.0}
\definecolor{columbiablue}{rgb}{0.61, 0.87, 1.0}

\algnewcommand{\Initialize}[1]{%
  \State \textbf{Initialize:}  
  \parbox[t]{.8\linewidth}{\raggedright #1}
}

\theoremstyle{plain}
\newtheorem{corollary}{Corollary}[section]

\DeclarePairedDelimiter\floor{\lfloor}{\rfloor}

\usepackage[sectionbib]{bibunits}
\defaultbibliographystyle{chicago}
\defaultbibliography{biblio}

\title{\huge Bayesian nonparametric modeling of \\ heterogeneous populations of networks
}
\date{}
\author[1]{Francesco Barile}
\author[2]{Sim\'on Lunag\'omez} 
\author[1]{Bernardo Nipoti}

\affil[1]{Department of Economics, Management and Statistics, University of Milano-Bicocca, Italy\\
\href{mailto:francesco.barile@unimib.it}{\nolinkurl{francesco.barile@unimib.it}}; \href{mailto:bernardo.nipoti@unimib.it}{\nolinkurl{bernardo.nipoti@unimib.it}} }
\affil[2]{Instituto Tecnol\'ogico Aut\'onomo de M\'exico, CDMX 01080, Mexico\\
\href{mailto:simon.lunagomez@itam.mx}{\nolinkurl{simon.lunagomez@itam.mx}}
}

\begin{document}

\maketitle

\begin{bibunit}

\begin{abstract}
The increasing availability of multiple network data has highlighted the need for statistical models for heterogeneous populations of networks. A convenient framework  
makes use of metrics to measure similarity between networks. In this context, we propose a novel Bayesian nonparametric model that identifies clusters of networks characterized by similar connectivity patterns. Our approach relies on a location-scale Dirichlet process mixture of centered Erdős--Rényi kernels, with components parametrized by a unique network representative, or mode, and a univariate measure of dispersion around the mode. We demonstrate that this model has full support in the Kullback--Leibler sense and is strongly consistent. An efficient Markov chain Monte Carlo scheme is proposed for posterior inference and clustering of multiple network data. The performance of the model is validated through extensive simulation studies, showing improvements over state-of-the-art methods. Additionally, we present an heuristic strategy to extend the application of the proposed model to datasets with a large number of nodes. We illustrate our approach with the analysis of human brain network data. 
\end{abstract}

\noindent%
\vfill

\newpage


\section{Introduction}\label{section1}
In recent years, multiple network link relations on the same set of nodes have become prominent in many fields of application. For instance, in neuroscience, interconnections among brain regions are collected to characterize a population of individuals suffering from a neurological disorder \citep{Nelson17}; in computer science, human mobility is studied by tracking individuals' movements in relation to intelligent displays \citep{humanmobility}. Multiple networks are also known as multiplex networks \citep{Mucha} and can be intended either as multiple link relations among the nodes of the network (replicated networks) or as a single link relation observed over different conditions, such as one network evolving over time (longitudinal networks). The developments proposed in this work are motivated by the problem of modeling heterogeneous populations of networks, with a key application of our modeling approach being the clustering of multiple network data. 
\begin{figure}[h]
\centering
\includegraphics[trim={6cm 9.5cm 5.6cm 9.3cm}, clip,width = 0.15\linewidth]{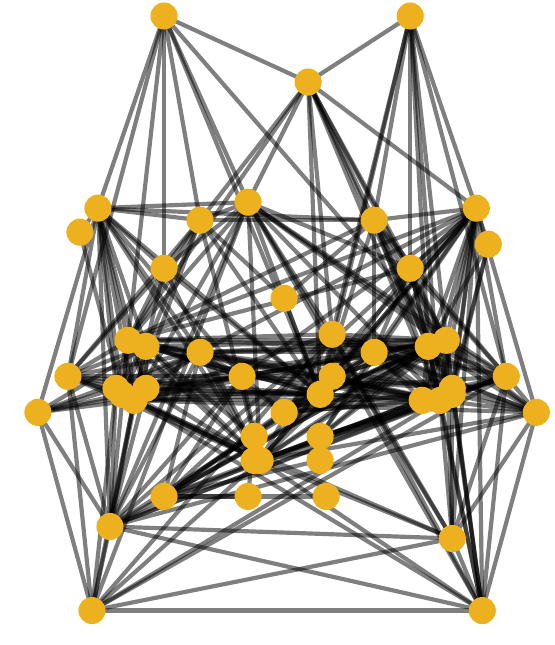}
\includegraphics[trim={6cm 9.5cm 5.6cm 9.3cm}, clip,width = 0.15\linewidth]{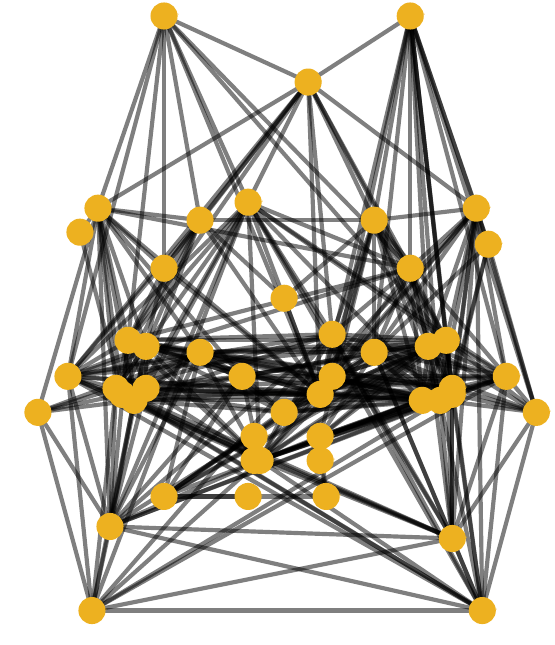}
\includegraphics[trim={6cm 9.5cm 5.6cm 9.3cm}, clip,width = 0.15\linewidth]{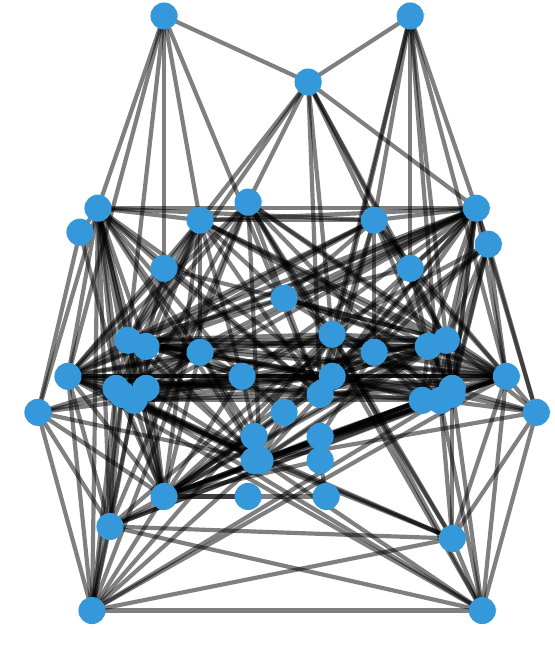}
\includegraphics[trim={6cm 9.5cm 5.6cm 9.3cm}, clip,width = 0.15\linewidth]{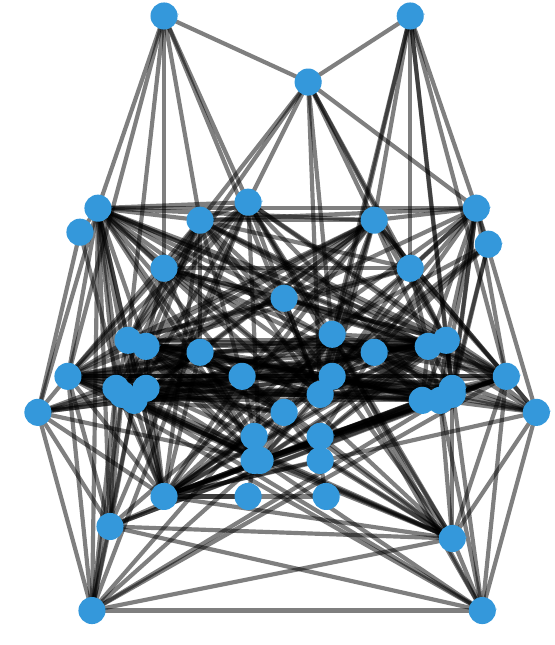}
\includegraphics[trim={6cm 9.5cm 5.6cm 9.3cm}, clip,width = 0.15\linewidth]{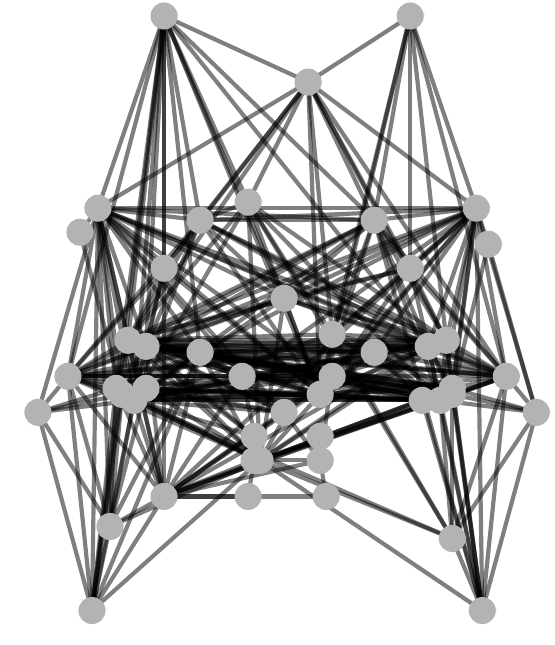}
\includegraphics[trim={6cm 9.5cm 5.6cm 9.3cm}, clip,width = 0.15\linewidth]{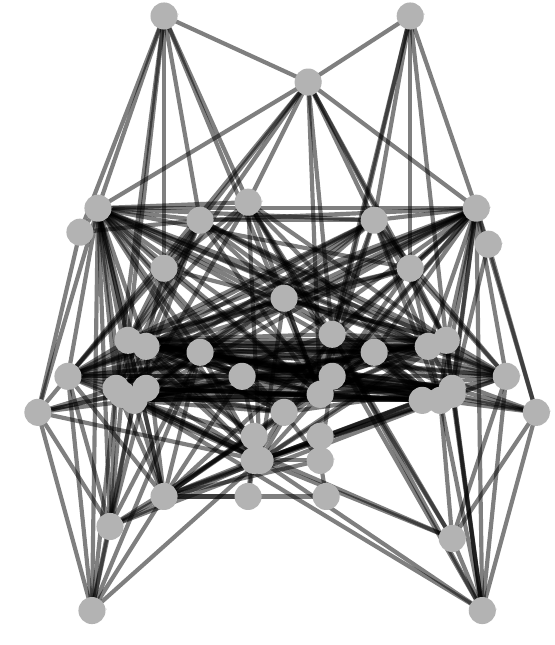}
\caption{\small Top-down projection of a sample of six network observations extracted from the Human Brain Networks dataset (see \Autoref{sec:section5} for details). The nodes of each network are colored according to the network cluster assignments, as inferred by the proposed method.
}\label{fig:intro_3}
\end{figure}
For instance, \Autoref{fig:intro_3} illustrates a sample of observations from the Human Brain Network dataset analyzed in  \Autoref{sec:section5}, with networks colored based on the clusters inferred by our method. Notably, the approach successfully identifies clusters of similar networks, even when differences between networks are not visually evident.

\noindent Recent studies have extended modeling strategies originally designed for a single network observation to multiple network data.
Three different, not necessarily disjoint, frameworks for multiple network data have been considered in the literature.  
The first one is defined through latent space models, 
with the occurrence of an edge between two nodes depending on the positions of the nodes in a latent space \citep{gollini, durante2017, nielsen2018multiple, dangelo19,  Arroyo2021,  Jing2021,Wang2021, dangelo_fop}. 
A second one is based on the use of metrics to measure similarities among networks \citep{donnat2018tracking}. Central to this framework is the notion of representative network, defined with respect to a probabilistic model induced by a suitable choice of metric on the space of networks, as a reference for a population \citep{hypo_test, Kolaczyk2017_averages_unlabeled, CERCER}. Third, measurement error models account for noise in network data by addressing the uncertainty associated with falsely observing edges or non-edges in a network \citep{Young_2022, anastasia}. In this setting, \cite{Le2018} propose a two-stage algorithm that assumes the network population can be represented by a single network with community structure, while accounting for a measurement error process that prevents an accurate observation of the representative network. 
\cite{Sta16} and \cite{Mukherjee2017} are among the first works to address the problem of clustering multiple network data. The former propose a method that clusters networks into groups sharing a common community structure and then detects node communities within each group using a shared stochastic block model, whereas the latter define a graphon mixture model. \cite{Scarpa} propose a clustering approach that involves specifying an ad hoc measure of similarity between networks and implementing an agglomerative method to cluster the networks based on these similarities. \cite{Signorelli} provide a model-based approach for clustering multiple network data with a predefined number of clusters, accounting for the presence of covariates. 
In line with \cite{Le2018}, \cite{anastasia} relax the unimodality assumption to accommodate heterogeneous network data by defining a mixture of measurement error models with a fixed and known number of components and a stochastic block model structure for the representative networks. A similar approach is proposed by \cite{Young_2022}, with the key difference being that the representative networks are not required to form communities in their structural topology. \cite{Yin2019} propose a finite mixture with components assumed to belong to the exponential random graph class of models. Finally, \cite{Durante_Gaffi2025} 
recently introduced a comprehensive framework for a broad class of multidimensional networks, including the multiple network data considered here, to accommodate block connectivity structures both within and across networks.\\
We propose a structure-free modeling approach for multiple network data, where no constraints are imposed on the topology characterizing the data generative process. Our strategy accounts for the heterogeneity that may exist in a population of networks, without imposing rigid assumptions on the number of network subgroups driving the heterogeneity. Such flexibility is achieved by adopting a Bayesian nonparametric approach. 
The literature on nonparametric inferential methods for network-valued observations in heterogeneous populations is relatively young and rapidly evolving.
\cite{durante2017} introduce a Bayesian nonparametric approach for the joint modeling of edge distributions through a flexible mixture representation. \cite{Guha_Guhaniyogi2024} propose a model for clustering networks via covariate-dependent coefficients. \cite{josephs2025} develop a framework that simultaneously clusters nodes and networks in the case of unlabeled graphs. Focusing exclusively on the structure of the nodes, \cite{Amini24} propose a method for community detection in multiplex networks.
Other relevant contributions to network clustering include \cite{reyes2016} and \cite{REN2023}, which are based on stochastic block models and exponential random graph models, respectively. 
In a supervised setting, \cite{Josephs_Lin2023} and \citet{Guha_Dinov2024} address the problem of network classification. \cite{Guha_Rodriguez2021, Guha_Rodriguez2023} propose methodologies in which networks serve as covariates.
To our knowledge, no existing nonparametric methods have been devised for clustering multiple networks without imposing structural assumptions on the generative process. We address this gap by proposing a Bayesian nonparametric distance-based model that combines flexibility and tractability.

\subsection{Our contribution}
We propose a location-scale Dirichlet process mixture of centered Erdős--Rényi kernels to model heterogeneous populations of networks. This kernel choice favors both analytical and computational tractability. The flexibility of the model is ensured by its full topological support, in the Kullback--Leibler sense, over the space of network distributions on a given set of nodes. Additionally, the model's location-scale structure aids in the interpretability of posterior inference. We investigate the properties of the proposed model, present a strategy for posterior computation, demonstrate its effectiveness in various inferential tasks, and explore its applicability to large-dimensional data. The key contributions of this work can be summarized as follows:
\begin{enumerate}
\item[i)] We prove that the proposed model possesses desirable theoretical features, including full support on the space of probability distributions on networks and posterior consistency.\\[-15pt]
    \item[ii)] We develop an efficient Gibbs sampler, relying on the availability of closed-form full conditional distributions for the model's elements.\\[-15pt]
    \item[iii)] We show the model is effective for various inferential tasks, such as clustering networks, estimating probability mass functions, and making predictions.\\[-15pt]
    \item[iv)] Through simulations, we evaluate our method’s performance against existing approaches from the literature in tasks such as clustering and probability mass function estimation.\\[-15pt]
    \item[v)] We apply our method to a dataset from the HNU1 study \citep{Zuo}, which includes diffusion magnetic resonance imaging (dMRI) from multiple subjects, to demonstrate its practical application.\\[-15pt]
    \item[vi)] To handle clustering in populations of networks with many nodes, we introduce an approximate solution called consensus subgraph clustering.
\end{enumerate}
The remainder of the article is organized as follows. In \Autoref{sec:section2}, we introduce a new model for multiple network data and study its main theoretical properties. \Autoref{sec:section3} outlines a strategy for posterior computation and discusses its implementation. \Autoref{sec:section4} presents extensive simulation studies comparing our model with existing methods. In \Autoref{sec:section5}, our method is applied to the analysis of the HNU1 human brain network dataset. 
\Autoref{sec:section6} introduces an approximate computational strategy for clustering large-dimensional datasets. Concluding remarks are presented in \Autoref{sec:section7}. Proofs and additional results are provided as Supplementary Material \citep{DPM_CER_supp}.

\section{Modeling strategy}\label{sec:section2}
\subsection{Preliminaries}\label{subsec:prel}
A simple undirected labeled binary graph $\mathcal{G}=(\mathcal{V}, \mathcal{E})$ consists of a set of labeled vertices $\mathcal{V}$ 
and a set of edges  $\mathcal{E}\subseteq\{(v_1,v_2)\in \mathcal{V}\times \mathcal{V}\,:\,v_1\neq v_2\}$,
that is a subset of the set of pairs of distinct nodes. Given $\mathcal{V}$, we let 
$\mathscr{G}_{\mathcal{V}}$  
denote the set of all 
simple undirected labeled binary graphs with nodes $\mathcal{V}$,  
or graph space. While $\mathscr{G}_{\mathcal{V}}$ is the main focus of this work, it is worth noting that both the modeling and computational strategies we introduce can be readily extended to directed graphs and graphs with self-relations. Given this, and for simplicity, we will henceforth refer to the elements of $\mathscr{G}_{\mathcal{V}}$ simply as graphs or networks, with a slight abuse of terminology.
We observe that, if
$N=|\mathcal{V}|$, then $\left|\mathscr{G}_{\mathcal{V}}\right|=2^{M}$, where $M=\binom{N}{2}$ is the maximum number of edges that a 
graph with $N$ nodes may feature. A graph $\mathcal{G}$ can be represented by an $N \times N$ adjacency matrix $A_{\mathcal{G}}$ such that $A_{\mathcal{G}[ij]}=1$ if $\{i,j\}\in\mathcal{E}$, and $A_{\mathcal{G}[ij]}=0$ otherwise,
where the subscript $[ij]$ is used to indicate the element in position $(i,j)$ of a matrix. The symmetry of the adjacency matrix $A_{\mathcal{G}}$ follows from the fact that $\mathcal{G}$ is assumed undirected. Finally, 
we assume that, throughout this work, all random variables are defined on the same probability space $(\Omega,\mathcal{F},\P)$.\\
We consider a dataset ${\mathcal{G}}^{(1:n)}=\left\{ \mathcal{G}_1, \ldots, \mathcal{G}_n \right\}$ of multiple networks, that is a collection of multiple observations of networks with nodes $\mathcal{V}=\{1,\ldots,N\}$. 
In other terms, for any $l=1,\ldots,n$, we have $\mathcal{G}_l=(\mathcal{V}, \mathcal{E}_l)$. 
This type of data is common,  for instance, in medical imaging, with brain regions 
assigned to the nodes of a graph according to an atlas, and edges representing the connections recorded among regions. 
Modeling an observation  $\mathcal{G}_l$ is equivalent to modeling
the $M$-dimensional vector $\text{vech}(A_{\mathcal{G}_l})$ defined as the half-vectorization of $A_{\mathcal{G}_l}$, whose components coincide with the elements of the lower triangular half of $A_{\mathcal{G}_l}$. 
Given the finite dimensionality of \(\mathscr{G}_{\mathcal{V}}\), a set ${\mathcal{G}}^{(1:n)}$ of random graphs, each taking values in \(\mathscr{G}_{\mathcal{V}}\), can be modeled using a categorical distribution. This can be achieved, for example, by employing latent class models \citep{lcm_goodman}, 
which require the selection of an appropriate number of classes, or by resorting to the nonparametric tensor factorization model for multivariate unordered categorical data proposed by \cite{Dunson_nonparam_bayes}.
We pursue a different strategy by building upon the notion of a representative category to capture the presence of common underlying structures shared among different network configurations. 
The model we propose is based on the idea that distributions on the graph space are conveniently parameterized in terms of a mean, induced by a specified metric, and a measure of the distribution's dispersion around this mean. To this end, the notion of Fréchet mean \citep{{Fréchet1948}} is particularly useful as it generalizes the first moment to non-Euclidean settings, providing a tool to identify a measure of central tendency with respect to the specified metric. Assessing similarities among networks based on their global or local characteristics naturally allows us to evaluate variability in the space of graphs. Our modeling strategy thus requires the specification of a metric on the graph space to appropriately define network structural similarity \citep{donnat2018tracking} and to map non-Euclidean objects to Euclidean spaces.
The Hamming distance \citep{Hamming50}, a special instance of the broader class of
graph-edit distances, is arguably the simplest distance metric between two graphs. Defined as 
\begin{equation*}
d_\text{H}\left(\mathcal{G}_1, \mathcal{G}_2\right)=\sum_{i=1}^{N-1}\sum_{j=i+1}^N \mathbbm{1}_{\left\{A_{\mathcal{G}_1[ij]} \neq A_{\mathcal{G}_2[ij]}\right\}},
\end{equation*}
the Hamming distance between two graphs measures the number of edge deletions and insertions necessary to transform one graph into another, thus capturing local changes. 
Working with the Hamming distance implies that all additions and deletions are assumed to have equivalent importance. 
When compared to other distances, e.g. spectral distances, the Hamming distance offers the notable advantage of allowing the use of standard combinatorial tools. This proves useful for both specifying families of probability distributions for graphs and carrying out efficient posterior inference. An example of a probability distribution for unordered categorical data defined using the Hamming distance is provided in \cite{ArgientoPaci24}. In the network literature,
a flexible distribution for random graphs, defined using the Hamming distance, is the centered Erdős–Rényi (CER) \citep{CERCER}. The CER distribution arises as the product of independent but not identically distributed Bernoulli probability mass functions. The probability of an edge connecting two given nodes is either $\alpha$ or $1-\alpha$, depending on whether an edge is connecting the same pair of nodes of a central graph \(\mathcal{C}\), which can be interpreted as graph mode. As a result,  
the scale of variation parameter $\alpha$ drives the variability around $\mathcal{C}$. Formally, for any $i<j$, 
$\P(A_{\mathcal{G}[ij]} \neq A_{\mathcal{C}[ij]})=1- \P(A_{\mathcal{G}[ij]} = A_{\mathcal{C}[ij]})=\alpha$. The joint distribution of the $M$ components of $\text{vech}(A_{\mathcal{G}})$ thus leads to an equivalent probability mass function for $\mathcal{G}$, that is
\begin{align}\label{eq:eq_CER}
    p_{\text{CER}}\left(\mathcal{G} ; \mathcal{C}, \alpha\right)
    &=\alpha^{d_\text{H}\left(\mathcal{G}, \mathcal{C}\right)}(1-\alpha)^{M-d_\text{H}\left(\mathcal{G}, \mathcal{C}\right)}.
\end{align}
A random graph $\mathcal{G}$, taking values in $\mathscr{G}_{\mathcal{V}}$ and with probability mass function \eqref{eq:eq_CER}, is said to have CER distribution with location parameter $\mathcal{C}\in \mathscr{G}_{\mathcal{V}}$ and scale of variation parameter $\alpha\in(0,1/2)$. We use the notation $\mathcal{G} \sim \operatorname{CER}\left(\mathcal{C}, \alpha\right)$. Although \eqref{eq:eq_CER} is well defined for any $\alpha\in(0,1)$, restricting the set of values that $\alpha$ can take to $(0,1/2)$
ensures that the resulting distribution is unimodal, with mode at $\mathcal{C}$. That is, if 
$d_\text{H}\left(\mathcal{G}_1, \mathcal{C}\right) > d_\text{H}\left(\mathcal{G}_2, \mathcal{C}\right)$ then $p_{\text{CER}}\left(\mathcal{G}_2 ; \mathcal{C}, \alpha \right) > p_{\text{CER}}\left(\mathcal{G}_1 ; \mathcal{C}, \alpha \right)$, 
formalizing the idea that graphs closer to the graph mode are more likely. The CER distribution serves as the building block of the flexible Bayesian model that we introduce next.

\subsection{A Bayesian nonparametric model}
We introduce a Bayesian nonparametric model for networks, defined as a nonparametric location-scale mixture of CER kernels. The CER kernel function, denoted as $\psi(\cdot ; \cdot)$, is defined on $\mathscr{G}_{\mathcal{V}} \times \Theta$, with $\Theta = \mathscr{G}_{\mathcal{V}} \times \left(0, 1/2 \right)$. Specifically, $\psi(\mathcal{G}; \vartheta = (\mathcal{C}, \alpha)) = p_{\text{CER}}(\mathcal{G}; \mathcal{C}, \alpha)$, where $p_{\text{CER}}$ is defined in \eqref{eq:eq_CER}. 
For simplicity, we define the nonparametric mixture model using the distribution of a Dirichlet process (DP). However, the posterior computation strategies presented in this work are easily adapted to more general mixing measures, as long as their predictive distribution is available in closed form. This is the case, for example, with the class of Gibbs-type priors \citep{gibbs}, and, following the introduction of a suitable auxiliary random variable, with the class of normalized random measures with independent increments \citep{Reg03}.

\begin{definition}[Location-scale DP mixture of CER kernels]\label{def:def_DPM_CER}
The location-scale DP mixture of CER kernels on $\mathscr{G}_{\mathcal{V}}$ is the random probability mass function $\tilde{f}$ defined as
\begin{equation}\label{eq:def_DPM}
    \tilde{f}(\cdot)=\int_{\Theta} \psi\left(\cdot ; \vartheta\right) \drm \tilde{P} \left( \vartheta \right),
\end{equation}
where $\Theta=\mathscr{G}_{\mathcal{V}} \times \left( 0, 1/2 \right)$ and
$\tilde{P}$ is distributed as a DP with base measure $P_{0}$ on $\Theta$, and concentration parameter $c>0$.
\end{definition}

\noindent Following the introduction of the latent variables $\vartheta^{(1:n)}=\{\vartheta_1,\ldots,\vartheta_n\}$, the same model can be expressed in hierarchical form as
\begin{equation}\label{eq:model_hier}
\begin{aligned}
\mathcal{G}_{l} \mid 
\vartheta^{(1:n)}
& \stackrel{\text {ind}}{\sim} \psi\left(\mathcal{G}_{l} ; \vartheta_{l} \right)& l=1,\ldots,n  \\
\vartheta_{l}=\left( \mathcal{C}_{l}, \alpha_{l} \right) \mid \tilde{P} & \stackrel{\text { iid }}{\sim} \tilde{P}& l=1,\ldots,n  \\
\tilde{P} & \sim \text{DP}\left(c, P_{0}\right).&
\end{aligned}
\end{equation}
The DP mixture of CER kernels is completed by specifying the base measure $P_0$, which we define as the joint distribution of $\vartheta=(\mathcal{C},\alpha)$ for which 
\begin{equation}\label{eq:base_meas}
\begin{aligned}
    \alpha &\sim \operatorname{TBeta}\left(1/2; a, b\right)\\
\mathcal{C} \mid \alpha  &\sim \operatorname{CER}\left(\mathcal{G}_0, \alpha\right),
\end{aligned}
\end{equation}
for some hyperparameters $a,b>0$, and $\mathcal{G}_0\in\mathscr{G}_{\mathcal{V}}$. For $q\in(0,1)$,  $\text{TBeta}\left(q; a, b\right)$ in \eqref{eq:base_meas} denotes the Truncated-Beta distribution on $(0,q)$, whose probability density function is given by
\begin{align*}
    f_{\text{TBeta}}(\alpha; q, a, b )=\frac{\alpha^{a-1} \left(1-\alpha \right)^{b-1}}{\mathcal{B}\left( q; a, b\right)},
\end{align*}
where $\mathcal{B}\left( q; a, b \right)=\int_0^q \alpha^{a-1} \left(1-\alpha \right)^{b-1} \drm \alpha$ indicates the incomplete beta function. We emphasize that restricting the support of the component-specific dispersion parameters $\alpha_l$ to the interval $(0, 1/2)$ is crucial for ensuring that the CER kernel is unimodal, with its mode at $\mathcal{C}_l$. In the context of mixture modeling, this property is particularly appealing, as it makes the inferred clusters easily interpretable. We further note that a prior distribution can be assigned to the hyperparameters $\left(a, b, c, \mathcal{G}_0\right)$ appearing in the Truncated-Beta/CER specification of $P_0$.

\noindent
\begin{remark}
The proposed model constitutes a non-trivial generalization of the CER model in \citet{CERCER}. While the CER model is parametric and unimodal, our approach is explicitly designed to capture heterogeneity in populations of networks, without imposing a fixed number of modes. This is achieved by placing a nonparametric prior on both the location and scale parameters of the CER kernel and by introducing a Truncated-Beta/CER base measure, which differs from the prior in \citet{CERCER} and substantially improves computational tractability. Such tractability plays a central role in this work, as it allows for the analytical marginalization of the DP and the formulation of a marginal algorithm for posterior sampling. We also note a close connection between our approach and the mixture models of \citet{Young_2022} and \citet{anastasia}. Both works propose parametric, multimodal models with kernel structures closely related to the CER kernel considered here, differing mainly by the inclusion of two component-specific dispersion parameters. Moreover, in \citet{anastasia} the modes are constrained by a stochastic block model structure. In contrast, our approach neither imposes a specific structural form on the modes nor fixes their number. A distinctive feature of our proposal is that the number of mixture components is allowed to grow unboundedly with the number of observed graphs. This becomes evident when considering the infinite-sum representation of $\tilde{f}(\cdot)$ obtained by substituting the stick-breaking representation of $\tilde{P}$ \citep{Sethuraman1994} into \eqref{eq:def_DPM} and goes beyond a mere technicality, playing a central role in establishing the desirable theoretical properties of our model, discussed in the next section.\\
\end{remark}

\subsection{Kullback--Leibler property and posterior consistency}\label{subsec:teo_res}
We show that the location-scale DP mixture of CER kernels, introduced in \Autoref{def:def_DPM_CER} and with base measure \eqref{eq:base_meas}, has full support in the Kullback--Leibler sense. Specifically, for any $\varepsilon>0$, the prior induced by $\tilde{f}$ assigns positive probability to the Kullback--Leibler neighborhood 
$\mathbb{B}_{\varepsilon}(p_*)=\left\{ p \in \mathcal{P}_{\mathcal{G}_{\mathcal{V}}}\,  :\, \text{KL}(p_*; p) \leq \varepsilon \right\}$
of any probability mass function \( p_* \in \mathcal{P}_{\mathscr{G}_{\mathcal{V}}} \), where 
$\text{KL}(p_*; p)$
denotes the Kullback--Leibler divergence between the probability mass functions $p_*$ and 
$p$, and 
 \(\mathcal{P}_{\mathscr{G}_{\mathcal{V}}}\) denotes the space of all probability distributions on \(\mathscr{G}_{\mathcal{V}}\).
 This property, also known as the Kullback--Leibler property, is appealing as it formalizes the idea that any distribution in \(\mathcal{P}_{\mathscr{G}_{\mathcal{V}}}\) can be approximated arbitrarily well by a set of realizations of \(\tilde{f}\) with positive prior probability. For Bayesian modeling, it is convenient to use priors with the Kullback--Leibler property, hence nonparametric, especially when there is no conclusive prior information about the parametric shape of the distribution generating the data \citep{Wal04}. The Kullback--Leibler property is also key in studying the large $n$
behavior of the posterior distribution of $\tilde{f}$, given $\mathcal{G}^{(1:n)}$. Considering the finiteness of \(\mathscr{G}_{\mathcal{V}}\), the Kullback--Leibler property implies that the posterior distribution of $\tilde{f}$ 
is strongly consistent at any $p_*\in\mathcal{P}_{\mathscr{G}_{\mathcal{V}}}$. 
These properties are formalized in \Autoref{thm:full_supp} and \Autoref{cor:cons}, with proofs provided in the Supplementary Material \citep{DPM_CER_supp}.

\begin{theorem}\label{thm:full_supp}
The prior $\Pi$ induced by a location-scale DP mixture of CER kernels with base measure as in \eqref{eq:base_meas} has the Kullback--Leibler property. That is, for any $p_* \in \mathcal{P}_{\mathscr{G}_{\mathcal{V}}}$ and any $\varepsilon>0$, $\Pi\left( \mathbb{B}_{\varepsilon}(p_*)\right) > 0$.
\end{theorem}
\noindent Although \Autoref{thm:full_supp} explicitly refers to the base measure in \eqref{eq:base_meas}, its proof only relies on the fact that $P_0$ has full support on $\Theta$. Therefore, the Kullback--Leibler property extends to any specification of the base measure with full support on $\Theta$.

\begin{corollary}\label{cor:cons} The posterior distribution $\Pi_n(\cdot\mid\mathcal{G}^{(1:n)})$ of a location-scale DP mixture of CER kernels with base measure as in \eqref{eq:base_meas}, given $\mathcal{G}^{(1:n)}$, is strongly consistent at any $p_*\in\mathcal{P}_{\mathscr{G}_{\mathcal{V}}}$. That is, for any $p_*\in\mathcal{P}_{\mathscr{G}_{\mathcal{V}}}$, $\Pi_n(U_{p_*}^\text{c}\mid \mathcal{G}^{(1:n)})\rightarrow 0$ almost surely, as $n\rightarrow \infty$, for any neighborhood $U_{p_*}$ of $p_*$.
\end{corollary}
\noindent It should be noted that, while the literature on the asymptotic properties of network models has traditionally focused on the regime where the number of nodes $N \rightarrow \infty$ within a single network \citep[see, e.g.,][]{Tang}, another interesting question arises when $N$ is fixed and $n \rightarrow \infty$. In this setting, \citet{Josephs_Lin2023} recently established posterior consistency for their graph classifier. \Autoref{cor:cons} contributes to this line of research by addressing the consistent estimation of the entire population distribution for a network-valued random variable.
This theoretical result is complemented by the simulation study presented in \Autoref{subsec:sim_cons}, which investigates how the posterior estimate concentrates around its true value as a function of sample size $n$.

\section{Posterior computation}\label{sec:section3}
\begin{figure}[t!]
\centering
\includegraphics[height=4cm]{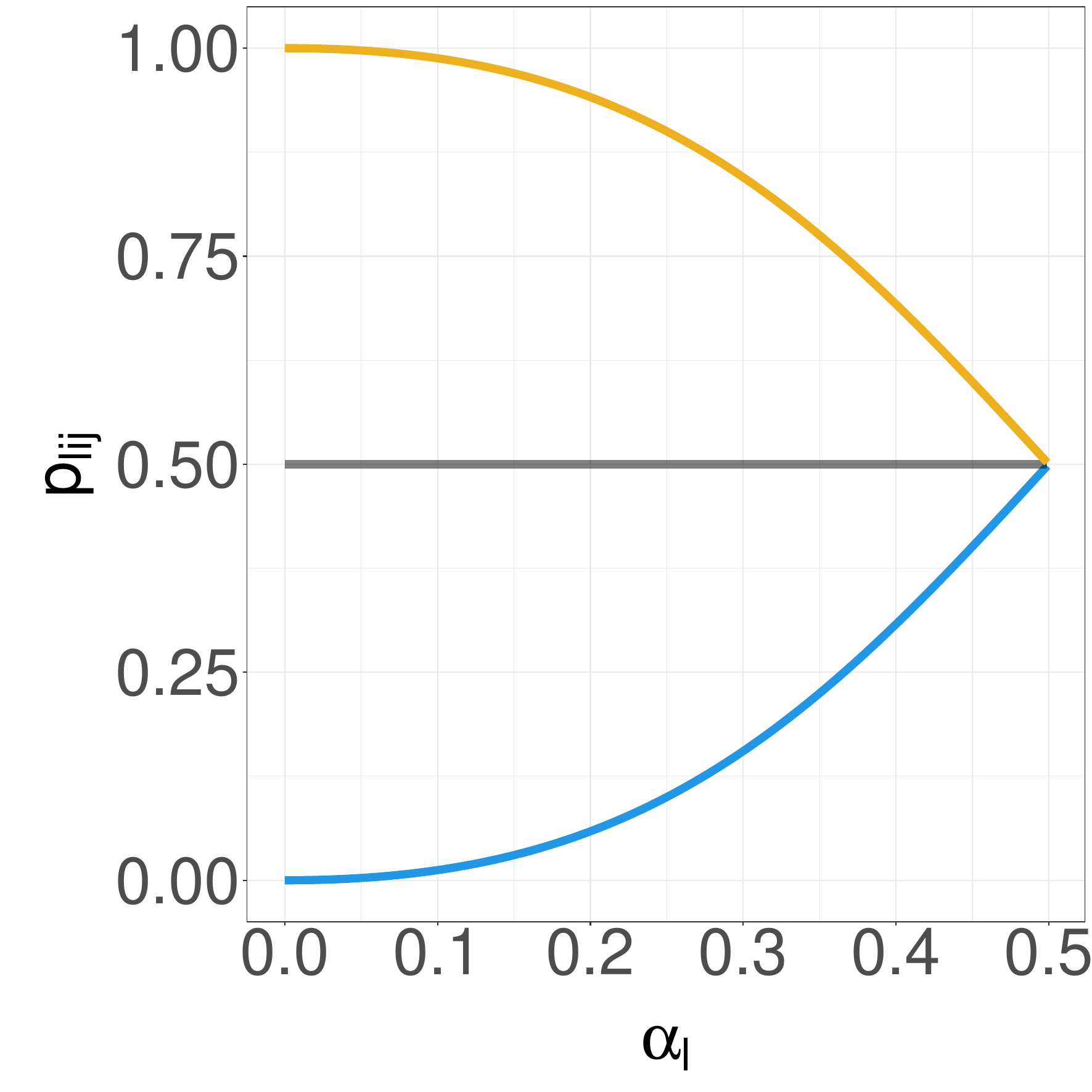}\hspace{1.5cm}\includegraphics[height=4cm]{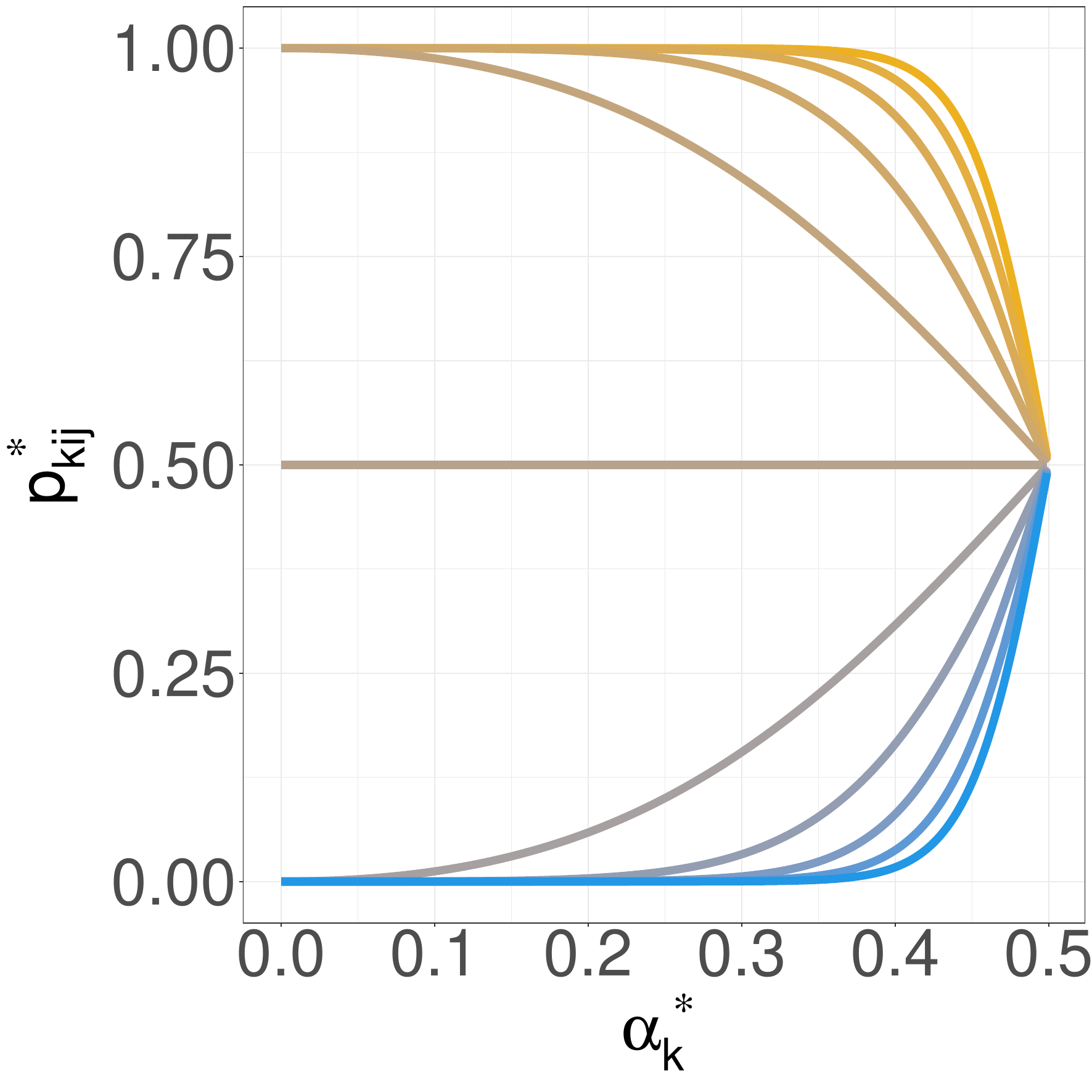}
\caption{\small Left panel: 
probability $p_{lij}$ in \eqref{eq:prob_edge1}, with $A_{\mathcal{G}_0[ij]}+A_{\mathcal{G}_l[ij]}\in\{0,1,2\}$ (blue for 0, gray for 1, and yellow for 2) and for $\alpha_l$ ranging in $(0,1/2)$. Right panel: 
probability $p_{kij}^*$ in \eqref{eq:prob_edgek}, with $n_{ij}^{(k)}\in\{0,1,\ldots,n_k+1\}$, $n_k+1=10$, (blue for low, yellow for high) and for $\alpha_k^*$ ranging in $(0,1/2)$. 
}\label{fig:prob_n_2_n10}
\end{figure}

We adapt the marginal algorithm from \citet{Escobar_West1995}, originally introduced for location-scale DP mixtures of univariate Gaussian kernels, to mixtures of CER kernels, as specified in \autoref{def:def_DPM_CER} and \eqref{eq:base_meas}. Posterior sampling is achieved through a Gibbs sampler, following the analytical marginalization of the DP $\tilde{P}$. Despite the inherently complex structure of $\mathscr{G}_{\mathcal{V}}$, the distributions involved in the algorithm are available in closed form, which conveniently simplifies posterior sampling. 
The algorithm consists of sequential Gibbs updates of the individual location-scale parameters $\vartheta_{l} = \left( \mathcal{C}_{l}, \alpha_{l} \right)$, for $l=1, \ldots, n$, from their full conditional distribution
\begin{align}\label{eq:polya_urn_prob}
\P\left(\vartheta_{l} \in \cdot \mid \vartheta^{(1:n)}_{(-l)}, \mathcal{G}^{(1:n)} 
\right)  = \pi_{l0} \; P_{l}(\cdot) + \sum_{k=1}^{K_{(-l)}} \pi_{lk} \; \delta_{{\vartheta_{k(-l)}^{*}}}(\cdot),
\end{align}
where the subscript $(-l)$ denotes quantities computed after removing $\vartheta_l$ from $\vartheta^{(1:n)}$. \Autoref{eq:polya_urn_prob} represents the celebrated generalized Pólya urn scheme of \citet{Blackwell_MacQueen}, formalizing the idea that $\vartheta_l$ can either coincide with any of the distinct values in $\vartheta^{(1:n)}_{(-l)}$, that is $\vartheta_{k(-l)}^*$, with probability $\pi_{lk}$, for $k = 1, \ldots, K_{(-l)}$, or take a new value with probability $\pi_{l0}$. The probabilities in \eqref{eq:polya_urn_prob} are given, up to a proportionality constant, by
\begin{align}
\pi_{l0} 
& \propto
c  \sum_{r=0}^{M-d_l} w_{lr} \frac{\mathcal{B}( 1/2; a_{lr}, b_{lr} )}{\mathcal{B}\left( 1/2; a, b\right)} & \label{eq:prob_new} \\
\pi_{lk}&\propto n_{k(-l)} \psi\left(\mathcal{G}_{l} ; {\vartheta_{k(-l)}^{*}}\right) & k=1,\ldots,K_{(-l)}, \label{eq:prob_k}
\end{align}
where $d_l=d_{\text{H}}\left(\mathcal{G}_0, \mathcal{G}_{l}\right)$, $w_{lr}=2^{d_l} \binom{M-d_l}{r}$,
$a_{lr}=a+2r+d_l$ and $b_{lr}=b+2M-2r-d_l$.
See Section~S2.1 in the Supplementary Material \citep{DPM_CER_supp} for the derivation of \eqref{eq:prob_new}.
The distribution $P_l$ of new values for $\vartheta_l$, conditionally on $\mathcal{G}_l$, is proportional to $\psi\left(\mathcal{G}_l; \vartheta \right) \mathrm{d} P_{0}(\vartheta)$. Sampling from $P_l$ translates into sampling from a mixture of Truncated-Beta distributions, and from $M$ independent Bernoulli distributions. Specifically, $\vartheta_l = (\mathcal{C}_l, \alpha_l)\mid \mathcal{G}_l \sim P_l$ can be expressed as
\begin{align}\label{eq:dist_new1}
    \alpha_l\mid \mathcal{G}_l&\sim \sum_{r=0}^{M-d_l}\varphi_{lr} \text{TBeta}(1/2;a_{lr},b_{lr})&\\
    A_{\mathcal{C}_l[ij]}\mid \alpha_l, \mathcal{G}_l &\simind \text{Bern}(p_{lij})&i<j,\label{eq:dist_new2}
\end{align}
where $\varphi_{lr}\propto w_{lr} \, \mathcal{B}( 1/2; a_{lr}, b_{lr})$ and
\begin{equation}\label{eq:prob_edge1}
    p_{lij}=\left(1 + \left( \frac{\alpha_l}{1-\alpha_l}\right)^{2(A_{\mathcal{G}_0[ij]}+A_{\mathcal{G}_l[ij]}-1)} \right)^{-1}.
\end{equation}
The probability $p_{lij}$ of generating a graph mode with an edge connecting the nodes $\{i,j\}$ thus depends on whether $\mathcal{G}_0$ and $\mathcal{G}_l$ display such an edge. The left panel of \autoref{fig:prob_n_2_n10} illustrates the dependence of $p_{lij}$ on  $A_{\mathcal{G}_0[ij]}+A_{\mathcal{G}_l[ij]}$ and $\alpha_l$. The derivation of the characterization of \(P_l\) in \eqref{eq:dist_new1}–\eqref{eq:prob_edge1} is reported in Section~S2.3 of the Supplementary Material \citep{DPM_CER_supp}.

To improve the mixing of the algorithm, it is key to introduce a reshuffling step that independently updates the cluster labels, i.e., the values $\vartheta_k^* = (\mathcal{C}^{*}_k, \alpha_k^*)$ for $k = 1, \ldots, K$, after each Gibbs sampling iteration \citep{Bush_MacEachern_96}. We let $\mathcal{D}_k = \{ l \in \{ 1, \dots, n \} : \vartheta_l = \vartheta_k^* \}$ be the index set of observations belonging to the $k$-th cluster, with $|\mathcal{D}_k| = n_k$, and define $\mathcal{D}^\dagger_k = \mathcal{D}_k \cup \{ 0 \}$. For any index set $\mathcal{D}\subseteq\{0,1,\ldots,n\}$, we introduce the notation $\mathcal{G}^{(\mathcal{D})}=\{\mathcal{G}_l:l\in\mathcal{D}\}$, and we let $n_{ij}^{(k)}= \sum_{l \in \mathcal{D}^\dagger_k} A_{\mathcal{G}_l[ij]}$ denote the number of graphs in $\mathcal{G}^{(\mathcal{D}^\dagger_k)}$ that present an edge connecting the nodes $\{i, j\}$. Similarly to $P_l$, updating $\vartheta_k^*$ from its full conditional distribution translates into sampling from a mixture of Truncated-Beta distributions, and from $M$ independent Bernoulli distributions. Specifically, after introducing the quantities $D_k^*=\sum_{i=1}^{N-1} \sum_{j=i+1}^{N} \max\{n_{ij}^{(k)}, n_k+1-n_{ij}^{(k)}\}$ and $d_k^*=\sum_{i=1}^{N-1} \sum_{j=i+1}^{N} \min\{n_{ij}^{(k)}, n_k+1-n_{ij}^{(k)}\}$, we can write
\begin{align}\label{eq:accell_1}
    \alpha_k^*\mid \mathcal{G}^{(\mathcal{D}_k)}& \sim \sum_{r=0}^{D_k^*-d_k^*} \varphi_{kr}^{*}\, \text{TBeta}(\alpha^*_k; 1/2, a_{kr}^*, b_{kr}^* )&\\
    A_{\mathcal{C}_k^{*}[ij]}\mid \alpha_k^*, \mathcal{G}^{(\mathcal{D}_k)}&\simind \text{Bern}(p_{kij}^*),&i<j.\label{eq:accell_2}
\end{align}
The Bernoulli parameters in \eqref{eq:accell_2} are defined as
\begin{equation}\label{eq:prob_edgek}
p_{kij}^*=\left(1 + \left( \frac{\alpha_k^*}{1-\alpha_k^*}\right)^{ 2\left(n_{ij}^{(k)}-(n_k+1)/2\right)} \right)^{-1}.
\end{equation}
The probability $p_{kij}^*$ to generate a graph mode with an edge connecting the nodes $\{i,j\}$ is increasing in the number $n_{ij}^{(k)}$ of graphs in $\mathcal{D}_k^\dagger$ displaying such an edge. Moreover, $p_{kij}^*$ is increasing in $\alpha_k^*$ if $n_{ij}^{(k)}< (n_k+1)/2$ and decreasing if $n_{ij}^{(k)}>(n_k+1)/2$. 
The right panel of \autoref{fig:prob_n_2_n10} shows that as $n_{ij}^{(k)}$ increases, the probability of $A_{\mathcal{C}^{*}_k[ij]} = 1$ also rises, with this effect being more pronounced for small values of $\alpha_k^*$.
The mixture weights in  \eqref{eq:accell_1} 
are given, up to a proportionality constant, by \begin{equation}\label{eq:weight1}
\varphi_{kr}^*\propto w_{kr}^* \mathcal{B}( 1/2; a_{kr}^*, b_{kr}^*),
\end{equation}
where $a_{kr}^*=a+d_{k}^*+r$ and $b_{kr}^*=b+( n_k+1) M -d_{k}^*-r$.
In turn, the coefficients $w_{kr}^*$ in \eqref{eq:weight1} result from a generating function defined by a product of polynomials, embedding a subset-sum problem. Specifically, we have
\begin{equation}\label{eq:weight2}
w_{kr}^*=\begin{cases}
\sum_{\mathcal{S}_{kr}}\prod_{h=0}^{n_k/2} \binom{M_{kh}}{s_h}& \text{ if $n_k$ is even}\\
\sum_{\mathcal{R}_{kr}} 2^{{m}_{k(\floor{n_k/2}+1)}} \prod_{h=0}^{\floor{n_k/2}} \binom{M_{kh}}{s_h}& \text{ if $n_k$ is odd},
\end{cases}
\end{equation}
where $\floor{x}$ denotes the integer part of $x$, the sums in \eqref{eq:weight2} are taken over the sets
\begin{align*}
\mathcal{S}_{kr}&=\Big\{(s_0,\ldots,s_{n_k/2}):s_h\in\{0,\ldots,M_{kh}\} \,\forall h,\, 
\sum_{h=0}^{n_k/2}\gamma_{kh}\left(s_h\right)-d_k^*=r\Big\},\\
\mathcal{R}_{kr}&=\Big\{(s_0,\ldots,s_{\floor{n_k/2}}): s_h\in\{0,\ldots,M_{kh}\}\,\forall h,\,\\
&\;\;\,\quad\qquad\qquad \sum_{h=0}^{\floor{n_k/2}}\gamma_{kh}\left(s_h\right)+(\floor{n_k/2}+1)m_{k(\floor{n_k/2}+1)}-d_k^*=r\Big\},
\end{align*}
$m_{kh}=\#\big\{ \{ i, j \} \in \mathcal{V}^2 : n^{(k)}_{ij}=h \big\}$ 
indicates the number of pairs of nodes that are connected by an edge in exactly $h$ graphs in $\mathcal{G}^{(\mathcal{D}_k^\dagger)}$, $M_{kh}=m_{kh} + m_{k(n_k + 1 - h)}$, and $\gamma_{kh}\left(s_h\right)=\left( n_k+1-2h  \right) s_h + h M_{kh}$. 
We note that the conditions defining the sets $\mathcal{S}_{kr}$ and $\mathcal{R}_{kr}$, for any $k=1,\ldots,K$ and $r=0,\ldots,D_k^*-d_k^*$, represent linear Diophantine equations with the decision variables 
 $\{s_0, \ldots, s_{\floor{n_k/2}}\}$ constrained by $s_h\leq M_{kh}$, for $h=1,\ldots, \floor{n_k/2}$.
 Moreover, as expected, when $n_k=1$, that is $\mathcal{D}_k=\{l\}$ for some $l=1,\ldots,n$, the distribution in \eqref{eq:accell_1} and \eqref{eq:accell_2} simplifies to the distribution $P_l$, in \eqref{eq:dist_new1} and \eqref{eq:dist_new2}, for $\vartheta_l$, conditionally on $\mathcal{G}_l$ and given that it takes a new value. 
The steps of the algorithm are summarized in \Autoref{alg:gibbs}. See Section~S2.2 in the Supplementary Material \citep{DPM_CER_supp} for the derivation of the full conditional distribution of $\vartheta_k^*$ involved in the reshuffling step.
Additionally, Section~S3 of the Supplementary Material \citep{DPM_CER_supp}
provides closed-form expressions for the cluster-specific posterior distribution of $\mathcal{C}_k^{*}$ and the cluster-specific $m$-step-ahead posterior predictive distribution, both obtained by building on the conditional distribution of $\vartheta_k^*$ in \eqref{eq:accell_1} and \eqref{eq:accell_2}.

\begin{algorithm}[!ht]
\caption{Gibbs sampler for DP mixture of CER kernels}\label{alg:gibbs}
\KwIn{Data $\mathcal{G}^{(1:n)}$; hyperparameters $a,b,c, \mathcal{G}_0$; number of iterations $T$; number of burn-in iterations $T_0$;}
\KwOut{Sample from the posterior of location-scale parameters:
$\{\vartheta_{[t]}^{(1:n)}\}_{t=T_0+1}^T$}

\BlankLine
\textbf{Initialise} 
$\vartheta^{(1:n)}$ randomly\;
\For{$t \gets 1$ \KwTo $T$}{
    \For{$l \gets 1$ \KwTo $n$}{
        Remove the $l$th component of $\vartheta^{(1:n)}$ to obtain 
        $\vartheta_{(-l)}^{(1:n)}$\;
        Let $K_{(-l)}$ be the number of distinct values in $\vartheta_{(-l)}^{(1:n)}$\;
        Let $(\vartheta_{1(-l)}^*,\ldots,\vartheta^*_{K_{(-l)}(-l)})$ be the set of distinct values in $\vartheta_{(-l)}^{(1:n)}$\;
        Compute $\pi_{l0}$, up to a constant, as in \eqref{eq:prob_new}\;
        \For{$k \gets 1$ \KwTo $K_{(-l)}$}{
            Compute $\pi_{lk}$, up to a constant, as in \eqref{eq:prob_k}\;
        }
        Normalize all probabilities to get $\pi_{l}=\big(\pi_{l0}, \pi_{l1}, \ldots \pi_{lK_{(-l)}}\big)$\;
        Sample $\emph{category}$ from $\operatorname{Categorical}(K_{(-l)}+1,\pi_{l})$\;
        \If{$category==1$}{
            Draw parameter $\vartheta_l$ from $P_l$, as in \eqref{eq:dist_new1}--\eqref{eq:prob_edge1}\;}
        \ElseIf{$category==k\in\{2,\ldots,K_{(-l)}+1\}$}{
        Set $\vartheta_{l}$ equal to $\vartheta_{k-1(-l)}^*$\;}
        }
        Let $K$ be the number of distinct values in $\vartheta^{(1:n)}$\;
        Let $(\vartheta_{1}^*,\ldots,\vartheta^*_{K})$ be the set of distinct values in $\vartheta^{(1:n)}$\;
    \For{$k \gets 1$ \KwTo $K$}{
    Update 
        $\vartheta_k^{*}$, sampling from \eqref{eq:accell_1}--\eqref{eq:prob_edgek}\;
    }
    Set $\vartheta_{[t]}^{(1:n)}$ equal to $\vartheta^{(1:n)}$\;
}
\Return{$\{\vartheta_{[t]}^{(1:n)}\}_{t=T_0+1}^T$}\;
\end{algorithm}

\section{Simulation study}\label{sec:section4}
We explore the behavior of the DP mixture of CER kernels through the analysis of synthetic data. The study has two objectives: (i) assessing the model's ability to cluster multiple network data with a known partition structure, under data-generating processes characterized by varying levels of variability; and (ii) investigating the impact of sample size on the accuracy of posterior estimates. 
To facilitate graphical presentation, we focus on networks with $N = 20$ nodes. Observations are sampled from a mixture of four CER components $p_*(\cdot)=\sum_{k=1}^4 0.25p_{\text{CER}}(\cdot;\mathcal{C}_{0k},\alpha_{0k})$, where well-defined component-specific network structures are defined through the modes $\mathcal{C}_{0k}$. Specifically, and in the same spirit as \cite{durante2017}, each component is centered around a network configuration, or centroid, with distinct structures: scale-free \mbox{\citep{scale-free}} for $\mathcal{C}_{01}$, small-world \citep{small_world} for $\mathcal{C}_{02}$, stochastic block model \citep{sbm} for $\mathcal{C}_{03}$, and Erdős--Rényi \citep{Erdos_Reny} for $\mathcal{C}_{04}$. This choice is designed to assess whether the proposed model can effectively cluster and estimate the distribution of a collection of multiple network data with heterogeneous underlying structures. The generated centroids are displayed in the first row of \autoref{fig:graph_modes}. The specification of the parameters for the four models used to generate the centroids is summarized in Table~S1 in the Supplementary Material \citep{DPM_CER_supp}. 
The study consists of two main parts: in the first one, multiple network data are generated by considering various component-specific scales of variation while keeping the sample size fixed; in the second one, the sample size varies while the scale parameters of the data-generating models remain fixed. For both parts, we specify the parameters of the base measure \eqref{eq:base_meas} as follows.
The centroid $\mathcal{G}_0$ is set, by using an empirical Bayes approach, as an element of the 
sample Fréchet mean set \citep[see][]{CERCER}. Specifically, $\mathcal{G}_0$ is the network that has an edge between the nodes $\{i,j\}$ if and only if that edge is present in at least $50\%$ of the networks in the dataset. The parameters $a$ and $b$ are both set equal to one, thus implying the prior model for the scale of variation parameter is centered at a uniform distribution on $\left( 0, 1/2\right)$. Finally, the concentration parameter $c$ is set equal to one.\\ 
For each scenario in the two simulation studies, 100 datasets are generated. Each dataset is analyzed by running 1,200 Gibbs iterations, with the first 200 discarded as burn-in. 
The results produced by our model are compared with those from \cite{durante2017}, \cite{anastasia}, \cite{Signorelli}, and \cite{josephs2025}, with the latter included only in the first part of the study.
Although not originally intended for this purpose, the method of \cite{durante2017}
is readily extended to address clustering problems for multiple network data. Further details on the implementation of the competing methods
are provided in Section~S5 of the 
Supplementary Material \citep{DPM_CER_supp}. 
This study suggests that, overall, our model performs comparably to or better than state-of-the-art methods in two key aspects: effectively modeling a population of networks with heterogeneous characteristics and accurately clustering the elements of a network population. An additional simulation experiment is reported in Section~S4 of the Supplementary Material \citep{DPM_CER_supp}, where the data-generating process is again specified as a mixture model, but with one component characterized by a core-periphery structure. This setting induces more intricate connectivity patterns and may arise, for instance, from a non-assortative stochastic block model. The performance of our model appears robust to this more complex scenario.

\subsection{Data-generating models with varying scales of variation}\label{sec:study1}
\begin{figure}[t!]
\centering
\includegraphics[clip,trim=2.3cm 0.2cm 2.3cm 0.1cm,width = 0.94\linewidth]
{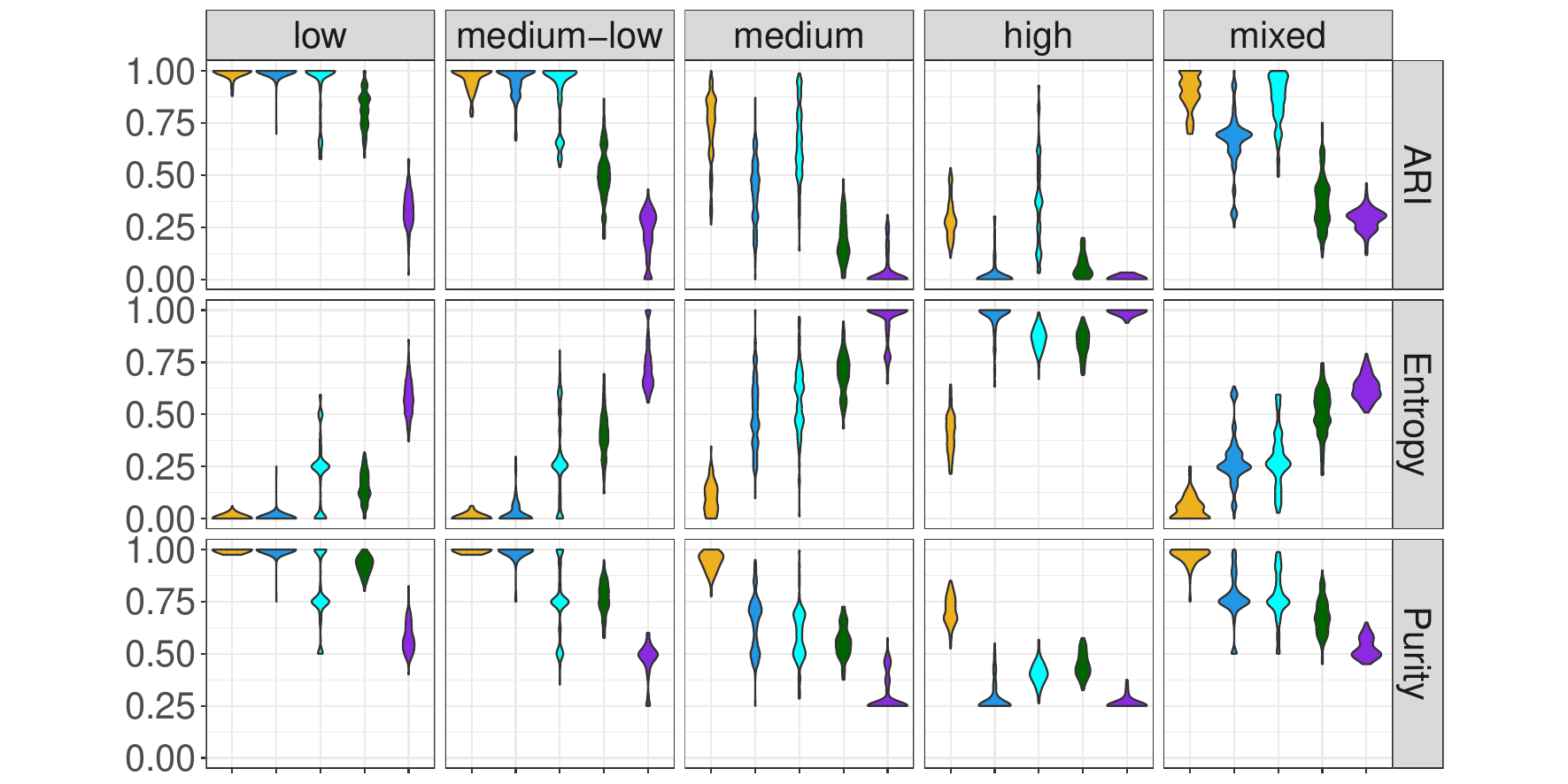}
\caption{\small Adjusted Rand index, entropy and purity, for our method (yellow violins) and the methods of \citet{durante2017} (blue violins),
\citet{anastasia} (cyan violins),
\citet{Signorelli} (green violins) and \citet{josephs2025} (violet violins).
Columns refer to the scenarios of \autoref{tab:table_var}. Distributions are estimated based on the analysis of 100 datasets. See \Autoref{sec:study1}.
}\label{fig:fig_variability}
\end{figure}

To assess the ability of a method to cluster multiple network data, we compare the estimated partition to the true partition, which reflects the four-component mixture structure of the data-generating model. We resort to three metrics: the adjusted Rand index (ARI), clustering entropy and clustering purity. A point estimate for the data partition is obtained from the posterior samples produced by \Autoref{alg:gibbs}, 
by minimizing the posterior expected Variation of Information, as implemented in the \texttt{Salso} R package \citep{salso}.  
We investigate the robustness of our method in clustering multiple network data generated from models characterized by different levels of variability. To this end, we fix a sample size of $n=40$ and focus on four scenarios with increasing scale of variation parameters shared across all components. We also consider a more realistic scenario with different scales of variation for each of the four components of the data-generating model. The values of the component-specific scales of variation
for these scenarios are reported in \autoref{tab:table_var}.
\begin{table}[b!]
\centering
\begin{tabular}{l c c c c}
Level of variability & $\alpha_{01}$ & $\alpha_{02}$ & $\alpha_{03}$ & $\alpha_{04}$ \\
\hline\hline
\text{low} & 0.25 & 0.25 & 0.25 & 0.25 \\
\text{medium-low} & 0.30 & 0.30 & 0.30 & 0.30 \\
\text{medium} & 0.35 & 0.35 & 0.35 & 0.35 \\
\text{high} & 0.40 & 0.40 & 0.40 & 0.40\\
\text{mixed} & 0.25 & 0.35 & 0.30 & 0.40 \\
\end{tabular}
\caption{\small Definition of five simulation scenarios through the specification of the scale of variation parameters $\{\alpha_{01},\ldots,\alpha_{04}\}$ of the four CER components of $p_*$. See \Autoref{sec:section4}.}
\label{tab:table_var}
\end{table}
The results of our investigations are displayed in \autoref{fig:fig_variability}. According to the considered metrics, our model outperforms all the competing methods
across all scenarios, showing higher values for ARI and clustering purity and lower values for clustering entropy. As expected, scenarios characterized by higher levels of variability are more challenging for all methods. 
We also observe the unsatisfactory performance of \cite{josephs2025}, even in relatively simple scenarios. This is likely due to the method overlooking node correspondence across layers, as it is designed for multiplex networks in the unlabeled setting. Consequently, we chose not to include this method as a competitor in the subsequent illustrations.
\begin{figure}[t!]
\centering
\includegraphics[width = 0.21\linewidth]{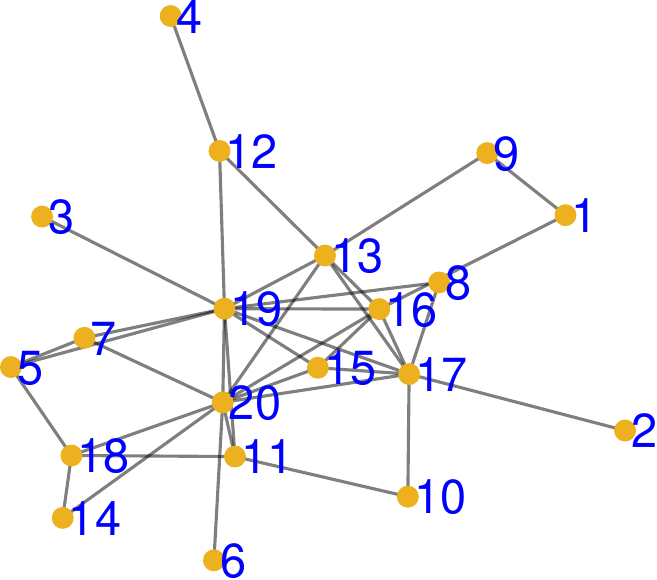}\hspace{0.5cm}
\includegraphics[width = 0.21\linewidth]{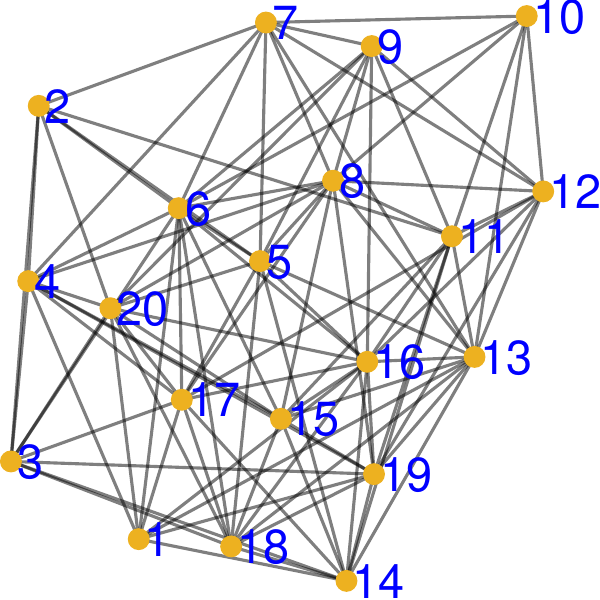}\hspace{0.5cm}
\includegraphics[width = 0.21\linewidth]{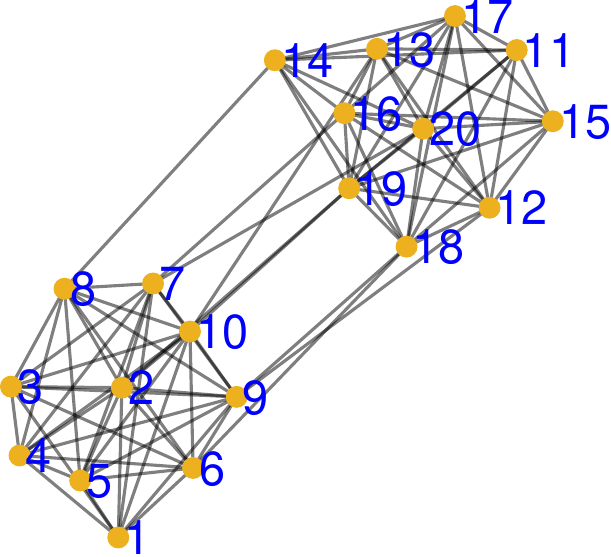}\hspace{0.5cm}
\includegraphics[width = 0.21\linewidth]{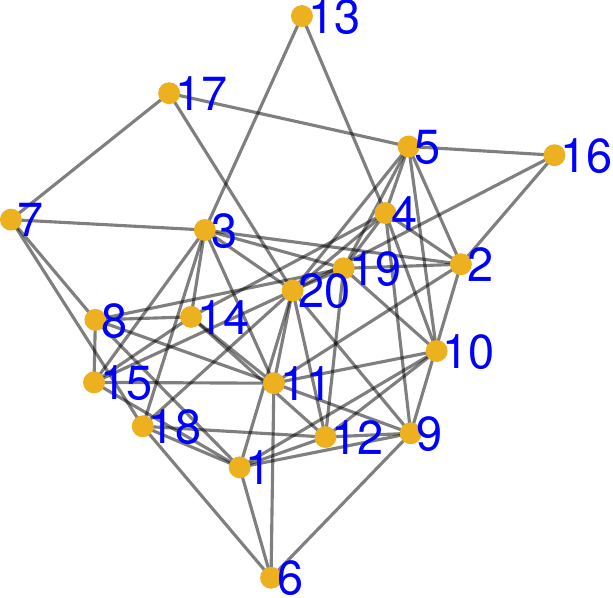}\\[8pt]
\includegraphics[width=0.21\linewidth]{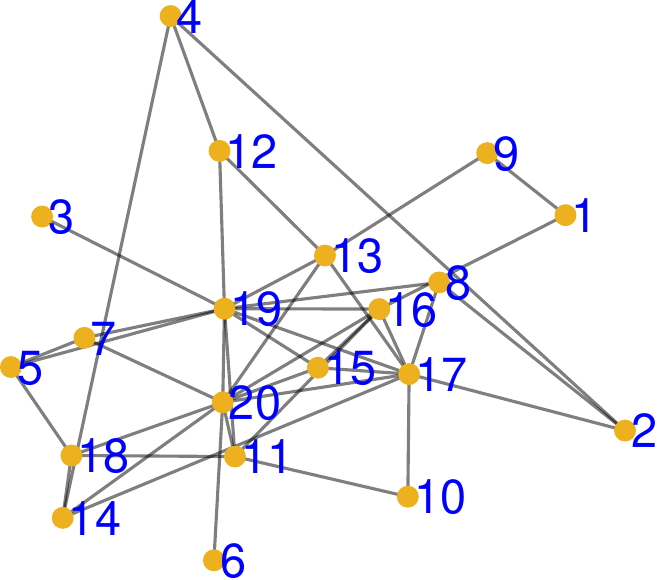}\hspace{0.5cm}
\includegraphics[width=0.21\linewidth]{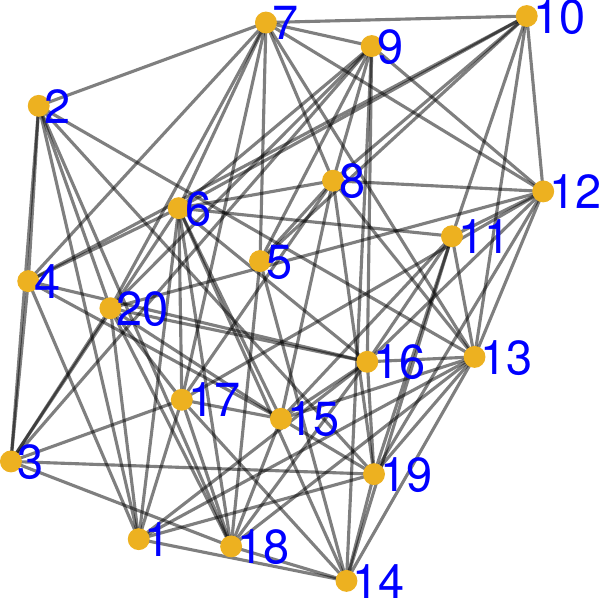}\hspace{0.5cm}
\includegraphics[width=0.21\linewidth]{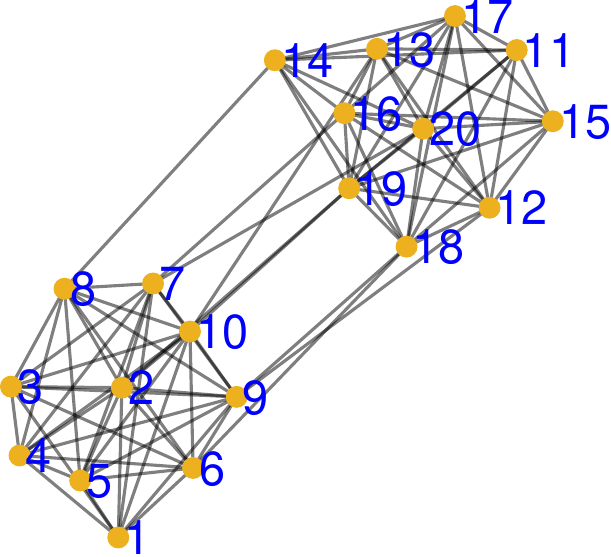}\hspace{0.5cm}
\includegraphics[width=0.21\linewidth]{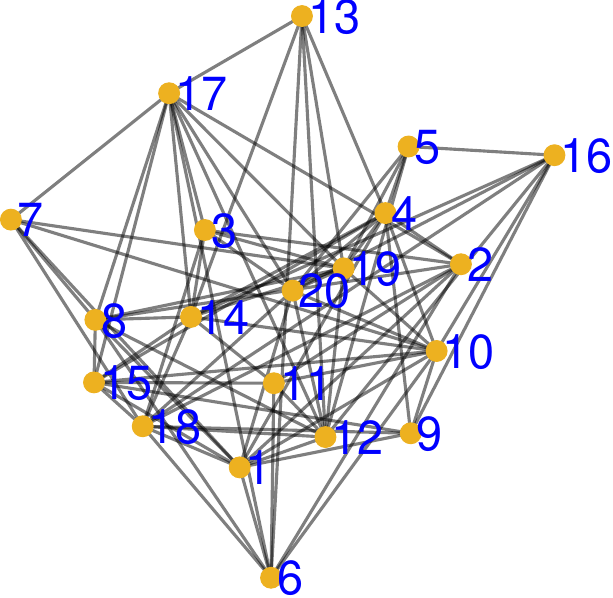}\\
\caption{\small Top row: centroids with Scale-free ($\mathcal{C}_{01}$), Small-world ($\mathcal{C}_{02}$), Stochastic Block Model ($\mathcal{C}_{03}$), and Erdős--Rényi ($\mathcal{C}_{04}$) structures (from left to right). 
Bottom row: 
posterior Fréchet means
for the four clusters estimated based on a dataset generated from the mixed level of variability scenario, with sample size $n=40$. See \Autoref{sec:section4}.}\label{fig:graph_modes}
\end{figure}
\noindent It is also instructive to explore the properties of the clusters identified by our method. Given the estimated partition and denoting by  $\hat{K}$ the corresponding number of clusters, we can produce cluster-specific point estimates for the centroids by looking at the Fréchet mean of the posterior distribution of $\mathcal{C}_k^{*}$, for $k=1,\ldots,\hat{K}$. 
Specifically, we sample from the posterior distribution of $\vartheta_{k}^*$ in \eqref{eq:accell_1} and \eqref{eq:accell_2}, and consider the sample Fréchet mean of $\mathcal{C}_k^{*}$.  
We henceforth refer to this as to the cluster-specific posterior Fréchet mean. Alternatively, one can resort to Equation S27 in the Supplementary Material \citep{DPM_CER_supp}.
For illustrative purposes, we focus on a randomly selected dataset generated from the mixed variability scenario. The second row of \autoref{fig:graph_modes} shows the posterior Fréchet mean for centroids of the four estimated clusters. The topological structures of these estimates align with those of the four-component data-generating mixture model, with the correspondence between true and estimated components identified based on the frequency of observations generated from a given component in the estimated clusters. Moreover, the point estimates of the cluster-specific scale parameters, that is \(\{0.262, 0.337, 0.295, 0.397\}\), reflect the heterogeneity in variability levels that characterize the data-generating model.

\subsection{Varying sample size}\label{subsec:sim_cons}
We study the accuracy of the posterior mean $\hat{f} = \mathbbm{E}[\tilde{f} \mid \mathcal{G}^{(1:n)}]$ as an estimator of the true data-generating distribution $p_*$, with $\hat{f}$ evaluated based on the posterior sample generated from \Autoref{alg:gibbs}.
Specifically, we investigate how this accuracy changes for different sample sizes $n$. This study aims to provide a finite-sample analogue to the strong consistency property of the DP mixture of CER kernels, as reported in \Autoref{cor:cons}, which states that for any $\varepsilon > 0$ and a given metric $d$ on $\mathcal{P}_{\mathscr{G}_{\mathcal{V}}}$, $\P(d(p_*,\tilde{f}) > \varepsilon \mid \mathcal{G}^{(1:n)}) \longrightarrow 0$, almost surely, as $n \rightarrow \infty$. Using the Kullback--Leibler divergence, we study the distribution of the distance between $p_*$ and $\hat{f}$ for finite samples of size $n \in \{40, 80, 120, 200\}$. The evaluation of $\text{KL}(p_*; \hat{f})$ requires summation over the graph space $\mathscr{G}_{\mathcal{V}}$, which is prohibitive even for moderate $N$. Thus, we propose an importance-sampling approximation of $\text{KL}(p_*; \hat{f})$: 
\begin{align*}
\text{KL}(p_*; \hat{f}) = \sum_{\mathcal{G} \in \mathscr{G}_{\mathcal{V}} } p_*(\mathcal{G}) \log \left( \frac{p_*(\mathcal{G})}{\hat{f}(\mathcal{G})} \right) = \mathbbm{E}_{p_*} \left[ \log \left( \frac{p_*(\mathcal{G})}{\hat{f}(\mathcal{G})} \right) \right] \approx \frac{1}{L} \sum_{l=1}^L \log \left( \frac{p_*(\mathcal{G}_l)}{\hat{f}(\mathcal{G}_l)} \right),    
\end{align*}
with $\mathcal{G}_l \simiid p_*$, for $l = 1, \ldots, L$. 
\begin{figure}[t!]
\centering
\includegraphics[clip,trim=0cm 3.45cm 0cm 3.45cm,width = 0.95\linewidth]{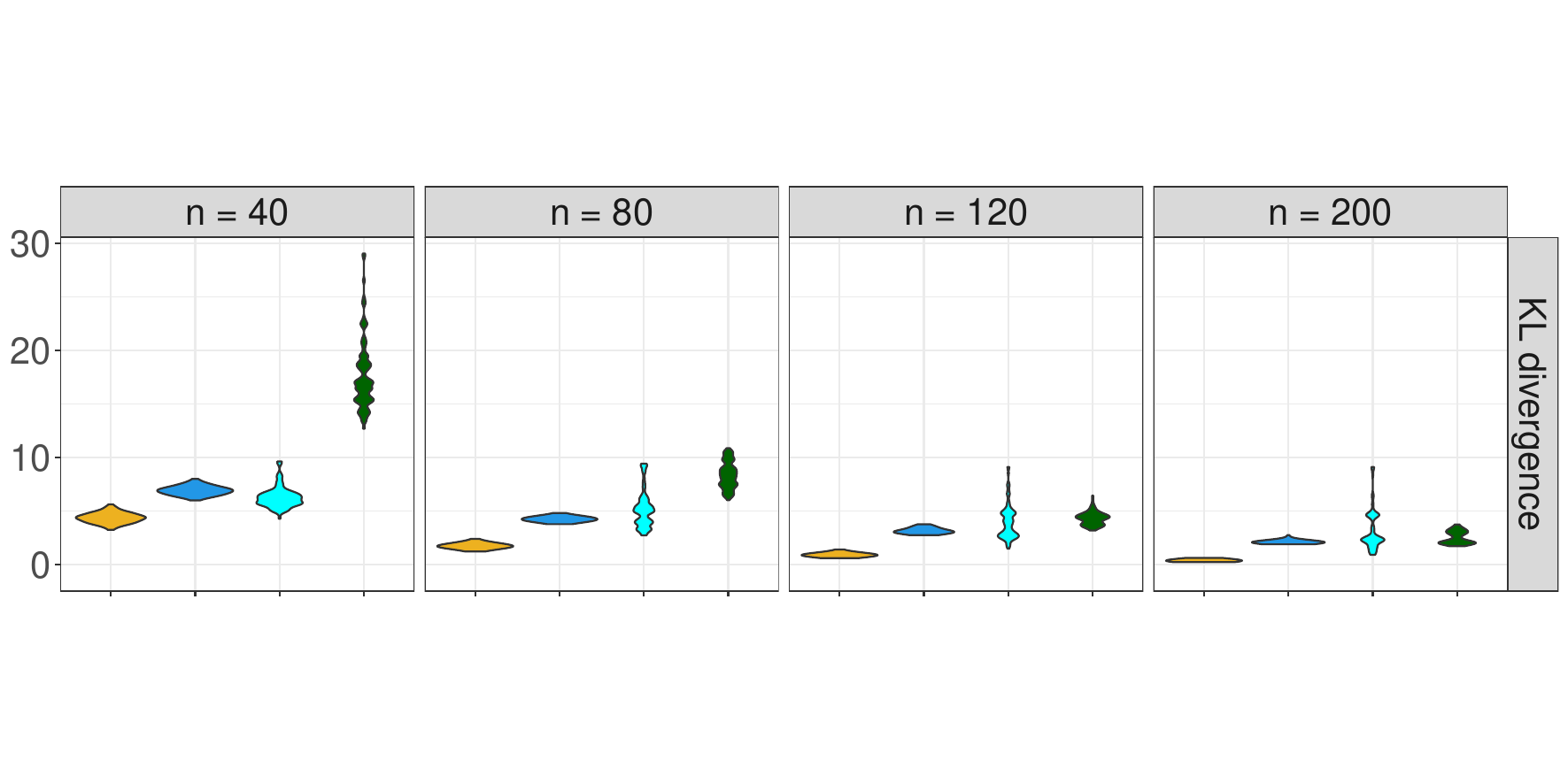}
\caption{\small Importance-sampling approximate distributions of $\text{KL}(p_*;\hat{f})$ for our method (yellow violins), and the methods of \citet{durante2017} (blue violins), \citet{anastasia} (cyan violins) and
\citet{Signorelli} (green violins).
Distributions are estimated based on the analysis of 100 datasets.  See \Autoref{subsec:sim_cons}. }\label{fig:fig_consistency}
\end{figure}
The results of our study are illustrated in \autoref{fig:fig_consistency}, which shows that the posterior estimate $\hat{f}$ gets closer to $p_*$ as the sample size increases. Our model appears to converge to $p_*$ faster than the models proposed by \cite{durante2017}, \cite{anastasia} and \cite{Signorelli}. 
Similar results, focusing on the $\mathbb{L}^1$ distance between $p_*$ and $\hat{f}$, are presented in Section~S4 of the Supplementary Material \citep{DPM_CER_supp}.

\section{Analysis of human brain networks data}\label{sec:section5}
\begin{figure}[t!]
\centering
\includegraphics[clip,trim=0.5cm 0.7cm 0cm 0.2cm,width = 0.95\linewidth]{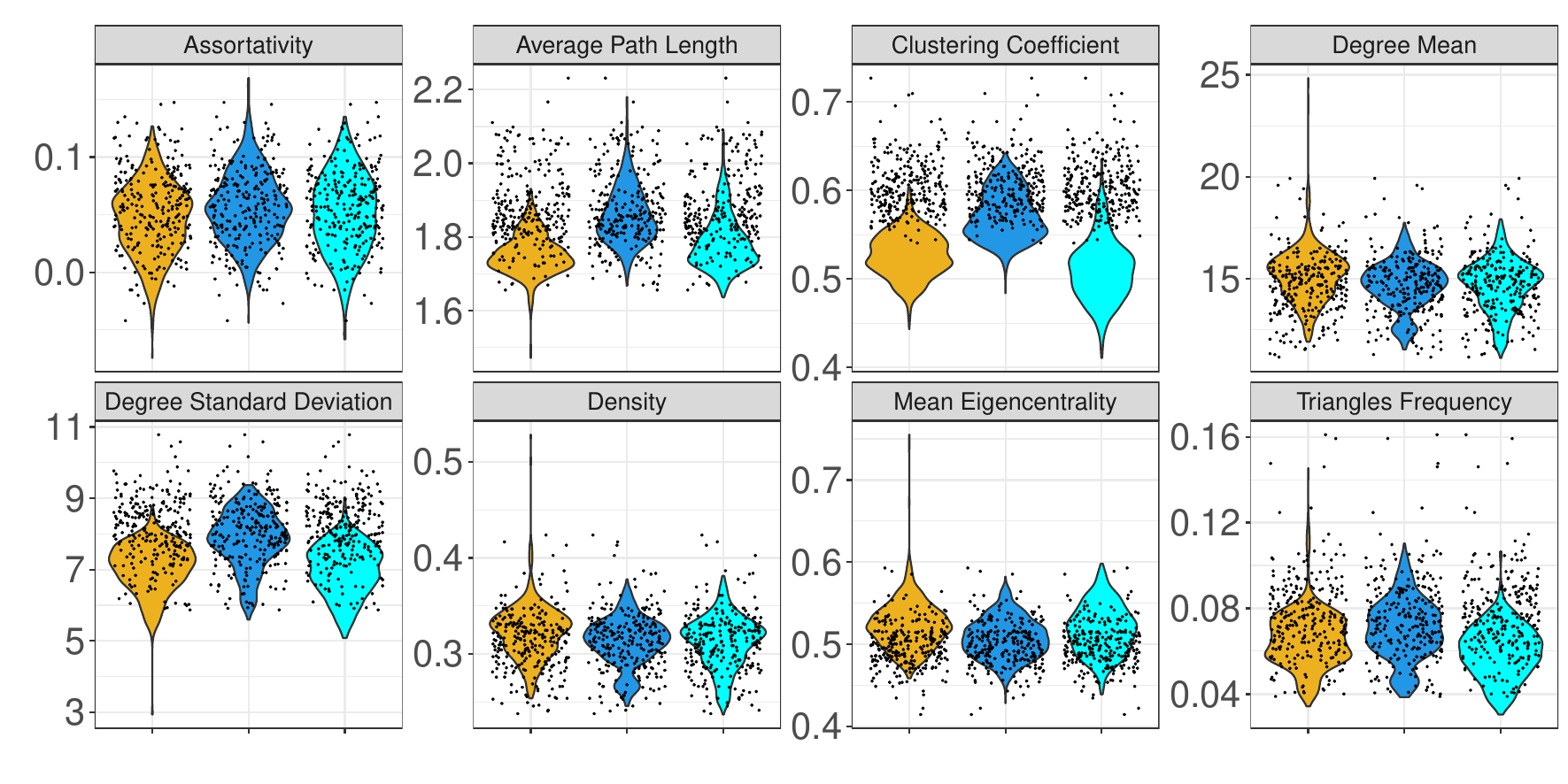}
\caption{\small Posterior predictive checks. Posterior predictive distribution for selected network summary measures, for our method (yellow violins) and the methods of \citet{durante2017} (blue violins) and \citet{anastasia} (cyan violins).
Jittered dots represent the network summary measures computed for the observed brain network data. See \Autoref{sec:section5}.
}\label{fig:post_pred_checks_young}
\end{figure}

We analyze the popular HNU1 human brain dataset, publicly available at \url{https://networks.skewed.de/net/human_brains} \citep{brain_repo}, from the Consortium for Reliability and Reproducibility (CoRR) repository \citep{Zuo}. Connectivity patterns across different brain regions were measured for 30 healthy individuals at rest. Up to 10 measurements per individual were taken using diffusion magnetic resonance imaging (dMRI) over a month, totaling $n=266$ network observations. 
These measurements are represented as labeled networks with $N=48$ nodes corresponding to fixed brain regions of interest (ROI), defined by the JHU ICBM DTI-81 atlas \citep{MORI2005}, and edges denoting connections among these regions. Two regions are considered connected if at least one white matter fiber links them. Importantly, fiber-tracking pipelines are subject to measurement errors. \Autoref{fig:intro_3} displays a sample of six observations from this dataset. The same dataset, though with different node granularity, has been discussed by \cite{Zuo}, \cite{Arroyo2021}, \cite{CERCER}, and \cite{anastasia}. The latter three studies analyze the data from a modeling perspective. \cite{Arroyo2021} investigate their method's ability to identify individual differences based on network communities. \cite{CERCER} assume unimodality in the network generation process and infer a representative network for the population, while \cite{anastasia} focus on detecting outlier networks.
\noindent Our analysis aims to characterize differences in brain connectivity between subjects in the dataset. 
We compare our model's results with those obtained using the methods of \citet{durante2017} and \cite{anastasia}.
We do not include the method of \cite{Signorelli} here, as the code made available by the authors restricts mixtures to a maximum of $7$ components, making it unsuitable for the current problem.
Methods are first compared using posterior predictive checks, to assess their ability to recover the generative mechanism underlying the observed graphs for selected network summary measures. We simulate networks from the posterior predictive distribution, 
which for our model is given in Equation S21 in the Supplementary Material \citep{DPM_CER_supp},  
and compute network summary measures for these simulations. If the model lacks flexibility, we expect the observed data's network measures to fall in the tails of their corresponding posterior predictive distributions. 
\autoref{fig:post_pred_checks_young} shows that the DP mixture of CER kernels, the method of \citet{durante2017}, and the approach of \citet{anastasia} are all sufficiently flexible to capture the variability of functionals of the posterior predictive distribution. Some network summary measures, however, such as transitivity and average path length, indicate a comparatively better fit for the model proposed by \citet{durante2017}.

Since edges represent physical connections between ROIs, the partition of observations based on the 30 subjects in the study can serve as a proxy for the true clustering structure of the data. We thus investigate whether brain scans of the same subject tend to be assigned to the same cluster and thus can be considered similar, a relevant question for researchers in neuroscience.
We do this by comparing the estimated partition of the sample of network data, with the partition implied by the presence of 30 subjects. Clustering metrics reported in \autoref{tab:table_cls} clearly indicate that our model detects similarities among the brain scans of the same individual, further validating its effectiveness in consistently clustering networks. Our model outperforms the methods of \cite{durante2017} and \cite{anastasia} in terms of clustering accuracy. 
\begin{table}[b!]
\centering
\begin{tabular}{l c c c c }
Model & $\hat{K}$ & Adjusted Rand Index & Entropy & Purity    \\
\hline
\hline
DPM-CER & 50 & 0.8065 & 0.0065 & 0.9925 \\
Durante et al. & 23 & 0.6822 & 0.1418 & 0.7143  \\
Mantziou et al. & 57 & 0.7508 & 0.0278 & 0.9511 \\
\end{tabular}
\caption{\small Clustering results of the Human Brain data set in terms of number of inferred clusters and clustering metrics with respect to the natural partition implied by individuals.  Findings are compared with the models of \cite{durante2017} and \cite{anastasia}. See \Autoref{sec:section5}. 
}
\label{tab:table_cls}
\end{table} 
The optimal partition identified by our model consists of 50 clusters, thus exceeding the number of subjects in the study. Only two clusters contain networks from different individuals, while for four subjects, the networks corresponding to the same subject are distributed across multiple clusters. These results hold potential biological significance, offering valuable insights for further investigation. \\
We conclude our analysis by evaluating whether the clusters identified by our model display features with neuroscientific interpretability. A similar question is addressed by \citet{anastasia}, who identify a subgroup of individuals with brain connectivity patterns distinct from the majority, based on network summary measures of interest to neuroscientists.
We focus on average path length and clustering coefficient, two metrics of neuroscientific significance as human brains are known to typically exhibit a small-world structure, characterized by short average path lengths and high clustering coefficients \citep{Bassett}. 

\begin{figure}[t!]
\centering
\includegraphics
[trim={6.8cm 0cm 6.8cm 0cm},clip,width=0.31\linewidth]{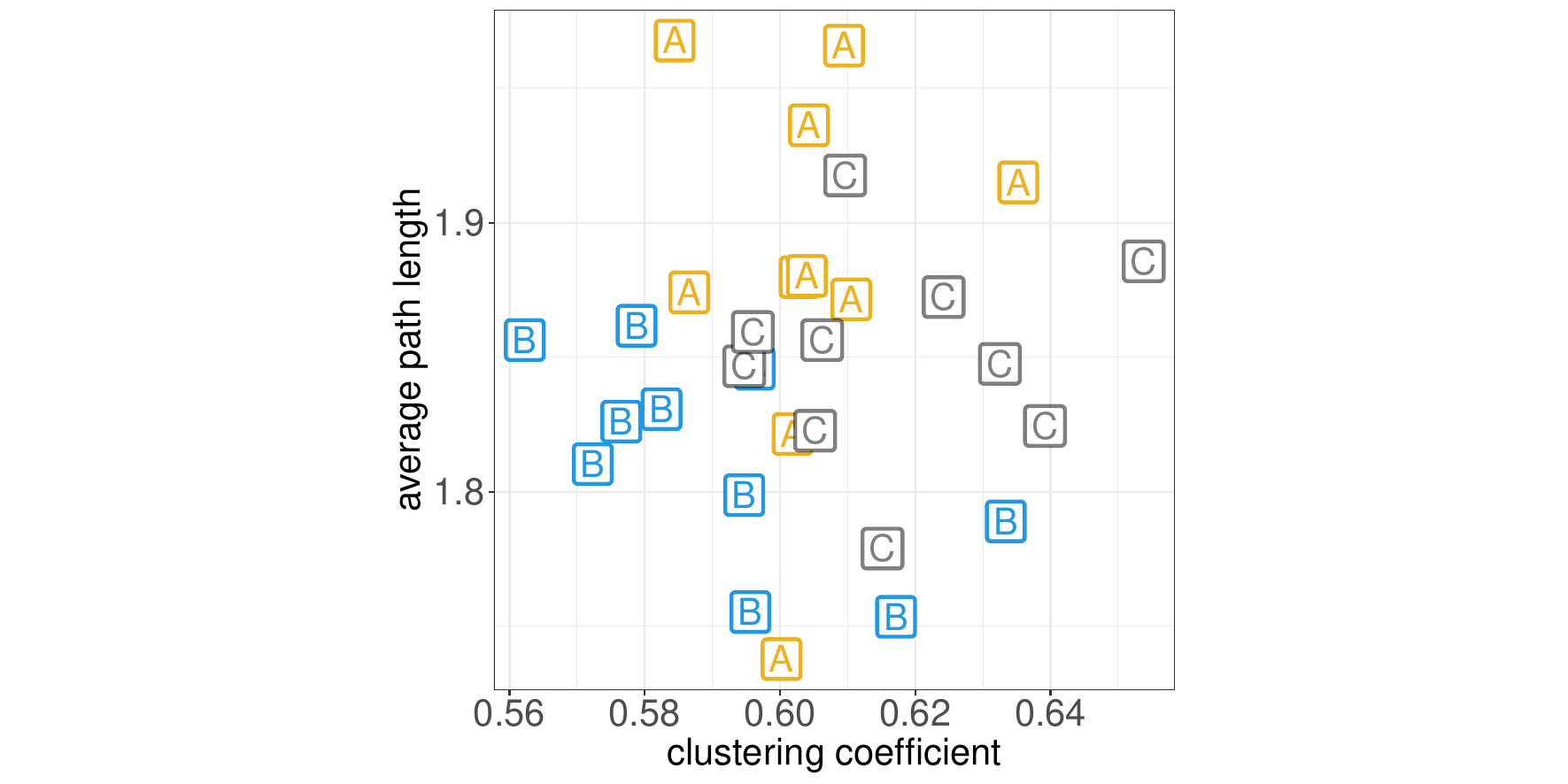}
\hspace{0.2cm}
\includegraphics[trim={6.8cm 0cm 6.8cm 0cm},clip,width=0.31\linewidth]{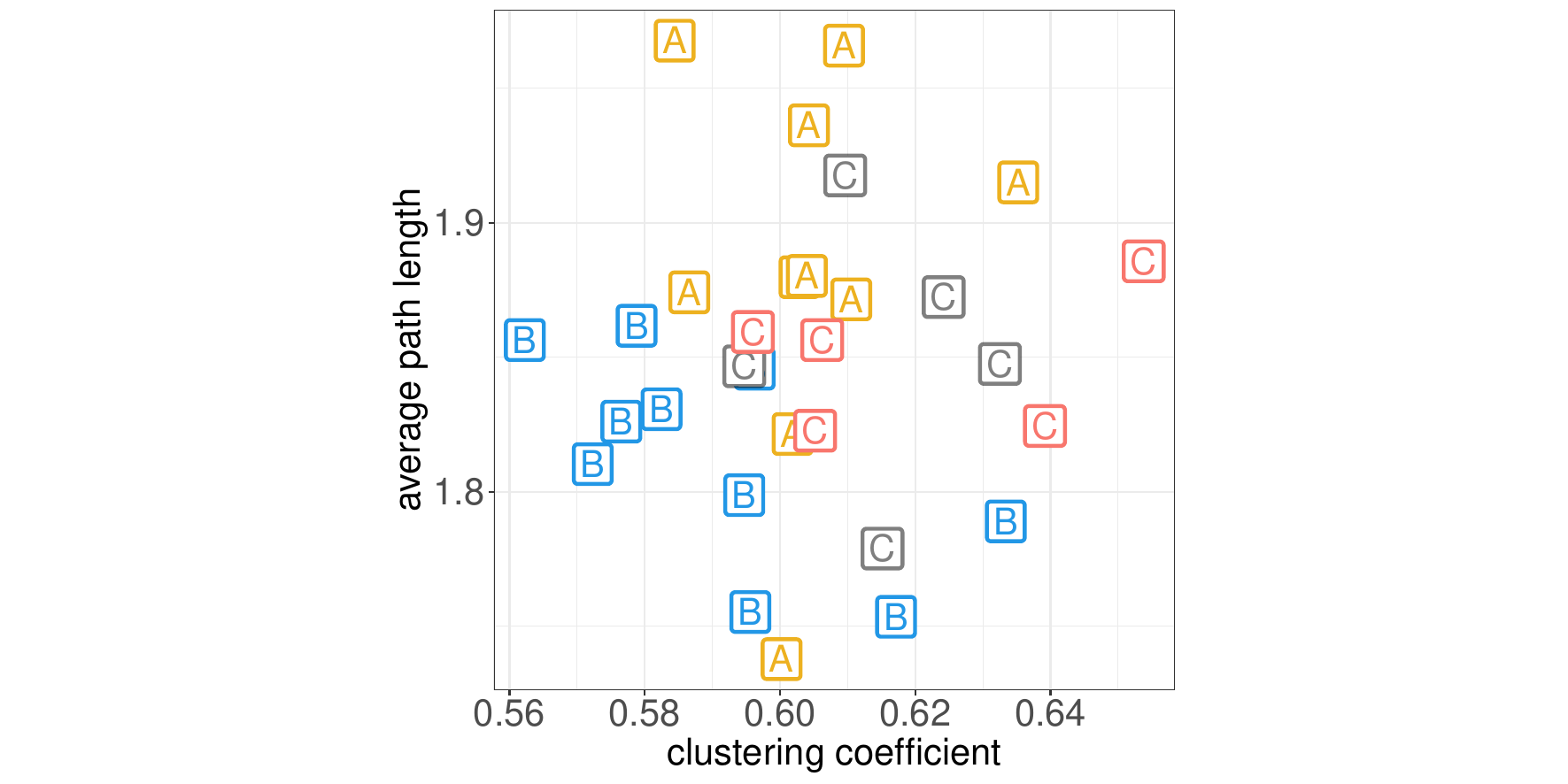}\hspace{0.2cm}
\includegraphics[trim={6.8cm 0cm 6.8cm 0cm},clip,width=0.31\linewidth]{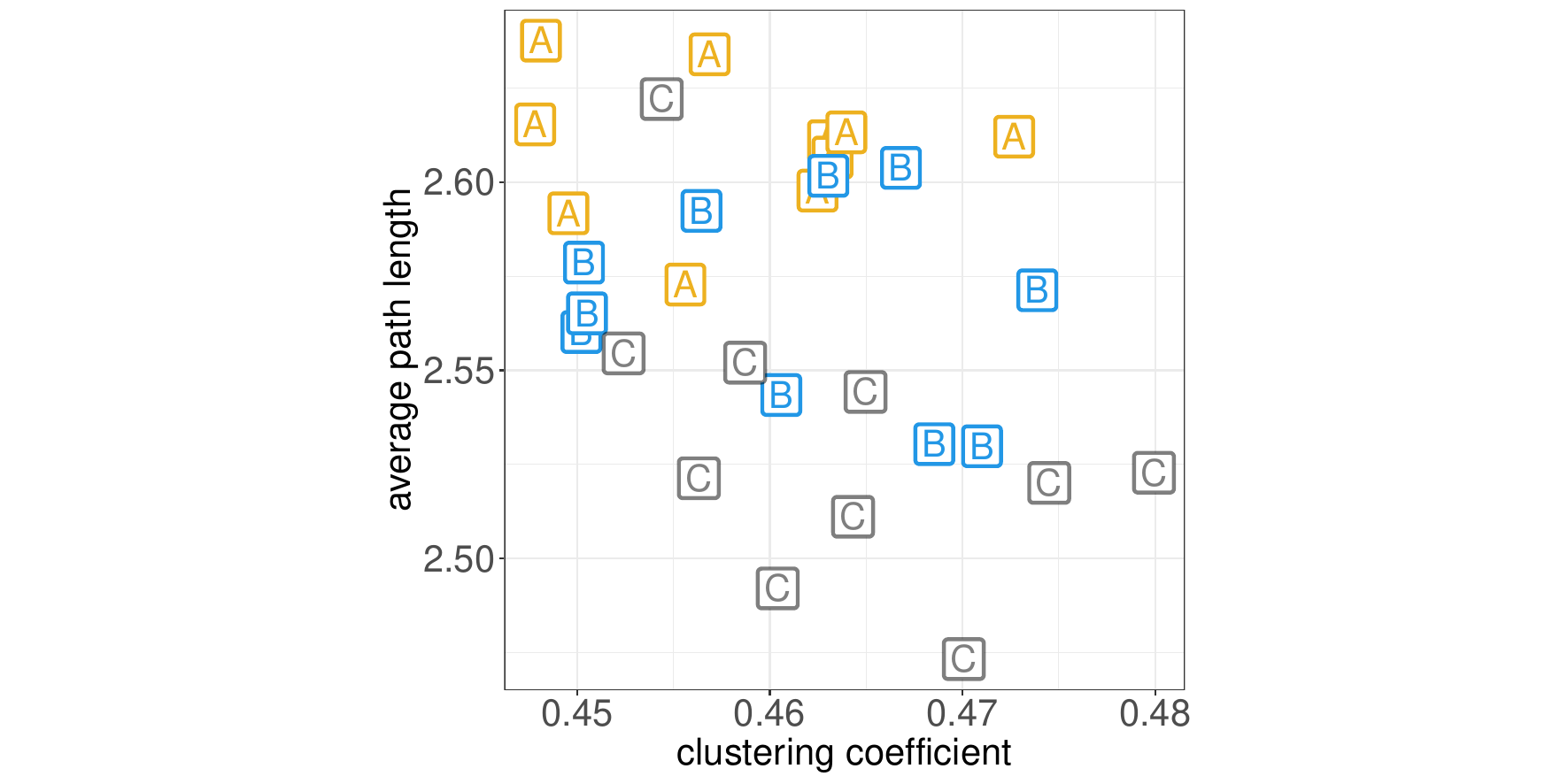}
\caption{\small  Scatter plots for the small-world properties of brain networks for three subjects in the dataset. Colors indicate the cluster membership, letters refer to the subject ID in the dataset, namely 0025443 (A), 0025445 (B) and 0025446 (C). Left panel: $N=48$, partition estimated via DP mixture of CER kernels (see \Autoref{sec:section5}). Central and right panels: $N=48$ and $N=200$ respectively, partition estimated via consensus subgraph clustering (see \Autoref{sec:section6}). 
}\label{fig:human_brain}
\end{figure}

\noindent The left panel of \autoref{fig:human_brain} shows the average shortest path length and clustering coefficient for the brain networks, for the three largest clusters of the estimated partition. The three clusters have size $10$ and coincide with the scans of three subjects in the dataset. There is a clear distinction between the brain scans in each cluster with respect to these two network properties. This confirms our model's ability to identify differences among data characterized by topological structures indicative of small-world behavior.

\section{Consensus subgraph clustering for large networks} 
\label{sec:section6}
Probabilistic models for network data often become computationally infeasible when the number of nodes is large \citep{large_network}. This issue is exacerbated when dealing with multiple network data. Our method, for instance, becomes computationally intensive as $N$ grows, requiring numerical evaluations that significantly slow down \Autoref{alg:gibbs}.
In addition, the computational burden increases with $n$. 
To address the challenge of clustering elements within a population of networks when the number of nodes is large, we propose a heuristic approach inspired by consensus clustering techniques \citep{consensus}, which we call \emph{consensus subgraph clustering}. Similar to variational methods, our approach breaks down some dependencies between nodes and measures similarity among networks based on their local characteristics. This is achieved by running our model-based clustering method in parallel on subgraph observations. Subgraphs are created by 
partitioning the $N$ nodes into 
blocks of at most $N_{\text{sub}}$ nodes. 
This step is akin to assuming a block structure at the vertex level, with block memberships possibly 
assigned based on available information, e.g. spatial, on the nodes. Each Gibbs sampler produces a sample of partitions of the multiple network data from the posterior distribution of the model, conditional on the subgraphs obtained by restricting the original data to specific node blocks. The subgraph-specific posterior samples are then pooled into a unique sample, from which we identify a representative partition by minimizing the posterior expected Variation of Information. We illustrate this strategy through the analysis of 
a version of the human brain network data analyzed in \Autoref{sec:section5}, with $N=200$ and thus characterized by finer node granularity. 

\subsection{Brain network data with finer granularity}\label{sec:brain_finer}
Constructed from the same $n=266$ dMRI scans of 30 healthy individuals already analyzed in \Autoref{sec:section5}, the dataset we consider is based on the CC200 human brain atlas \citep{atlas200N}, which includes $N=200$ ROIs, and is available in the same repository. 
Compared to the 48 ROIs considered in \Autoref{sec:section5}, this version presents a substantially higher dimensionality, posing a greater computational challenge. 
We implement the described consensus subgraph clustering approach and start by investigating the effect of $N_{\text{sub}}$ on the estimated data partition. We consider subgraphs defined on mutually exclusive vertex sets with cardinality at most equal to $N_{\text{sub}} \in \{5, 10, 15, 20, 25, 30, 35, 40, 50\}$, which leads, respectively, to $m_{\text{sub}} \in \{40, 20, 14, 10,  8, 7,  6,\allowbreak   5,  4\}$ distinct datasets of multiple network data of dimension at most $N_{\text{sub}}$. 
Nodes are divided into 
$m_{\text{sub}}$ blocks based on the physical distance between ROIs, using a balanced clustering technique \citep{bal_clust}, as implemented in the \texttt{anticlust} R package \citep{bal_clust_2}.
This approach reduces dependencies between the most distant nodes, as illustrated in Figure S8 in the Supplementary Material \citep{DPM_CER_supp}.
For each subgraph we thus have $n=266$ observations, corresponding to the restriction of the original network data to a subset of nodes. Conditionally on each sample of subgraphs, we run \Autoref{alg:gibbs} for $1,200$ iterations, of which the first $200$ are discarded as burn-in.

\begin{figure}[b!]
\centering
\includegraphics[trim={0.5cm 4cm 0cm 4cm},clip,width=1\linewidth]
{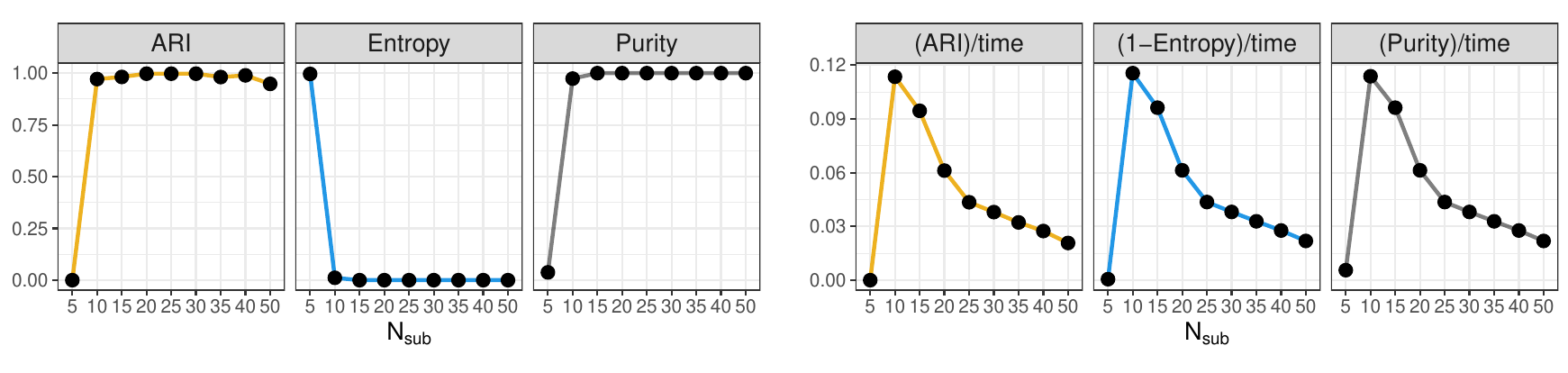}\caption{\small Clustering metrics comparing the partition estimated based on the consensus subgraph approach, with the one 
implied by the 30 individuals in the study, for the Human Brain dataset based on $200$ ROIs, for $N_{sub}$ ranging in $\{5,10,\ldots,50\}$. See \Autoref{sec:brain_finer}. 
}\label{fig:large_N_onerow}
\end{figure}

\noindent As in \Autoref{sec:section5}, we assess the accuracy of the estimated partition by comparing it to the one \textcolor{blue}implied by the presence of 30 subjects in the study. 
The three panels on the left of \autoref{fig:large_N_onerow} show the values of three summary metrics for the considered values of $N_{\text{sub}}$. The consensus subgraph clustering approach struggles to correctly discriminate individual brain scans based on local characteristics when subgraphs with $5$ nodes are used. However, when the analysis is conducted using subgraphs with at least  $10$ nodes, the method successfully captures the heterogeneity inherent in the brain networks of the 30 individuals. This experiment suggests that, as long as the subgraphs are not too small, examining local differences in connectivity patterns of subregions of the brain may be sufficient to detect overall similarities and differences across brain network data.
Selecting an optimal $N_{\text{sub}}$ involves balancing accuracy and computational time. The three right panels of \autoref{fig:large_N_onerow} display the ratios of the three clustering summary metrics already considered, relative to the computational time. To improve interpretability, we used 
$1-
\text{Entropy}$ instead of rescaling the \text{Entropy} directly. For each value of $N_{\text{sub}}$, the computational time is defined as the maximum time taken to analyze any of the  $m_{\text{sub}}$ datasets. Since these datasets can be analyzed in parallel, this definition of computational time represents the total time needed by a machine with unlimited cores.  
It is apparent that, for $N_{\text{sub}} \ge 15$,  the extra computational cost is not rewarded in terms of accuracy.
Therefore, it seems reasonable to select a value for $N_{\text{sub}}$ by looking at where the time-rescaled clustering metrics are maximized, which for this dataset is $N_{\text{sub}}=10$.  
For comparison, we conducted a similar study on the effectiveness of the consensus subgraph clustering approach using the Human Brain dataset based on 48 ROIs, dataset for which we can compare the approximate posterior distribution with the exact one, as studied in \Autoref{sec:section5}. The results, displayed in Figure S9 in the Supplementary Material \citep{DPM_CER_supp},
lead to selecting $N_{\text{sub}} = 15$. 

\noindent \autoref{tab:table_cls_largeN} compares the results of the consensus subgraph clustering method applied to the human brain datasets with 48 and 200 ROIs,
where $N_{\text{sub}}$ was set equal to $15$ and $10$ respectively, and the summary metrics computed by comparing the estimated partition with that one implied by the presence of 30 subjects in the study. It can be appreciated that the summary metrics indicate a better performance of our method when analyzing the dataset with finer granularity. The results referring to the case $N=48$ 
appear slightly worse than those obtained by applying our model on the entire graph observations, presented in \autoref{tab:table_cls}. While it is clear that exploring only local properties of the graphs might reduce the ability to detect global properties of the graphs, 
these results indicate that the consensus subgraph clustering approach might be considered a valid alternative to cluster multiple network data when the number of nodes is large.
Finally and for simplicity of illustration, we focus on the cluster allocation of the 30 observations referring to the three subjects assigned to the three largest clusters in the analysis run in \Autoref{sec:section5}. The central and right panels of \autoref{fig:human_brain} display the cluster allocation of these 30 observations, obtained by resorting to the consensus subgraph clustering approach to analyze the human brain datasets with $N=48$ and $N=200$, respectively, and highlight the topological properties of the identified clusters. 
Cluster allocation for the two cases resembles the results obtained in \Autoref{sec:section5} when analyzing the complete dataset with \(N=48\) ROIs, as shown in the left panel of \autoref{fig:human_brain}. The only notable difference is that the consensus subgraph clustering applied to the dataset with 48 ROIs separates the scans of the subject labeled \say{C} into two distinct clusters. 
\begin{table}[t!]
\centering
\begin{tabular}{c c c c c }
$N$ ($N_{\text{sub}}$) & $\hat{K}$ & Adjusted Rand Index & Entropy & Purity \\
\hline\hline
$48\; (15)$ & 34 & 0.6710 &  0.1407 & 0.7932 \\
$200 \; (10)$ & 30 & 0.9714 & 0.0115 & 0.9737  \\
\end{tabular}
\caption{\small Human brain dataset. Estimated number of clusters and clustering metrics with respect to the partition implied by the presence of 30 individuals in the study. See \Autoref{sec:brain_finer}.
}
\label{tab:table_cls_largeN}
\end{table}

\section{Discussion}\label{sec:section7}
We introduced a novel Bayesian nonparametric approach to model heterogeneous populations of networks. The model's location-scale structure favors interpretability while offering appealing theoretical properties, such as full support in the space of labeled graphs and posterior consistency. A key feature of our approach is that the proposed algorithm samples from distributions that are available in closed form. These distributions are derived using standard combinatorial arguments, which is made possible by the use of the Hamming distance to detect structural similarities among networks. 
As shown in \Autoref{sec:section4} and the brain network data analysis in \Autoref{sec:section5}, our model offers notable flexibility and demonstrates overall performance improvements over existing methods in the literature.
The implementation of the model results in a per-iteration computational cost that is quadratic in the number of nodes $N$. 
As a result, implementing our model can become computationally intensive when large values for $N$ are considered. To address this challenge, we proposed a heuristic approach, named consensus subgraph clustering, designed to handle large graphs
efficiently. Our analysis of human brain data with finer node granularity, as presented in \Autoref{sec:section6}, demonstrates that this method performs well with larger networks and shows promise for scaling to the analysis of massive networks. Interestingly, while the method was presented using the spatial locations of the ROIs represented by network nodes, promising results were also obtained when the nodes were randomly partitioned, disregarding the available spatial information, as illustrated in the Supplementary Material \citep{DPM_CER_supp}.
In addition, alternative strategies for node partitioning could be explored to implement the preliminary step of the consensus subgraph clustering approach. For instance, the concept of signal-subgraphs introduced by \citet{vog12} provides a promising option.\\
The modeling strategy we presented offers multiple avenues for extension, opening up new research directions. An intriguing one involves exploring alternatives to the Hamming distance. Although the Hamming distance stands out for its tractability, it falls short in capturing the broader structural changes within a graph \citep{donnat2018tracking}. For instance, in brain networks, spectral distances can better assess global changes in connectivity. A natural extension of our work in this direction would be to employ the diffusion distance, which is based on the graph Laplacian and treats networks in a functional manner, focusing on changes that affect the global structure \citep{CERCER}. This approach, however, may not admit closed-form expressions, thus requiring alternative computational strategies. In this context, a particularly interesting research question is whether posterior consistency holds for kernels based on distances other than the Hamming distance, and how the choice of kernel affects the corresponding rates of convergence. Another interesting direction is to explore the possibility of introducing a stochastic block model structure through an appropriate specification of the base measure \(P_0\), thereby extending the scope of our model to allow for the simultaneous clustering of networks and nodes within each network. 
Finally, our model could serve as the building block for modeling related populations of networks in a partially exchangeable setting, in line with the approach of \citet{Durante_Gaffi2025}, first proposed in \cite{gaffi2023phd}.
This could be achieved by inducing dependence at the level of population-specific mixing measures, such as through the Dependent Dirichlet process \citep{Mac00}, thereby facilitating the sharing of information across models \citep[see][for a review]{Qui22}.


\section*{Acknowledgments}
The authors are grateful to the Associate Editor and the two anonymous Referees for their valuable comments and suggestions. Bernardo Nipoti gratefully acknowledges the Department of Statistics at ITAM, Mexico City, for its hospitality during his visit, when this work was initiated.

\section*{Funding}
Bernardo Nipoti acknowledges support of MUR – Prin 2022 – Grant no.
2022CLTYP4, funded by the European Union – Next Generation EU.




\end{bibunit}

\clearpage

\begin{bibunit}

\bigskip
\section*{}\label{sec:supp_mat}
\begin{center}
{\huge
\bf Supplementary Material for \\ \say{Bayesian nonparametric modeling of \\[0.5em] heterogeneous populations of networks}}
\end{center}

\allowdisplaybreaks

The Supplementary Material is organized as follows.
In \Autoref{sec:proof} we provide the proof of the results in Section 2.3 of the main paper. \Autoref{app:post_comp} provides additional details on posterior computations, including strategies to enhance the algorithm’s efficiency.
\Autoref{app:prediction} provides a closed-form expression for the cluster-specific one-step-ahead posterior predictive distribution, conditionally on an estimated partition of the sample.
This result is generalized to make a joint prediction on $m$ graphs, assuming that they all belong to a specific cluster of the estimated partition. In addition, the cluster-specific posterior distribution of $\mathcal{C}_k^{*}$ is also provided in closed-form. Finally, \Autoref{app:simulation}, \Autoref{app:data_analysis} and \Autoref{app:largeN} present further information on the simulation studies and the illustrations in Section 4 Section 5 and Section 6 of the main paper.

\section{Proofs of Theorem 2.1 and Corollary 2.1}\label{sec:proof}
\begin{proof}[Proof of Theorem 2.1]
$\mathscr{G}_{\mathcal{V}}$ consists of $\left|\mathscr{G}_{\mathcal{V}}\right|=2^{M}$ possible network configurations, with $M=\binom{N}{2}$. We can then name the elements of $\mathscr{G}_{\mathcal{V}}$ as $\mathscr{G}_{\mathcal{V}}=\left\{ 
\mathcal{H}_1, \mathcal{H}_2, \ldots, \mathcal{H}_{2^M} 
\right\}$. 
We observe that any probability mass function $p_* \in \mathcal{P}_{\mathscr{G}_{\mathcal{V}}}$ is characterized by a set of $2^M$ weights $p_{0l}=p_*(\mathcal{H}_l)$, with $l=1,\ldots,2^M$, as
$$p_*(\cdot)=\sum_{l=1}^{2^M} p_{0l}\delta_{\mathcal{H}_l}(\cdot).$$ 
By exploiting the stick-breaking representation of the DP  \citep{Sethuraman1994}, we can rewrite $\tilde{f}$ as $$\tilde{f}(\cdot)=\sum_{j=1}^\infty
\tilde{p}_j \psi( \cdot ; \tilde{\vartheta}_j  ), 
$$
where $\tilde{\vartheta}_j= ( \tilde{\mathcal{C}}_j, \tilde{\alpha}_j ) \stackrel{\text { iid }}{\sim} P_{0}$ and the $\tilde{p}_j\text{'s}$
are positive weights with Griffiths-Engen-McCloskey distribution with parameter $c$ \citep[see, e.g.,][]{Ewe90}, such that $\sum_{j=1}^\infty
\tilde{p}_j = 1$ almost surely. We observe that, as done for $p_*$, also $\tilde{f}$ can be written as a finite sum, that is 
$$\tilde{f}(\cdot)=\sum_{l=1}^{2^M}
\tilde{q}_l \delta_{\mathcal{H}_l}(\cdot).
$$
For any $\omega\in\Omega$, we henceforth use the superscript $(\omega)$ to denote a realization of a random variable, e.g. $\tilde{f}^{(\omega)}$.

\noindent The remainder of the proof is organized as follows. For any $p_* \in \mathcal{P}_{\mathscr{G}_{N}}$ and any $\epsilon > 0$:  
\begin{itemize}  
    \item[Part 1.]  We define a set of conditions and show that, if a realization $\tilde{f}^{(\omega)}$ of $\tilde{f}$ satisfies these conditions, then $\tilde{f}^{(\omega)}$ belongs to $\mathbb{B}_\epsilon(p_*)$, the Kullback--Leibler neighborhood of $p_*$ of radius $\epsilon$. That is, $\text{KL}(p_*;\tilde{f}^{(\omega)}) \leq \epsilon$.  
    \item[Part 2.]  We define the events $A_1,A_2\subseteq \mathcal{P}_{\mathscr{G}_N}$ as
    \begin{align*}
        A_1&=\{\tilde{f}^{(\omega)}\text{ with }\omega\in\Omega\,:\, \text{\texttt{b1}, \texttt{b3} hold for $(\alpha_\star,\eta_\star)$ satisfying \texttt{a1}, \texttt{a2}}\},\\
        A_2&=\{\tilde{f}^{(\omega)}\text{ with }\omega\in\Omega\,:\, \text{\texttt{b2} holds for $(\alpha_\star,\eta_\star)$ satisfying \texttt{a1}, \texttt{a2}}\},
    \end{align*}
    and, by exploiting the result in Part 1, we show that 
$$\mathbb{B}_\epsilon(p_*)\supseteq A_1\cap A_2.$$
    \item[Part 3.] We show that $\Pi$ assigns positive probability to $A_1\cap A_2$ and, given the result in Part 2, to $\mathbb{B}_\epsilon(p_*)$.
\end{itemize}
\noindent \textbf{Part 1.} For any $\varepsilon>0$, we consider $(\alpha_\star,\eta_\star)$ such that 
\begin{enumerate}
    \item[\texttt{a1})] $0<\alpha_{\star}< 1-\exp\{-\varepsilon/M\}$;
    \item[\texttt{a2})] $0<\eta_{\star}< 1-\exp\{-\varepsilon\}/(1-\alpha_\star)^M$.
\end{enumerate}
Given $p_*\in \mathcal{P}_{\mathscr{G}_{N}}$, we let $\tilde{f}^{(\omega)}$ 
be such that, for any $l=1,\ldots,2^M$,
\begin{enumerate}
    \item[\texttt{b1})] $\tilde{\mathcal{C}}_l^{(\omega)}=\mathcal{H}_l$;
    \item[\texttt{b2})] $\tilde{p}_l^{(\omega)}\in[p_{0l}(1-\eta_\star),p_{0l}]$;
    \item[\texttt{b3})] $\tilde{\alpha}_l^{(\omega)}\in(0,\alpha_\star]$.
\end{enumerate} 
We observe that condition \texttt{a1} guarantees that the set of solutions $\{\eta_\star \, :\,\text{\texttt{a2} holds}\}$ is not empty. We next show that if $\tilde{f}^{(\omega)}$ satisfies \texttt{b1}, \texttt{b2}, and \texttt{b3}, then $\text{KL}(p_* ;\tilde{f}^{(\omega)})\leq \varepsilon$. 
We first observe that, for any $l=1,\ldots, 2^M$,

\begin{equation}\label{eq:ub_pi}
\tilde{q}_{l}^{(\omega)}\stackrel{\texttt{b1}}{\geq} \tilde{p}_l^{(\omega)}(1-\tilde{\alpha}^{(\omega)}_l)^M.
\end{equation}
Then, it follows that
\begin{align*}
    \text{KL}(p_*;\tilde{f}^{(\omega)})&=\sum_{l=1}^{2^M} p_{0l}\log\left(\frac{p_{0l}}{\tilde{q}_l^{(\omega)}}\right)\\
    &\stackrel{\eqref{eq:ub_pi}}{\leq} \sum_{l=1}^{2^M} p_{0l}\log\left(\frac{p_{0l}}{\tilde{p}_l^{(\omega)}(1-\tilde{\alpha}^{(\omega)}_l)^M}\right)\\
    &\stackrel{\texttt{b2}}{\leq} \sum_{l=1}^{2^M}p_{0l}\log\left(\frac{1}{(1-\eta_\star)(1-\tilde{\alpha}^{(\omega)}_l)^M}\right)\\
    &\stackrel{\texttt{b3}}{\leq} \sum_{l=1}^{2^M}
    p_{0l}\log\left(\frac{1}{(1-\eta_\star)(1-\alpha_\star)^M}\right)\\
    &=\log\left(\frac{1}{(1-\eta_\star)(1-\alpha_\star)^M}\right)\stackrel{\texttt{a2}}{\leq}\varepsilon.
\end{align*}
\textbf{Part 2.} We observe that
\begin{align*}
\mathbb{B}_{\varepsilon}(p_*) \supseteq& \{\tilde{f}^{(\omega)}\text{ with }\omega\in\Omega\,:\,\text{\texttt{b1}, \texttt{b2}, \texttt{b3} hold for $(\alpha_\star,\eta_\star)$ satisfying \texttt{a1}, \texttt{a2}}\}\\
=&\{\tilde{f}^{(\omega)}\text{ with }\omega\in\Omega\,:\, \text{\texttt{b1}, \texttt{b3} hold for $(\alpha_\star,\eta_\star)$ satisfying \texttt{a1}, \texttt{a2}}\}\\
&\cap \{\tilde{f}^{(\omega)}\text{ with }\omega\in\Omega\,:\, \text{\texttt{b2} holds for $(\alpha_\star,\eta_\star)$ satisfying \texttt{a1}, \texttt{a2}}\}=A_1\cap A_2.
\end{align*}
\textbf{Part 3.} The proof is completed by showing that, for any $\varepsilon>0$, $\Pi$ assigns positive probability to 
$\mathbb{B}_{\varepsilon}(p_*)$. 
Given the independence of weights $(\tilde{p}_j)_{j\geq 1}$ and  atoms $(\tilde{\vartheta}_j)_{j\geq1}$ in the definition of $\tilde{f}$, the events $A_1$ and $A_2$ are disjoint and thus $\Pi(A_1\cap A_2)=\Pi(A_1)\Pi(A_2)$. To prove that $\Pi(\mathbb{B}_{\varepsilon}(p_*))>0$, it then suffices to check that both $\Pi(A_1)$ and $\Pi(A_2)$ are positive. $\Pi(A_1)>0$ follows from the fact that $P_0$ has full support on $\Theta$. 
Moreover, $\Pi(A_2)>0$ as, for any $j=1,2,\ldots$, the distribution of $\tilde{p}_j$ for the DP has full support on $[0,1-\sum_{i=1}^{j-1}\tilde{p}_i]$. 
\end{proof}

\begin{proof}[Proof of Corollary 2.1]
A direct application of Example 6.21 in \citet{ghosal_van_der_vaart_2017}.
\end{proof}

\section{Derivation of posterior computations}\label{app:post_comp}
\subsection{Probability of a new value in the generalized Pólya urn scheme}\label{subsec:prob_new_polya}
We provide a detailed derivation of Equation 6 from the main paper, which gives the probability $\pi_{l0}$ of sampling a new pair $\vartheta_l = \left( \mathcal{C}_l, \alpha_l \right)$ in the generalized Pólya urn scheme. Specifically,
\begin{align}\label{eq:integral}
\pi_{l0} &  \propto c
\int \psi\left(\mathcal{G}_l; \vartheta_l \right) \mathrm{d} P_{0}(\vartheta_l)
\notag \\
& \propto
c \int_{0}^{1/2} \sum_{\mathcal{C}_l \in \mathscr{G}_{\mathcal{V}} } \alpha_l^{d_\text{H}\left(\mathcal{G}_l, \mathcal{C}_l\right)}(1-\alpha_l)^{M-d_\text{H}\left(\mathcal{G}_l, \mathcal{C}_l\right)} 
p_{\text{CER}}\left(\mathcal{C}_l ; \mathcal{G}_0, \alpha_l \right)  f_{\text{TBeta}}(\alpha_l;1/2,a,b) \mathrm{d}\alpha_l
\notag \\
& \propto c
 \int_{0}^{1/2}  f_{\text{TBeta}}(\alpha_l;1/2,a,b)  \times \\
  &\qquad \times \sum_{\mathcal{C}_l \in \mathscr{G}_{\mathcal{V}} }   \alpha_l^{d_\text{H}\left(\mathcal{G}_l, \mathcal{C}_l\right)} (1-\alpha_l)^{M-d_\text{H}\left(\mathcal{G}_l, \mathcal{C}_l\right)}  \alpha^{d_\text{H}\left(\mathcal{C}_l, \mathcal{G}_0\right)}(1-\alpha_l)^{M-d_\text{H}\left(\mathcal{C}_l, \mathcal{G}_0\right)} \, \mathrm{d}\alpha_l \notag \\
 & \propto c
 \int_{0}^{1/2}  f_{\text{TBeta}}(\alpha_l;1/2,a,b)  \times\\
  &\qquad \times\sum_{\mathcal{C}_l \in \mathscr{G}_{\mathcal{V}} }   \alpha_l^{d_\text{H}\left(\mathcal{G}_l, \mathcal{C}_l\right)+d_\text{H}\left(\mathcal{C}_l, \mathcal{G}_0\right)}(1-\alpha)^{2M-\left[ d_\text{H}\left(\mathcal{G}_l, \mathcal{C}_l\right) + d_\text{H}\left(\mathcal{C}_l, \mathcal{G}_0\right) \right] }  \, \mathrm{d}\alpha_l \notag \\
 & \propto c
 \int_{0}^{1/2}  f_{\text{TBeta}}(\alpha_l;1/2,a,b)  (1-\alpha_l)^{2M} \sum_{\mathcal{C}_l \in \mathscr{G}_{\mathcal{V}} }   \left(\frac{\alpha_l}{1-\alpha_l}\right)^{d_\text{H}\left(\mathcal{G}_l, \mathcal{C}_l\right)+d_\text{H}\left(\mathcal{C}_l, \mathcal{G}_0\right)}  \, \mathrm{d}\alpha_l. 
\end{align}
We note that
\begin{align}\label{eq:sum_2graphs}
    \sum_{\mathcal{C}_l \in \mathscr{G}_{\mathcal{V}} }   \left(\frac{\alpha_l}{1-\alpha_l}\right)^{d_\text{H}\left(\mathcal{G}_l, \mathcal{C}_l\right)+d_\text{H}\left(\mathcal{C}_l, \mathcal{G}_0\right)} = \sum_{h=0}^{2M} w_{lh} \left(\frac{\alpha_l}{1-\alpha_l}\right)^h,
\end{align}
where $w_{lh}$ determines how many graphs $\mathcal{C}_l \in \mathscr{G}_{\mathcal{V}}$ are such that $d_\text{H}\left(\mathcal{G}_l, \mathcal{C}_l\right)+d_\text{H}\left(\mathcal{C}_l, \mathcal{G}_0\right)=h$, for $h=0, 1, \ldots, 2M$, given that  $d_{\text{H}}\left(\mathcal{G}_0, \mathcal{G}_{l}\right)=d_l$. Conveniently, $w_{lh}$ coincides with
\begin{align}\label{eq:coef_int}
    w_{lh}= 
\begin{dcases} 0 \qquad \qquad \qquad \; \text{if} \; h<d_l \\
              0 \qquad \qquad  \qquad \; \text{if} \; h \geq d_l \; \text{and} \; h-d_l \; \text{is} \; \text{odd} \\
              2^{d_l} \binom{M-d_l}{\frac{h-d_l}{2}} \quad \text{if} \; h \geq d_l \; \text{and} \; h-d_l \; \text{is} \; \text{even.} \\
\end{dcases}
\end{align}
We observe that when $h-d_l$ is even, then $d_l \leq h \leq 2M-d_l$. Therefore, armed with \eqref{eq:coef_int} and setting $r=(h-d_l)/2$, the left hand side of \autoref{eq:sum_2graphs} can be written as
\begin{align}\label{eq:sum_2graphs_cont}
     \sum_{\mathcal{C}_l \in \mathscr{G}_{\mathcal{V}} }   \left(\frac{\alpha_l}{1-\alpha_l}\right)^{d_\text{H}\left(\mathcal{G}_l, \mathcal{C}_l\right)+d_\text{H}\left(\mathcal{C}_l, \mathcal{G}_0\right)} 
     & = \sum_{h=d_l}^{2M-d_l} w_{lh} \left(\frac{\alpha_l}{1-\alpha_l}\right)^h \notag \\
     & = \sum_{r=0}^{(M-d_l)} w_{l(2r+d_l)} \left(\frac{\alpha_l}{1-\alpha_l}\right)^{2r+d_l} \notag \\ 
    & = \sum_{r=0}^{(M-d_l)} 2^{d_l} \binom{M-d_l}{r} \left(\frac{\alpha_l}{1-\alpha_l}\right)^{2r+d_l}.
\end{align}
In turn, exploiting \eqref{eq:sum_2graphs_cont}, \autoref{eq:integral} can be rewritten as
\begin{align}\label{eq:integral_cont}
\pi_{l0} & \propto c \int_{0}^{1/2}  f_{\text{TBeta}}(\alpha_l;1/2,a,b) (1-\alpha_l)^{2M} \sum_{\mathcal{C}_l \in \mathscr{G}_{\mathcal{V}} }   \left(\frac{\alpha_l}{1-\alpha_l}\right)^{d_\text{H}\left(\mathcal{G}_l, \mathcal{C}_l\right)+d_\text{H}\left(\mathcal{C}_l, \mathcal{G}_0\right)}  \, \mathrm{d}\alpha_l \notag \\
& \propto c \int_{0}^{1/2}  f_{\text{TBeta}}(\alpha_l;1/2,a,b) (1-\alpha_l)^{2M} \sum_{r=0}^{(M-d_l)} 2^{d_l} \binom{M-d_l}{r} \left(\frac{\alpha_l}{1-\alpha_l}\right)^{2r+d_l} \, \mathrm{d}\alpha_l \notag \\
& \propto c \sum_{r=0}^{(M-d_l)} 2^{d_l} \binom{M-d_l}{r} \int_{0}^{1/2} \frac{\alpha_l^{a-1} (1-\alpha_l)^{b-1}}{\mathcal{B}( 1/2; a, b)} (1-\alpha_l)^{2M} \left(\frac{\alpha_l}{1-\alpha_l}\right)^{2r+d_l} \, \mathrm{d}\alpha_l \notag \\ 
& \propto c \sum_{r=0}^{(M-d_l)} 2^{d_l} \binom{M-d_l}{r} \int_{0}^{1/2} \frac{\alpha_l^{a+2r+d_l-1} (1-\alpha_l)^{b+2(M-r)-d_l-1}}{\mathcal{B}( 1/2; a, b)} \, \mathrm{d}\alpha_l \notag \\ 
& \propto c \sum_{r=0}^{(M-d_l)} w_{lr} 
\frac{\mathcal{B}( 1/2; a_{lr}, b_{lr})}{\mathcal{B}( 1/2; a, b)},
\end{align}
where $w_{lr}=2^{d_l} \binom{M-d_l}{r}$,
$a_{lr}=a+2r+d_l$ and $b_{lr}=b+2M-2r-d_l$.

\subsection{Reshuffling step} 
We provide an explicit derivation of the characterization of the full conditional distribution for each $\vartheta^*_k$, with $k=1,\ldots,K$, given in Equations 11, 12 and 13 of the main manuscript. We start by recalling that we let $\mathcal{D}_k = \{ l \in \{ 1, \dots, n \} : \vartheta_l = \vartheta_k^* \}$ denote the index set of observations belonging to the $k$-th cluster, with $|\mathcal{D}_k| = n_k$, and define $\mathcal{D}^\dagger_k = \mathcal{D}_k \cup \{ 0 \}$. For any index set $\mathcal{D}\subseteq\{0,1,\ldots,n\}$, we denote $\mathcal{G}^{(\mathcal{D})}=\{\mathcal{G}_l:l\in\mathcal{D}\}$, and we let $n_{ij}^{(k)}= \sum_{l \in \mathcal{D}^\dagger_k} A_{\mathcal{G}_l[ij]}$ denote the number of graphs in $\mathcal{G}^{(\mathcal{D}^\dagger_k)}$ that present an edge connecting the nodes $\{i, j\}$. Finally, throughout this section, we use $p(x)$ to denote the distribution of $x$ and $p(x \mid y)$ to denote the conditional distribution of $x$ given $y$.\\
The full conditional distribution of $\vartheta_k^*$ can be factorized as: 
\begin{align}\label{eq:full_cond_theta_star}
p(\vartheta_k^* \mid {\mathcal{G}}^{(\mathcal{D}_k)} )=p(\alpha_k^* \mid {\mathcal{G}}^{(\mathcal{D}_k)} ) p(\mathcal{C}^{*}_k \mid \alpha_k^*, {\mathcal{G}}^{(\mathcal{D}_k)} ).    
\end{align}
To study the distributions appearing in the right-hand side of \eqref{eq:full_cond_theta_star}, it is instructive to start from the joint distribution $p(\alpha_k^*, \mathcal{C}^{*}_k, {\mathcal{G}}^{(\mathcal{D}_k)} )$. Namely, 
\begin{align}\label{eq:joint_acc_step}
p(\alpha_k^*, \mathcal{C}^{*}_k, {\mathcal{G}}^{(\mathcal{D}_k)} ) & = p(\alpha_k^*, \mathcal{C}^{*}_k ) p( {\mathcal{G}}^{(\mathcal{D}_k)} \mid \alpha_k^*, \mathcal{C}^{*}_k) \notag \\
& 
= p(\alpha_k^*) p(\mathcal{C}^{*}_k \mid \alpha_k^*) 
\prod_{l \in \mathcal{D}_k } 
 p( \mathcal{G}_l \mid \alpha_k^*, \mathcal{C}^{*}_k)
\notag \\
& = p(\alpha_k^*) p_{\text{CER}}\left(\mathcal{C}^{*}_k ; \mathcal{G}_0, \alpha_k^* \right)  \prod_{l \in \mathcal{D}_k } 
 p_{\text{CER}}\left(\mathcal{G}_l ; \mathcal{C}^{*}_k, \alpha_k^* \right)
\notag \\ & 
 = p(\alpha_k^*)  
 { \alpha_k^*}^{d_\text{H}\left(\mathcal{G}_0, \mathcal{C}^{*}_k\right)}(1-{\alpha_k^*})^{M-d_\text{H}\left(\mathcal{G}_0, \mathcal{C}^{*}_k \right)}
 \prod_{l \in \mathcal{D}_k } { \alpha_k^*}^{d_\text{H}\left(\mathcal{G}_l, \mathcal{C}^{*}_k \right)}(1-{\alpha_k^*})^{M-d_\text{H}\left(\mathcal{G}_l, \mathcal{C}^{*}_k \right)} 
\notag \\ & 
  = p(\alpha_k^*)  \prod_{l \in \mathcal{D}_k^\dagger } { \alpha_k^*}^{d_\text{H}\left(\mathcal{G}_l, \mathcal{C}^{*}_k \right)}(1-{\alpha_k^*})^{M-d_\text{H}\left(\mathcal{G}_l, \mathcal{C}^{*}_k \right)} \notag \\ & 
  = p(\alpha_k^*) (1-{\alpha_k^*})^{\left(n_k+1\right)M} \left(  \frac{\alpha_k^*}{1-\alpha_k^*} \right)^{ \sum_{l \in \mathcal{D}_k^\dagger  } d_\text{H}\left(\mathcal{G}_l, \mathcal{C}^{*}_k \right)  }.
\end{align}
We next focus on the first distribution in the factorization on the right-hand side of \eqref{eq:full_cond_theta_star}.
\begin{align}\label{eq:alpha_acc_step}
p(\alpha_k^*, {\mathcal{G}}^{(\mathcal{D}_k)} ) & =\sum_{\mathcal{C}^{*}_k \in \mathscr{G}_{\mathcal{V}} }  p(\alpha_k^*, \mathcal{C}^{*}_k, {\mathcal{G}}^{(\mathcal{D}_k)} ) \notag \\ & =  p(\alpha_k^*) (1-{\alpha_k^*})^{\left(n_k+1\right)M} \sum_{\mathcal{C}^{*}_k \in \mathscr{G}_{\mathcal{V}} } \left(  \frac{\alpha_k^*}{1-\alpha_k^*} \right)^{ \sum_{l \in \mathcal{D}_k^\dagger  } d_\text{H}\left(\mathcal{G}_l, \mathcal{C}^{*}_k \right)  }.
\end{align}
We let $U_k= \sum_{l \in \mathcal{D}_k^\dagger  } d_\text{H}\left(\mathcal{G}_l, \mathcal{C}^{*}_k \right)$ and note that
\begin{align}\label{eq:sum_dist}
   U_k = \sum_{l \in \mathcal{D}_k^\dagger  } d_\text{H}\left(\mathcal{G}_l, \mathcal{C}^{*}_k \right) & = \sum_{i=1}^{N-1} \sum_{j=i+1}^{N} \left[ (n_k+1-n_{ij}^{(k)}) A_{\mathcal{C}^{*}_k[ij]} + n_{ij}^{(k)} (1-A_{\mathcal{C}^{*}_k[ij]})   \right] \notag \\ 
    & = \sum_{i<j} \left[ (n_k+1-2n_{ij}^{(k)}) A_{\mathcal{C}^{*}_k[ij]} + n_{ij}^{(k)}   \right],
\end{align}
has support $u_k \in \left\{d_k^*, d_k^*+1, \ldots, D_k^*-1, D_k^* \right\}$,
where 
\begin{align*}
    d_k^*=\sum_{i=1}^{N-1} \sum_{j=i+1}^{N} \min\{n_{ij}^{(k)}, n_k+1-n_{ij}^{(k)}\}, \quad D_k^*=\sum_{i=1}^{N-1} \sum_{j=i+1}^{N} \max\{n_{ij}^{(k)}, n_k+1-n_{ij}^{(k)}\}.
\end{align*}
We let $m_{kh}=\#\left\{ \{ i, j \} \in \mathcal{V}^2 : n^{(k)}_{ij}=h \right\}$ 
denote the number of pairs of distinct nodes that are connected by an edge in exactly $h$ graphs in $\mathcal{G}^{(\mathcal{D}_k^\dagger)}$ and define $M_{kh}=m_{kh} + m_{k(n_k + 1 - h)}$. In addition, we let $\gamma_{kh}\left(s_h\right)=\left( n_k+1-2h  \right) s_h + h M_{kh}$. At this stage, it is worth noticing that \eqref{eq:alpha_acc_step} involves a polynomial in the variable $x_k=\alpha_k^*/(1-\alpha_k^*)$ of the form
\begin{align}
   \mathcal{P}(x_k) & = \sum_{\mathcal{C}^{*}_k \in \mathscr{G}_{\mathcal{V}} }  x_k^{ \sum_{l \in \mathcal{D}_k^\dagger  } d_\text{H}\left(\mathcal{G}_l, \mathcal{C}^{*}_k \right) } = \sum_{\mathcal{C}^{*}_k \in \mathscr{G}_{\mathcal{V}} }  x_k^{ \sum_{i<j} \left[ (n_k+1-2n_{ij}^{(k)}) A_{\mathcal{C}^{*}_k[ij]} + n_{ij}^{(k)}   \right] } \notag \\
   & = \sum_{ A_{\mathcal{C}^{*}_k[12]} \in \{0,1 \}  } x_k^{ (n_k+1-2n_{12}^{(k)}) A_{\mathcal{C}^{*}_k[12]} + n_{12}^{(k)}   } \times
   \dots \times  \notag \\
   & \; \, \sum_{ A_{\mathcal{C}^{*}_k[(N-1)N]} \in \{0,1 \}  }  x_k^{ (n_k+1-2n_{(N-1)N}^{(k)}) A_{\mathcal{C}^{*}_k[(N-1)N]} + n_{(N-1)N}^{(k)}    } \notag \\
    & = \left( x_k^{ n_{12}^{(k)}   } + x_k^{ n_k+1-n_{12}^{(k)}  } \right) \times \dots \times \left( x_k^{ n_{(N-1)N}^{(k)}   } + x_k^{ n_k+1-n_{(N-1)N}^{(k)}  } \right)
    \notag \\
    & = \prod_{i=1}^{N-1} \prod_{j=i+1}^{N}  \left( x_k^{ n_{ij}^{(k)}   } + x_k^{ n_k+1-n_{ij}^{(k)}  } \right) = \prod_{h=0}^{n_k+1} \left( x_k^{ h   } + x_k^{ n_k+1-h  } \right)^{m_{kh}} \notag \\
& = \begin{cases} 
\prod_{h=0}^{n_k/2}  \sum_{s_h=0}^{M_{kh}} \binom{M_{kh}}{s_h} x_k^{\gamma_{kh}\left(s_h\right)} 
&  \qquad \quad \quad \; \, \, \text{ if $n_k$ is even} \\[8pt]
\left( 2x_k^{\floor{n_k/2}+1 } \right)^{{m}_{k(\floor{n_k/2}+1)}}  \prod_{h=0}^{\floor{n_k/2}} \sum_{s_h=0}^{M_{kh}} \binom{M_{kh}}{s_h} x_k^{\gamma_{kh}\left(s_h\right)}   
 &  \qquad \quad \quad \; \, \, \text{ if $n_k$ is odd}
\end{cases} \notag \\
& = 
\sum_{u=d_k^*}^{D_k^*} w_{k(u-d_k^*)}^* x_k^{u} = 
\sum_{r=0}^{D_k^*-d_k^*} w_{kr}^* x_k^{d_k^*+r}, \label{eq:sum_ngraphs}
\end{align}
where $\floor{z}$ denotes the integer part of $z$. The polynomial $\mathcal{P}(x_k)$ represents a generating-function with coefficients of the form
\begin{equation}\label{eq:weight2_app}
w_{kr}^*=\begin{cases}
\sum_{\mathcal{S}_{kr}}\prod_{h=0}^{n_k/2} \binom{M_{kh}}{s_h}& \text{ if $n_k$ is even}\\[8pt]
\sum_{\mathcal{R}_{kr}} 2^{{m}_{k(\floor{n_k/2}+1)}} \prod_{h=0}^{\floor{n_k/2}} \binom{M_{kh}}{s_h}& \text{ if $n_k$ is odd,}
\end{cases}
\end{equation}
where the sums in \eqref{eq:weight2_app} are taken over the sets
\begin{align*}
\mathcal{S}_{kr}&=\Big\{(s_0,\ldots,s_{n_k/2}):s_h\in\{0,\ldots,M_{kh}\} \,\forall h,\, 
\sum_{h=0}^{n_k/2}\gamma_{kh}\left(s_h\right)-d_k^*=r\Big\},\\
\mathcal{R}_{kr}&=\Big\{(s_0,\ldots,s_{\floor{n_k/2}}): s_h\in\{0,\ldots,M_{kh}\}\,\forall h,\\
&\qquad \sum_{h=0}^{\floor{n_k/2}}\gamma_{kh}\left(s_h\right)+(\floor{n_k/2}+1)m_{k(\floor{n_k/2}+1)}-d_k^*=r\Big\}.
\end{align*}
Thanks to \eqref{eq:sum_ngraphs} and \eqref{eq:weight2_app}, and recalling
 that each $\alpha_k^*$ is distributed a priori as a Truncated-Beta on $(0,1/2)$ with parameters $a,b>0$, \eqref{eq:alpha_acc_step} can be written as
\begin{align}\label{eq:alpha_acc_step_cont}
p(\alpha_k^*, {\mathcal{G}}^{(\mathcal{D}_k)} ) & =  p(\alpha_k^*) (1-{\alpha_k^*})^{\left(n_k+1\right)M} \sum_{\mathcal{C}^{*}_k \in \mathscr{G}_{\mathcal{V}} } \left(  \frac{\alpha_k^*}{1-\alpha_k^*} \right)^{ \sum_{l \in \mathcal{D}_k^\dagger  } d_\text{H}\left(\mathcal{G}_l, \mathcal{C}^{*}_k \right)  } \notag \\
& = p(\alpha_k^*) (1-{\alpha_k^*})^{\left(n_k+1\right)M} 
\sum_{r=0}^{D_k^*-d_k^*} w_{kr}^* \left(\frac{\alpha_k^*}{1-\alpha_k^*} \right)^{d_k^*+r} \notag \\
& = \frac{1}{\mathcal{B}( 1/2; a, b)} \sum_{r=0}^{D_k^*-d_k^*} w_{kr}^* {\alpha_k^*}^{a+d_k^*+r-1} (1-{\alpha_k^*})^{b+\left(n_k+1\right)M-d_k^*-r-1}. 
\end{align}
Marginalizing \eqref{eq:alpha_acc_step_cont} with respect to $\alpha_k^*$, we obtain the marginal distribution of ${\mathcal{G}}^{(\mathcal{D}_k)}$ as
\begin{align}\label{eq:mar_prior_ngraphs}
 p( {\mathcal{G}}^{(\mathcal{D}_k)} ) & = \int_0^{1/2} p(\alpha_k^*, {\mathcal{G}}^{(\mathcal{D}_k)} ) \mathrm{d} \alpha_k^* \notag  \\ 
 & =  \sum_{r=0}^{D_k^*-d_k^*}   \frac{w_{kr}^*}{\mathcal{B}( 1/2; a, b)} \int_0^{1/2}{\alpha_k^*}^{a+d_k^*+r-1} (1-{\alpha_k^*})^{b+\left(n_k+1\right)M-d_k^*-r-1} \mathrm{d} \alpha_k^* \notag \\
 & = \sum_{r=0}^{D_k^*-d_k^*} w_{kr}^* \frac{\mathcal{B}( 1/2; a_{kr}^*, b_{kr}^*)}{\mathcal{B}\left( 1/2; a, b\right)},
\end{align}
where $a_{kr}^*=a+d_{k}^*+r$ and $b_{kr}^*=b+( n_k+1) M -d_{k}^*-r$.\\
Combining \eqref{eq:alpha_acc_step_cont} and \eqref{eq:mar_prior_ngraphs}, the first distribution on the right hand side of \eqref{eq:full_cond_theta_star} can be written as
\begin{align}\label{eq:alpha_givenG}
p(\alpha_k^* \mid {\mathcal{G}}^{(\mathcal{D}_k)} ) = \frac{p(\alpha_k^*, {\mathcal{G}}^{(\mathcal{D}_k)} )}{p( {\mathcal{G}}^{(\mathcal{D}_k)} )} = \sum_{r=0}^{D_k^*-d_k^*}  \varphi_{kr}^{*} \, f_{\text{TBeta}}\left(\alpha^*_k; 1/2, a_{kr}^*, b_{kr}^*\right),
\end{align}
where the mixture weights $\varphi_{kr}^*$ are given by:
\begin{align}\label{eq:alpha_weight}
 \varphi_{kr}^*= \frac{w_{kr}^* \mathcal{B}( 1/2; a_{kr}^*, b_{kr}^*)}{ \sum_{r=0}^{D_k^*-d_k^*} w_{kr}^* \mathcal{B}( 1/2; a_{kr}^*, b_{kr}^*)}.
\end{align}
Focusing on the second distribution on the right hand side of \eqref{eq:full_cond_theta_star}, we
start from \eqref{eq:joint_acc_step} and use \eqref{eq:sum_dist} to get
\begin{align}\label{eq:full_cond_G_m}
p(\alpha_k^*, \mathcal{C}^{*}_k, {\mathcal{G}}^{(\mathcal{D}_k)} ) =  p(\alpha_k^*) (1-{\alpha_k^*})^{\left(n_k+1\right)M} \left(  \frac{\alpha_k^*}{1-\alpha_k^*} \right)^{ \sum_{i<j} \left[ (n_k+1-2n_{ij}^{(k)}) A_{\mathcal{C}^{*}_k[ij]} + n_{ij}^{(k)}   \right] }.
\end{align}
As mentioned in Section 2.1 of the main paper, 
modeling a graph  $\mathcal{G}$ is equivalent to modeling the $M$-dimensional vector $V_\mathcal{G}=\text{vech}(A_{\mathcal{G}})$ defined as the half-vectorization of $A_{\mathcal{G}}$, whose components coincide with the elements of the lower triangular half of $A_{\mathcal{G}}$. We let $V_{\mathcal{C}^{*}_k(-[ij])}$ denote the $(M-1)$-dimensional vector after removing the element encoding the binary relation $A_{\mathcal{C}^{*}_k[ij]}$ between the nodes $\{i,j\}$ from the
half-vectorization $V_{\mathcal{C}^{*}_k}$ of $A_{\mathcal{C}^{*}_k}$.
We note that \eqref{eq:full_cond_G_m} highlights the conditional independence property of the elements of $V_{\mathcal{C}^{*}_k}$, allowing for independent edge-specific distributions, conditionally on
$\alpha_k^*$ and ${\mathcal{G}}^{(\mathcal{D}_k)}$.
Thus,
for $g \in \{0, 1\}$, the full conditional of $A_{\mathcal{C}^{*}_k[ij]}$ is such that:
\begin{align}\label{eq:full_gm_ij}
p_{kij}^* &=  \P( A_{\mathcal{C}^{*}_k[ij]}=g \mid  \alpha_k^*, V_{\mathcal{C}^{*}_k(-[ij])},{\mathcal{G}}^{(\mathcal{D}_k)}) 
  \notag \\ &  =
  \P( A_{\mathcal{C}^{*}_k[ij]}=g \mid  \alpha_k^*,{\mathcal{G}}^{(\mathcal{D}_k)})  
\notag \\ &  =
  \frac{ \left(  \frac{\alpha_k^*}{1-\alpha_k^*} \right)^{(n_k+1-2n_{ij}^{(k)}) g + n_{ij}^{(k)} }  }{\left(  \frac{\alpha_k^*}{1-\alpha_k^*} \right)^{(n_k+1-2n_{ij}^{(k)}) g + n_{ij}^{(k)} } + \left(  \frac{\alpha_k^*}{1-\alpha_k^*} \right)^{(n_k+1-2n_{ij}^{(k)}) (1-g) + n_{ij}^{(k)} } } \notag \\
  & = \left[ 1 +  \left(\frac{\alpha_k^*}{1-\alpha_k^*} \right)^{-2\left(n_{ij}^{(k)}-(n_k+1)/2\right) \left( 1-2g\right)} \right]^{-1}.
\end{align}
The conditional distribution $p(\mathcal{C}^{*}_k \mid \alpha_k^*, {\mathcal{G}}^{(\mathcal{D}_k)} )$ in \eqref{eq:full_cond_theta_star} is fully determined.

\subsection{Distribution $P_l$ in the generalized Pólya urn scheme}
When $n_k=1$, that is $\mathcal{D}_k=\{l\}$ for some $l=1,\ldots,n$, the distribution in $(11)$ and $(12)$ simplifies to the distribution $P_l$ in $(8)$ and $(9)$ of the main paper, for $\vartheta_l$, conditionally on $\mathcal{G}_l$ and given that $\vartheta_l$ takes a new value. To see this, it can be easily verified from \eqref{eq:sum_dist} that, for $n_k=1$, $U_k$ has support $u_k \in \{d_k^*=d_l, \ldots, D_k^*=2M-d_l \}$. Moreover, $n_{ij}^{(k)}=A_{\mathcal{G}_0[ij]}+A_{\mathcal{G}_l[ij]}\in \{0,1,2\}$, 
$m_{k1}=d_l$, $M_{k0}=m_{k0}+m_{k2}=M -d_l$, $\gamma_{k0}\left(s_0\right)=2s_0$. 
Thus, for $r=u-d_l$, the coefficient in \eqref{eq:weight2_app}
\begin{align*}
 w_{k(u-d_l)}^*=2^{d_l} \binom{M-d_l}{(u-d_l)/2}=2^{d_l} \binom{M-d_l}{s_0},   
\end{align*}
as there is a constrained scalar decision variable $s_0 \in \{0, \ldots, M-d_l\}$ in the condition $2s_0=u-d_l$ defining the set $\mathcal{R}_{k(u-d_l)}$
 for fixed $u$ and $d_l$. Moreover, from \eqref{eq:mar_prior_ngraphs}, $a_{k(u-d_l)}^*=a+u=a+2s_0+d_l$ and $b_{k(u-d_l)}^*=b+2(M-d_l)-u=b+2M-2s_0-d_l$. 
That is, the coefficients $a_{kr}^*$, $b_{kr}^*$
in $(14)$ and $w_{kr}^*$ in $(15)$,
boil down respectively to $a_{ls_0}$, $b_{ls_0}$ and $w_{ls_0}$ in $(6)$ of the main paper. Finally, the distribution specified by $(9)$ and $(10)$ directly follows from \eqref{eq:full_gm_ij} by setting $n_k=1$. 

\subsection{Additional details on posterior computations}
The Gibbs sampling in Algorithm 1 allows us to sample from the posterior distribution of $\vartheta^{(1:n)}$ conditionally on $\mathcal{G}^{(1:n)}$.  
We discuss some details that can help improving its efficiency.\\
The first one is that in the first step of the sampler, generating a new $\vartheta_l$, for all $l$, has a computational complexity of the order of $\mathcal{O}\left( n M \right)$, which can become an issue for larger network dimensions and/or population sizes.
However, the unnormalized probability of sampling a new value for $\vartheta_l$, given in (6)
and represented by the r.h.s. of \eqref{eq:integral_cont},
and the distribution $P_l$ of new values for $\vartheta_l$, conditionally on $\mathcal{G}_l$, defined in (8)-(9),
are iteration-invariant, if the hyperparameter $\mathcal{G}_0$ is kept fixed. These quantities can thus be computed once before running the Gibbs sampler.\\  
Secondly, in the reshuffling step, the update of $\vartheta_k^*=(\mathcal{C}_k^{*}, \alpha_k^*)$ has a computational complexity of the order of $\mathcal{O}\left( K M \right)$, with $K$ being the number of clusters at a given iteration. Although, this step of the sampler may present a lower computational complexity, as $K\leq n$, 
calculations are more involved because the solution of a set of linear Diophantine equations is required for each $k$ to compute the coefficients $w_{kr}^*$ defined in (15)
and appearing in the 
mixture weights given in (14). 
The number of equations to solve for each $k$ depends on $N$ and  $n_k$ 
through $D_k^*-d_k^*$. The complexity of each equation directly depends on $n_k$.
Some considerations are worth noting. 
 Let $e_{kh}=n_k+1-2h$ for $h=0, \ldots, \floor{n_k/2}$, and let $g_k$ denote the common greatest divisor  of the vector $e_k=(e_{k0}, e_{k1}, \ldots, e_{k\floor{n_k/2}})$, namely $g_k=cgd(e_k)$. The linear Diophantine equation defining the set $\mathcal{S}_{kr}$
($\mathcal{R}_{kr}$)
 has no solution when $r+d_k^*-\sum_{h=0}^{n_k/2} h M_{kh}$ $\left(r+d_k^*-(\floor{n_k/2}+1)m_{k(\floor{n_k/2}+1)}-\sum_{h=0}^{\floor{n_k/2}} h M_{kh}\right)$ 
is not a multiple of $g_k$. In this case, $w_{kr}^*=0$. 
Moreover, the coefficient $w_{kr}^*$ is symmetric with respect to the index $r$, that is $w_{kr}^*=w_{k \Bar{r}}^*$, with $\Bar{r}=D_k^*-d_k^*-r$.
With these arguments, the overall computational time needed to define the set $\mathcal{S}_{kr}$
($\mathcal{R}_{kr}$) can be more than halved. From a practical perspective, this can be solved with the algorithm based on a generating function of \cite{subset_sum_prob}, used by \cite{subset_sum_prob_book} and implemented in the \texttt{nilde} R package \citep{nilde}, by imposing $\sum_{h=0}^{\floor{n_k/2}} s_h \leq M_{k}$, where $M_{k}=\sum_{h=0}^{\floor{n_k/2}} M_{kh}$ and retaining only the feasible solutions, that is those satisfying $s_h \in \left\{0, \ldots, M_{kh} \right\}$ $\forall h$. Yet, defining the set $\mathcal{S}_{kr}$
($\mathcal{R}_{kr}$) at each iteration for any $k=1,\ldots,K$ and $r=0,\ldots,D_k^*-d_k^*$ substantially increases the computational time required in the reshuffling step. An alternative and cheaper strategy consists in replacing the conditional distribution of $\alpha_k^*$ given $\mathcal{G}^{(\mathcal{D}_k)}$ in (11)
by its full conditional. That is 
$\alpha_k^* \mid \mathcal{C}^{*}_k, \mathcal{G}^{(\mathcal{D}_k)} \sim \operatorname{TBeta}\left(1/2; a_k^*, b_k^*\right)
$ where $a_k^*=a+\sum_{l \in \mathcal{D}^\dagger_k }d_\text{H}\left(\mathcal{G}_l, \mathcal{C}^{*}_k\right)$ and $b_k^*=b+\left(n_k+1\right)M-\sum_{l \in \mathcal{D}^\dagger_k }d_\text{H}\left(\mathcal{G}_l, \mathcal{C}^{*}_k\right)$. Sampling $\alpha_k^*$ from its full conditional finds justification in that, along the chain, it targets the distribution in (11).
Yet in this step, noting that the Bernoulli random variables given in (12)
are identically distributed for all pairs of nodes $\{i,j \}$ sharing the same value of $n_{ij}^{(k)}=h$, only $m_{kh} \leq M$ Bernoulli parameters $p^*_{kij}$ defined in (13) must be computed.
\section{Posterior prediction}\label{app:prediction}
\subsection{Cluster-specific posterior predictive distribution}\label{sec:clust_post_pred}
The posterior predictive distribution implied by the statistical model defined in (2.1) is
\begin{align}\label{eq:ppd}
  p(\mathcal{G}^\star \mid  {\mathcal{G}}^{(1:n)} )=\int_{\Theta} \psi \left(\mathcal{G}^\star ; \vartheta \right) \mathrm{d} p ( \vartheta  \mid {\mathcal{G}}^{(1:n)} )  
\end{align}
where $p ( \vartheta  \mid {\mathcal{G}}^{(1:n)} )$ denotes the posterior distribution of $\vartheta=\left({\mathcal{C}}, \alpha \right)$. Although the above integral is not analytically available, it is straightforward to simulate networks from the posterior predictive distribution exploiting MCMC samples for $\vartheta$ along with the predictive distribution structure of the underlying DP.
On the other hand, conditionally on an estimated partition of the observed graphs, the cluster-specific one-step-ahead posterior predictive distribution implied by our model is available in closed-form.
Predicting a graph $\mathcal{G}^\star$ from the posterior predictive distribution specific to the cluster of observations with indices in $\mathcal{D}_k$ translates into sampling $M$ independent Bernoulli distributions. Specifically,  
\begin{align}\label{eq:cls_post_pred}
A_{\mathcal{G}^\star[ij]} \mid  {\mathcal{G}}^{(\mathcal{D}_k)}  & \simind \text{Bern}(\tilde{p}_{kij}), & i<j. 
\end{align}
After introducing the quantities $T^\star_{kij}=\sum_{\{u,v\} \neq \{i,j\}} \max\{n_{uv}^{(k)}, n_k+1-n_{uv}^{(k)}\}$ and 
$t^\star_{kij}=\sum_{\{u,v\} \neq \{i,j\}} \min\{n_{uv}^{(k)}, n_k+1-n_{uv}^{(k)}\}$ where $n_{uv}^{(k)}= \sum_{l \in \mathcal{D}^\dagger_k} A_{\mathcal{G}_l[uv]}$, 
the Bernoulli parameters in  \eqref{eq:cls_post_pred} are given by:
\begin{equation}\label{eq:prob_edge_pred}
\tilde{p}_{kij} = \mathbbm{E}\left[ A_{\mathcal{G}^\star[ij]}  \mid {\mathcal{G}}^{(\mathcal{D}_k)} \right]= \frac{1}{p({\mathcal{G}}^{(\mathcal{D}_k)} )}
\sum_{r=0}^{T^\star_{kij}-t^\star_{kij}} \tilde{w}_{kr} 
\frac{
\left[ \mathcal{B}( 1/2; \tilde{a}_{kij}^{(r)}, \tilde{b}_{kij}^{(r)} ) + \mathcal{B}( 1/2; \tilde{c}_{kij}^{(r)}, \tilde{d}_{kij}^{(r)} ) \right]
}{\mathcal{B}\left( 1/2; a, b\right)}
,
\end{equation}
where $\tilde{a}_{kij}^{(r)}=a+n^{(k)}_{ij}+t^\star_{kij} + r +1$,   $\tilde{b}_{kij}^{(r)}= b + \left(n_k+1\right)M - ( n^{(k)}_{ij}+t^\star_{kij} + r )$, $\tilde{c}_{kij}^{(r)}= a + n_k+1 - n^{(k)}_{ij} + t^\star_{kij} + r$ and $\tilde{d}_{kij}^{(r)} =  b + \left(n_k+1\right) M  - ( n_k+1 - n^{(k)}_{ij}+t^\star_{kij} + r  )  +1$.
Thus, cluster-specific one-step-ahead edge prediction probability is a linear combination of two incomplete beta functions whose parameters reflect edge presence and edge absence, respectively. 
In turn, the expression of the coefficient $\tilde{w}_{kr}$ in \eqref{eq:prob_edge_pred} can be retrieved from (15) with minor modifications. Specifically, it suffices to replace: $D^*_k$ with $T^\star_{kij}$, $d^*_k$ with $t^\star_{kij}$, $m_{kh}$ with $q_{khij}=\#\left\{ \{ u,v \} \in \mathcal{V}^2 : \{u,v\} \neq \{i,j\}, n^{(k)}_{uv}=h \right\}$ and $M_{kh}$ with $Q_{khij}=q_{khij} + q_{k(n_k + 1 - h)ij}$. Finally, the marginal likelihood $p({\mathcal{G}}^{(\mathcal{D}_k)} )$ in 
\eqref{eq:prob_edge_pred} 
is analytically available and results from the joint distribution of ${\mathcal{G}}^{(\mathcal{D}_k)}$ and $A_{\mathcal{G}^\star[ij]}$ by
marginalizing the latter. It is given by:
\begin{equation}\label{eq:prop_cons_pred}
    p({\mathcal{G}}^{(\mathcal{D}_k)} )=
    \sum_{r=0}^{D_k^*-d_k^*} w_{kr}^* \frac{\mathcal{B}( 1/2; a_{kr}^*, b_{kr}^*)}{\mathcal{B}\left( 1/2; a, b\right)},
\end{equation}
with $a_{kr}^*$ and $b_{kr}^*$ defined in (14) and $w_{kr}^*$ in (15).
It is worth noting that the marginal prior $p\left(\mathcal{G}^\star \right)=\int \psi\left(\mathcal{G}^\star; \vartheta \right) \mathrm{d} P_{0}(\vartheta)$, appearing in $\pi_{l0}$ in (6), can be easily retrieved for generative purposes as an instance of \eqref{eq:cls_post_pred} when $n_k=0$, that is $\mathcal{D}_k=\varnothing$. 
In this case, $p({\mathcal{G}}^{(\varnothing) }
)=1$ and  \eqref{eq:prob_edge_pred} boils down to the prior expectation:
\begin{align*}
\mathbbm{E}\left[ A_{\mathcal{G}^\star[ij]}  \right]=\frac{\mathcal{B}( 1/2; a+1+A_{\mathcal{G}_0[ij]}, b+1-A_{\mathcal{G}_0[ij]} ) + \mathcal{B}( 1/2; a+1-A_{\mathcal{G}_0[ij]}, b+1+A_{\mathcal{G}_0[ij]} )}{\mathcal{B}\left( 1/2; a, b\right)}.
\end{align*}

\begin{figure}[h!]
\centering
\includegraphics[height=4.3cm]{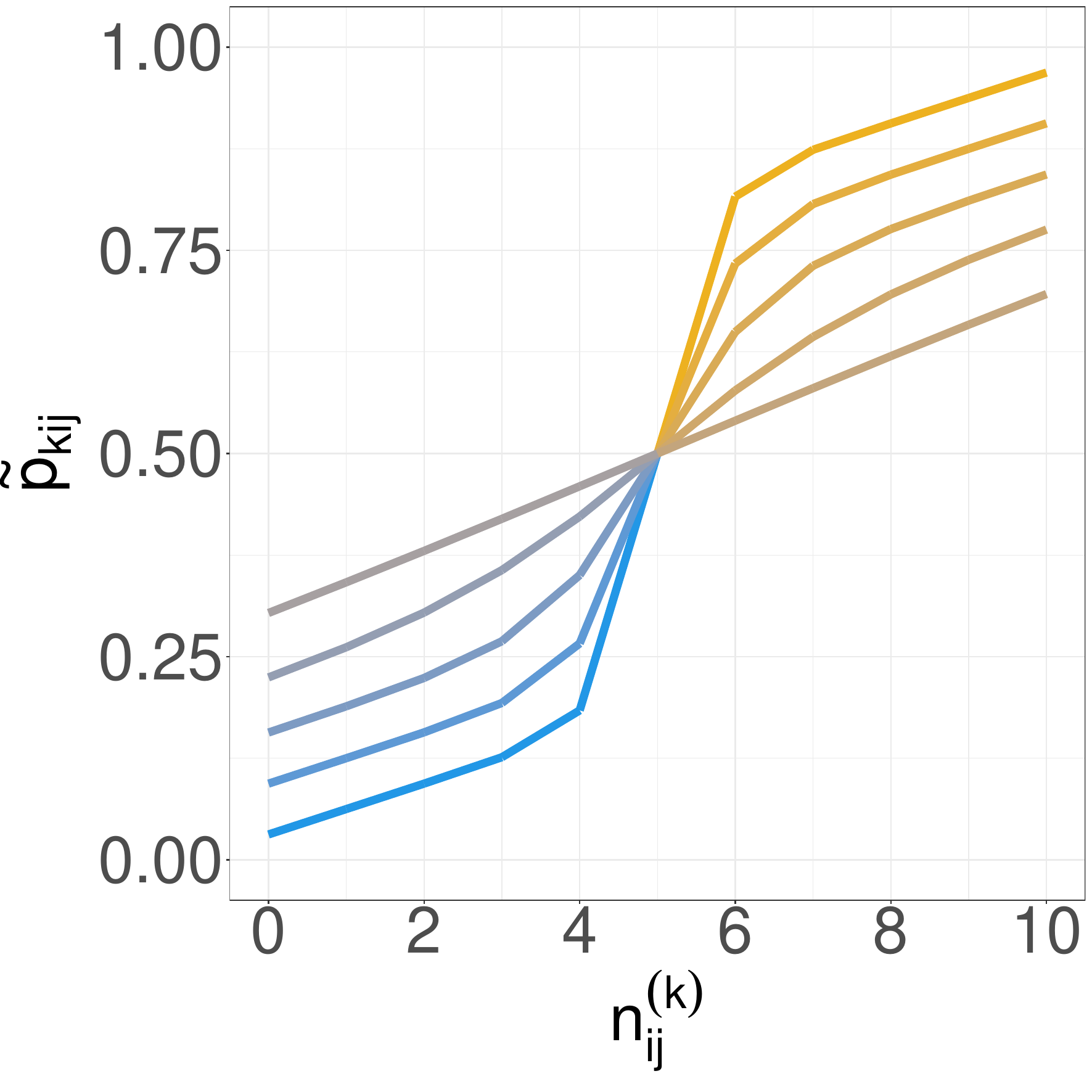}
\includegraphics[height=4.3cm]{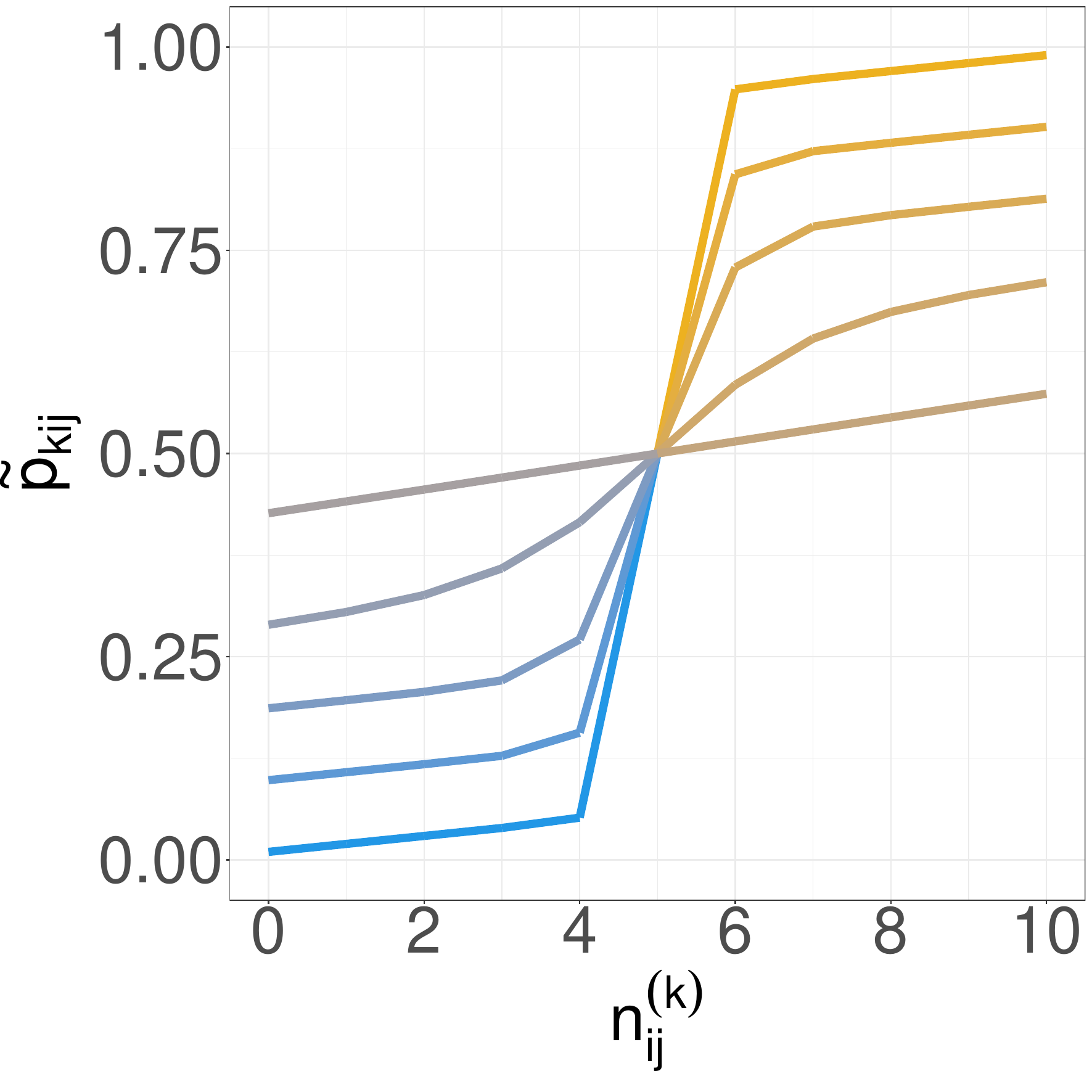}
\includegraphics[height=4.3cm]{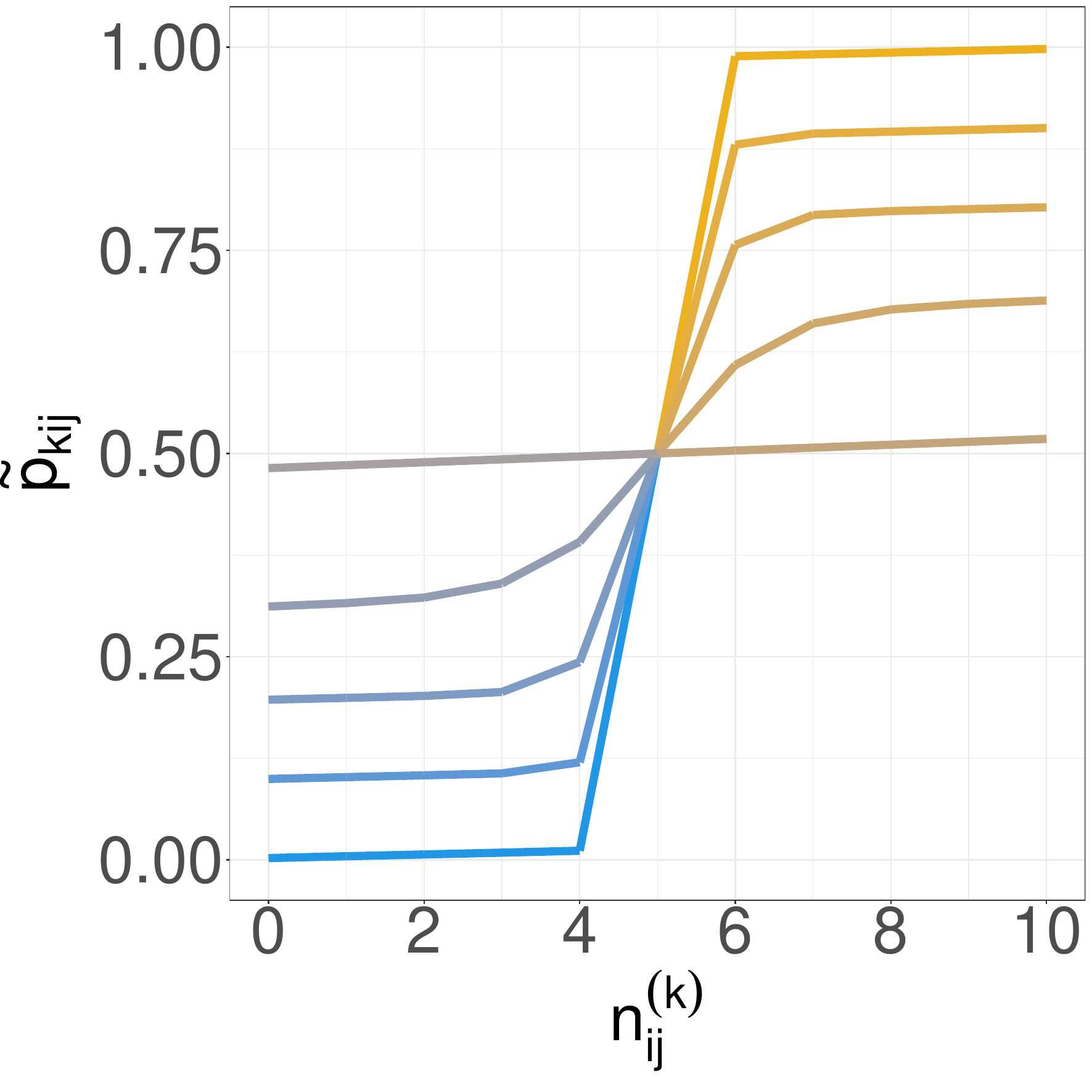}
\caption{\small  
Probability $\tilde{p}_{kij}$ in \eqref{eq:prob_edge_pred}, with $n_{uv}^{(k)} = z$ $\forall \{u,v\} \neq \{i,j\}$, with $z \in \{0,1, \ldots, n_k+1\}$  (blue for low and yellow for high) and for $n_{ij}^{(k)} \in \{0,1, \ldots, n_k+1\}$, with $n_k+1=10$, and for $N \in \{ 3, 5, 10 \}$ (from left to right).
}\label{fig:cls_post_pred_prob}
\end{figure}

\begin{figure}[h!]
\centering
\includegraphics[height=4.3cm]{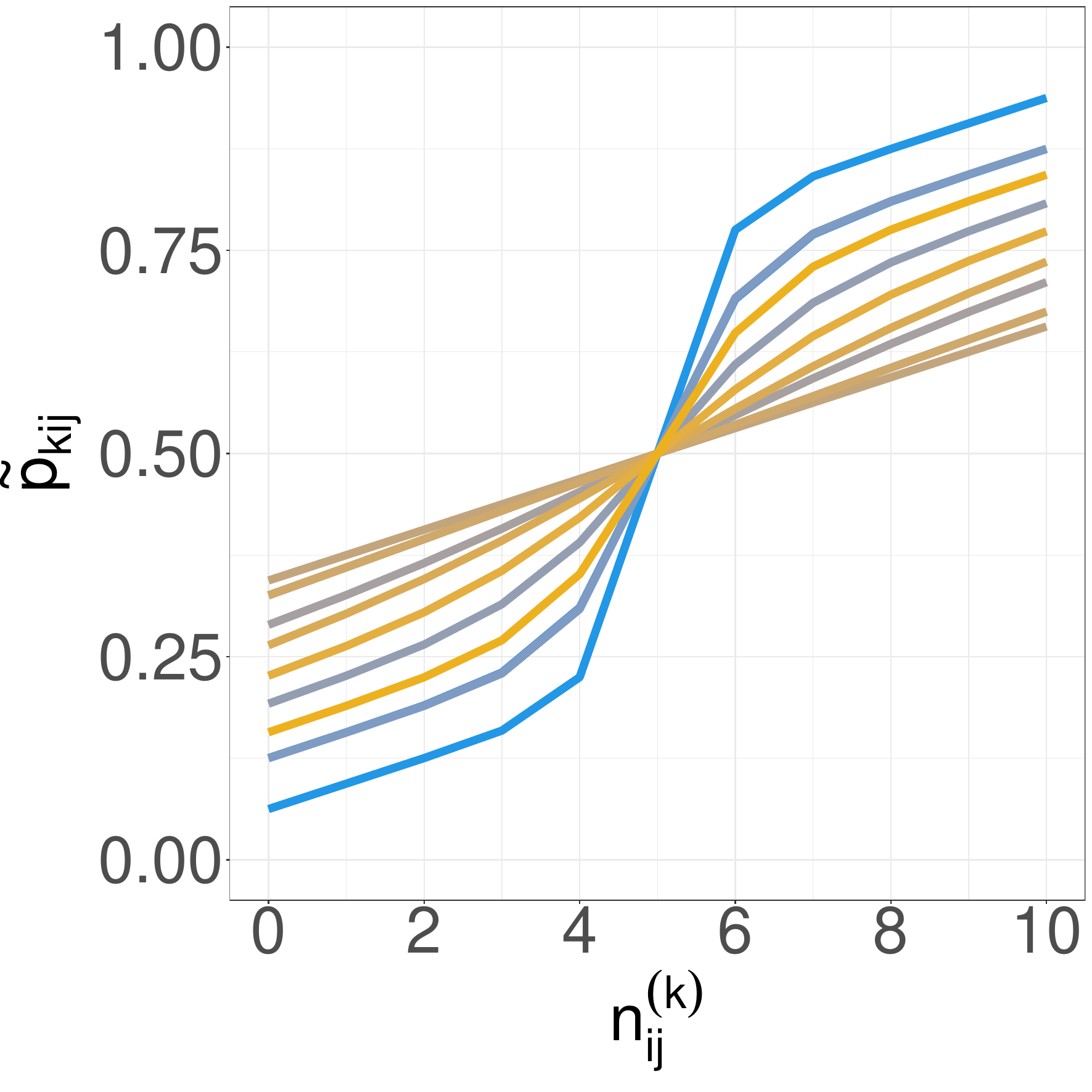}
\includegraphics[height=4.3cm]{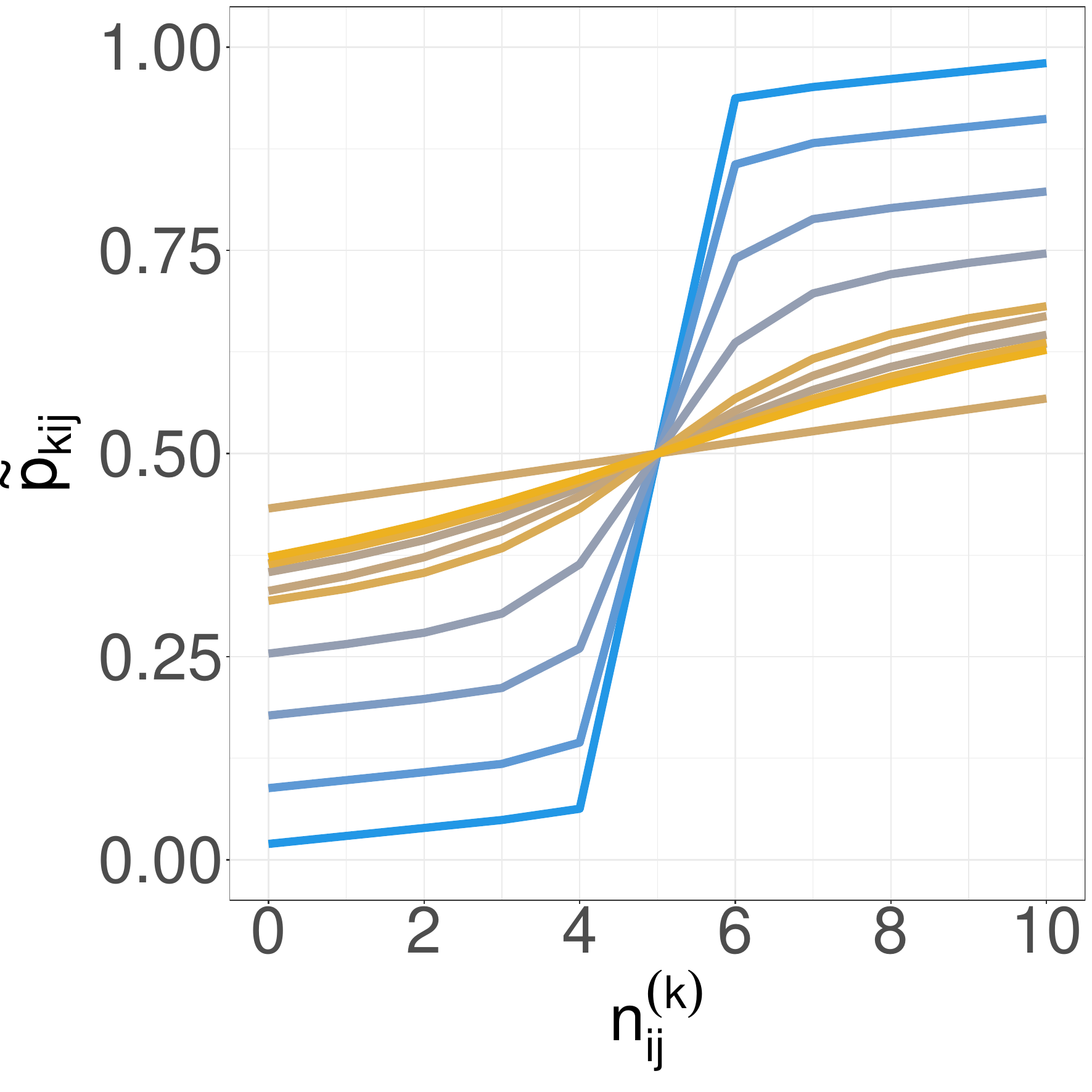}
\includegraphics[height=4.3cm]{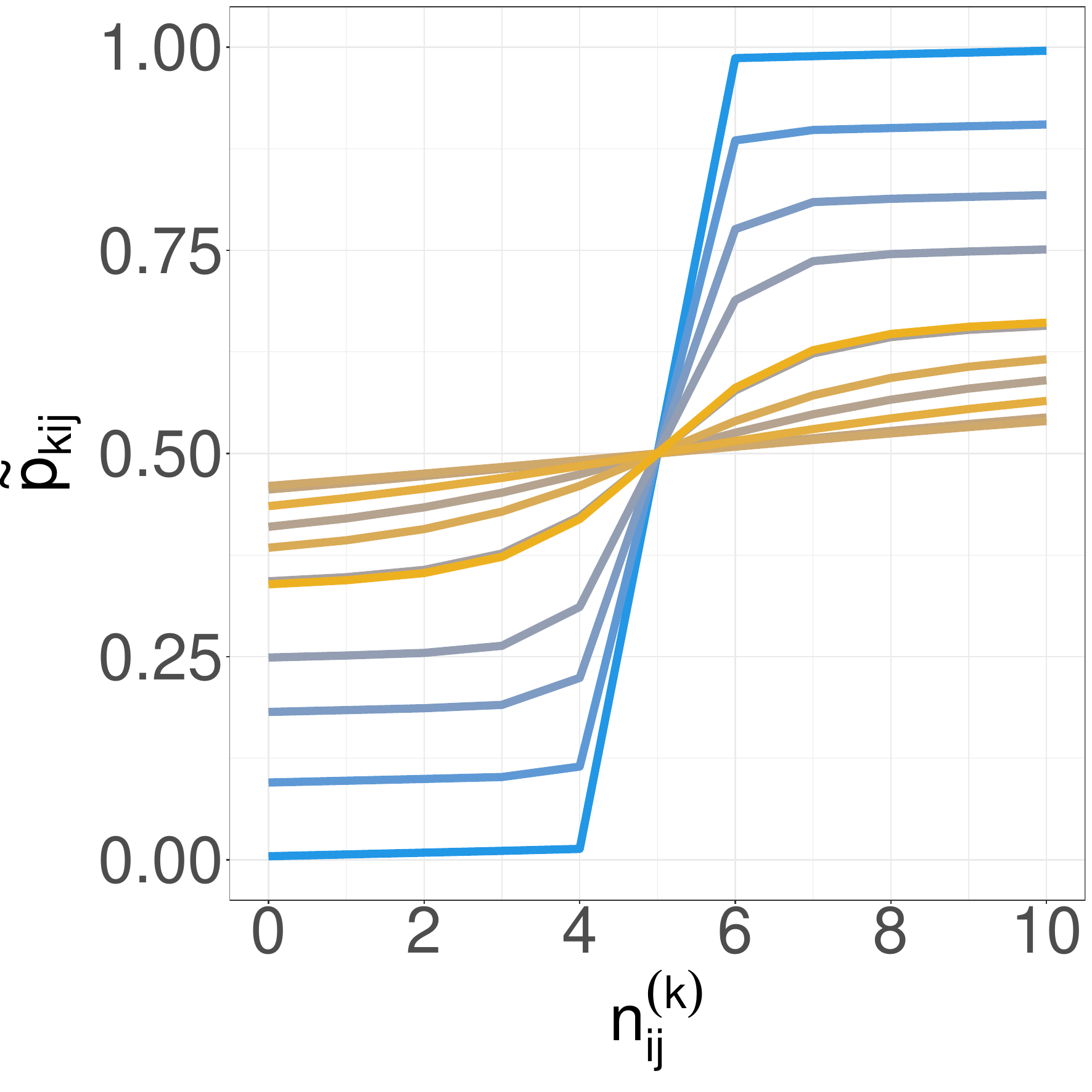}
\caption{\small  
Probability $\tilde{p}_{kij}$ in \eqref{eq:prob_edge_pred}, with $\sum_{\{u,v\} \neq \{i,j\}} n_{uv}^{(k)} = z$, with $z$ taking $11$ equally-spaced values in $\{0,1, \ldots, (n_k+1)(M-1)\}$  (blue for low and yellow for high) and for $n_{ij}^{(k)} \in \{0,1, \ldots, n_k+1\}$, with $n_k+1=10$, and for $N \in \{ 3, 5, 10 \}$ (from left to right).
}\label{fig:cls_post_pred_prob_sparse}
\end{figure}

\autoref{fig:cls_post_pred_prob} shows how the cluster-specific one-step-ahead posterior predictive probability in \eqref{eq:prob_edge_pred} varies based on how frequently the edge between node $\{ i,j\}$ appears in the graphs belonging to the $k$-th cluster and the prior graph $\mathcal{G}_0$, with $\tilde{p}_{kij}=1/2$ for $n_{ij}^{(k)} = (n_k+1)/2$. Moreover, as highlighted by the mixed color of each line, $\tilde{p}_{kij}$ is symmetric with respect to the (equal) frequency of all the other edges $\{u,v\} \neq \{i,j\}$, meaning that the values $n_{uv}^{(k)} = z$ and $n_{uv}^{(k)} = n_k+1-z$ share the same curve. In addition, it is worth noting how $\tilde{p}_{kij}$ is an odd function with center shifted at $\left((n_k+1)/2, 1/2\right)$, meaning that $f(x)=1-f(n_k+1-x)$ for $\tilde{p}_{kij}$ function of $n_{ij}^{(k)}$.
While in \autoref{fig:cls_post_pred_prob} we aim at isolating the effect of differences among the $n_{uv}^{(k)}$'s on $\tilde{p}_{kij}$, \autoref{fig:cls_post_pred_prob_sparse}, shows, instead, how $\tilde{p}_{kij}$ varies as a function of $n_{ij}^{(k)}$, for $z=\sum_{\{u,v\} \neq \{i,j\}} n_{uv}^{(k)}$ ranging in 
$\{0,1, \ldots, (n_k+1)(M-1)\}$,
where $n_{uv}^{(k)}$ can change across $\{ u,v\}$. In this more realistic case, symmetries can only happen based on the value of $q_{khij}$.

\subsubsection{Cluster-specific $m$-step-ahead posterior predictive distribution}
\begin{figure}[h!]
\centering
\includegraphics[height=4.3cm]{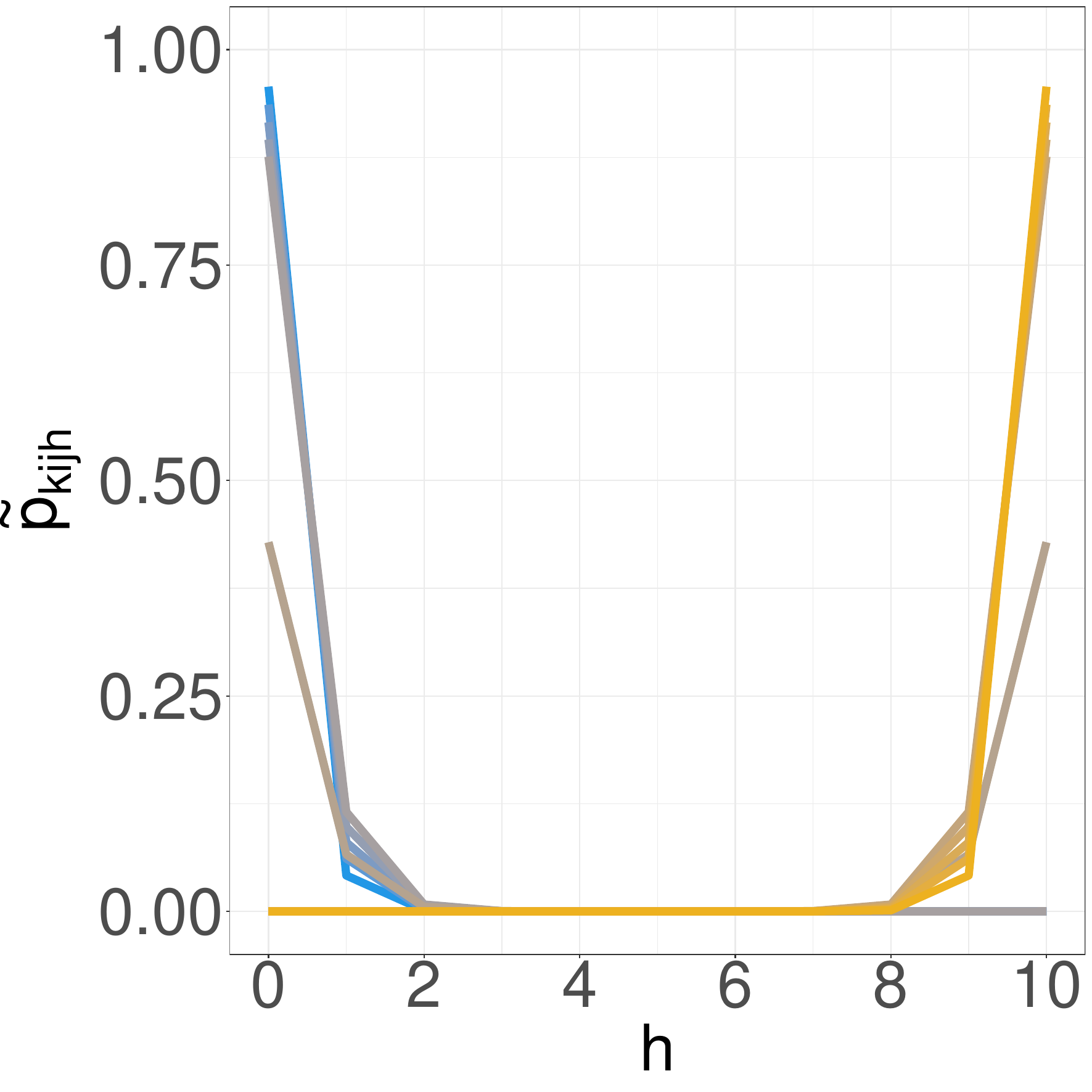}
\includegraphics[height=4.3cm]{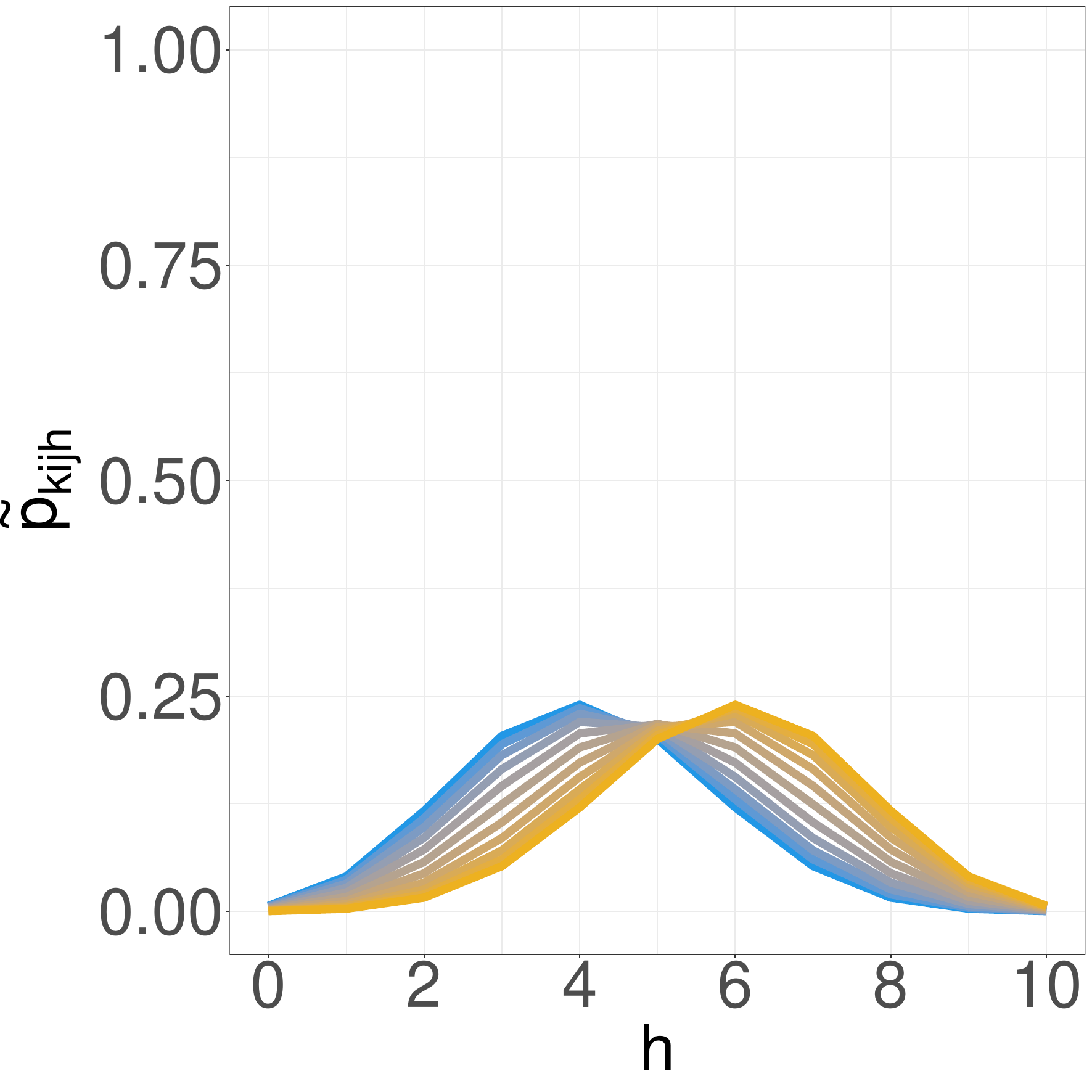}
\includegraphics[height=4.3cm]{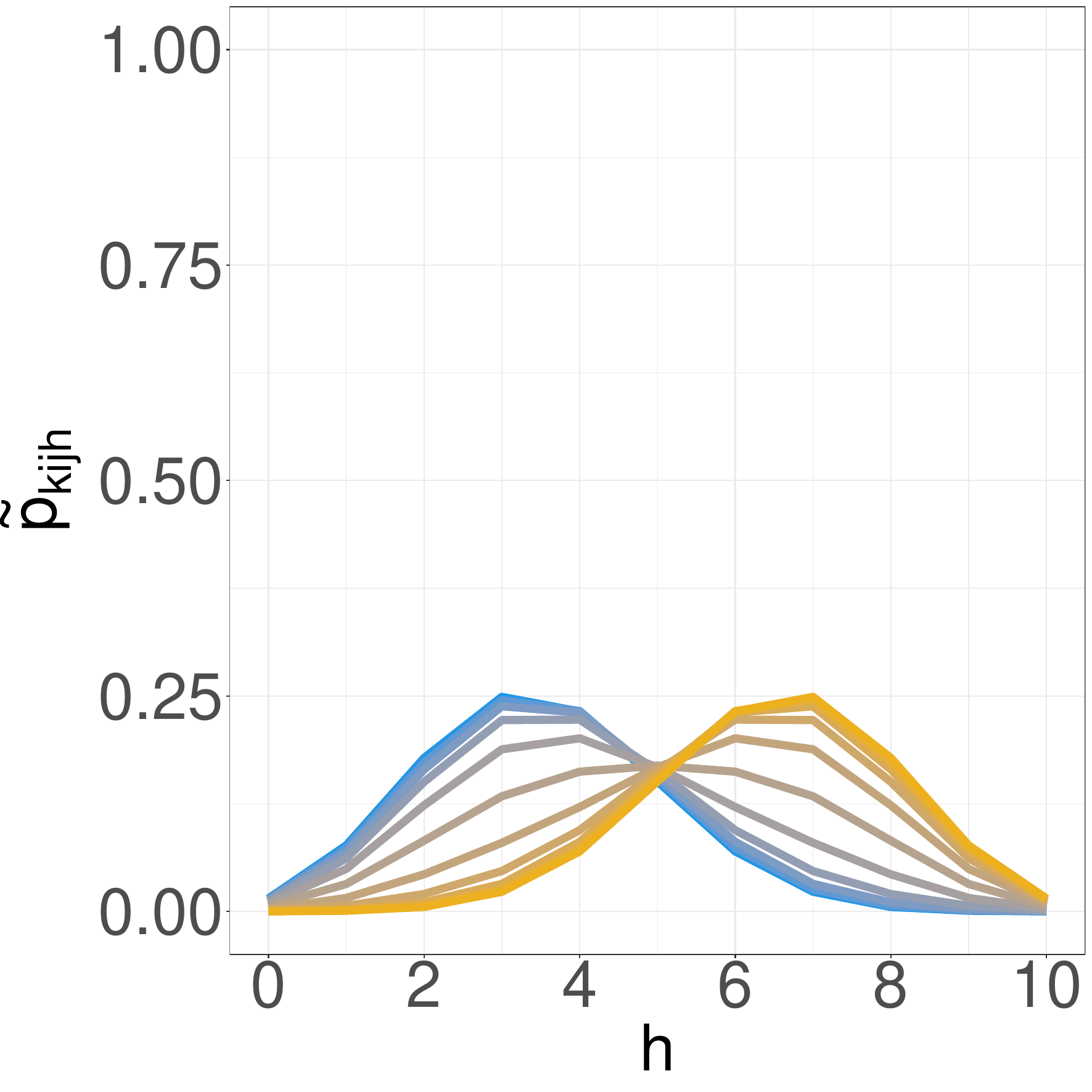}
\caption{\small  
Probability $\tilde{p}_{kijh}$ in \eqref{eq:prob_edge_pred_m_step}, for $h \in \{0, \ldots, m\}$, with $m=10$, and $n_{ij}^{(k)} \in \{0,1, \ldots, n_k+1\}$, with $n_k+1=10$ (blue for low and yellow for high), with $N=10$, and $\sum_{\{u,v\} \neq \{i,j\}} n_{uv}^{(k)} = z$, with $z \in \{0, (\frac{n_k+1}{2})(M-1), (n_k+1)(M-1)\}$ (from left to right). 
}\label{fig:cls_post_pred_prob_joint_m}
\end{figure}

The distribution given in \eqref{eq:cls_post_pred}--\eqref{eq:prop_cons_pred} can be generalized to make predictions on, say, $m$ graphs jointly, conditionally  on the estimated partition and on the fact that they belong to the same cluster, say the $k$-th one. 
Such distribution, denoted by $p(\mathcal{G}^\star_1, \ldots, \mathcal{G}^\star_m \mid {\mathcal{G}}^{(\mathcal{D}_k)} )$, is defined on the $m$-dimensional cartesian product $\mathscr{G}_{\mathcal{V}} \times \ldots \times \mathscr{G}_{\mathcal{V}}$ 
and 
it turns out that we can make edge-specific predictions independently, conditionally on ${\mathcal{G}}^{(\mathcal{D}_k)}$. It is thus sufficient to study the conditional distribution of 
$\mathcal{H}^{(1:m)}_{[ij]}=\sum_{l=1}^m A_{\mathcal{G}^{\star}_l[ij]}$, given $\mathcal{G}^{(\mathcal{D}_k)}$, for which we get
\begin{align*}
\mathcal{H}^{(1:m)}_{[ij]} \mid  {\mathcal{G}}^{(\mathcal{D}_k)}  & \simind \text{Cat}( \tilde{p}_{kij0},  \tilde{p}_{kij1}, \tilde{p}_{kij2}, \ldots, \tilde{p}_{kijm} ), & i<j 
\end{align*}
where, for $h=0,\ldots, m$,
\begin{align}\label{eq:prob_edge_pred_m_step}
\tilde{p}_{kijh}& =\P\left( \mathcal{H}^{(1:m)}_{[ij]} = h \mid  {\mathcal{G}}^{(\mathcal{D}_k)} \right) \notag \\
& =\frac{\binom{m}{h}}{p({\mathcal{G}}^{(\mathcal{D}_k)} )}
\sum_{r=0}^{T^\star_{kij}-t^\star_{kij}} \tilde{w}_{kr} 
\frac{
\left[ \mathcal{B}( 1/2; \tilde{a}_{kijh}^{(r)}, \tilde{b}_{kijh}^{(r)} ) + \mathcal{B}( 1/2; \tilde{c}_{kijh}^{(r)}, \tilde{d}_{kijh}^{(r)} ) \right]
}{
 \mathcal{B}( 1/2; a, b )  
}
\end{align}
where $\tilde{a}_{kijh}^{(r)}=a+n^{(k)}_{ij}+t^\star_{kij} + r + h$,   
$\tilde{b}_{kijh}^{(r)}= b + \left(n_k+1\right)M - ( n^{(k)}_{ij}+t^\star_{kij} + r ) + m-h$, $\tilde{c}_{kijh}^{(r)}= a + n_k+1 - n^{(k)}_{ij} + t^\star_{kij} + r + m-h$ and $\tilde{d}_{kijh}^{(r)} =  b + \left(n_k+1\right) M  - ( n_k+1 - n^{(k)}_{ij}+t^\star_{kij} + r  ) +h$, with $t^\star_{kij}$, $T^\star_{kij}$, $ \tilde{w}_{kr}$ and $p({\mathcal{G}}^{(\mathcal{D}_k)} )$ defined in \Autoref{sec:clust_post_pred}. \Autoref{fig:cls_post_pred_prob_joint_m} shows how $\tilde{p}_{kijh}$ varies as function of $h$, for different values of $n_{ij}^{(k)}$ and $n_{uv}^{(k)}$.

\subsection{Cluster-specific posterior distribution of $\mathcal{C}_k^{*}$}\label{sec:cls_spec_post_mode}
For each $k=1,\ldots,\hat{K}$, the conditional distribution of $\mathcal{C}^{*}_k$ given $\alpha_k^*$ and ${\mathcal{G}}^{(\mathcal{D}_k)}$ is given in (12) and (13). 
Here we study the conditional distribution of $\mathcal{C}^{*}_k$ given ${\mathcal{G}}^{(\mathcal{D}_k)}$, that we obtain from the latter by marginalizing with respect to $\alpha_k^*$. It turns out that: 
\begin{align}\label{eq:post_G_m_mar}
A_{\mathcal{C}^{*}_k[ij]} \mid  {\mathcal{G}}^{(\mathcal{D}_k)}  & \simind \text{Bern}(p^m_{kij}), & i<j 
\end{align}
where the Bernoulli parameters in  \eqref{eq:post_G_m_mar} are given by:
\begin{align}\label{eq:post_G_m_mar_prob}
    p^m_{kij} =  \mathbbm{E}\left[ A_{\mathcal{C}^{*}_k[ij]} \mid {\mathcal{G}}^{(\mathcal{D}_k)} \right]=
    \frac{1}{p({\mathcal{G}}^{(\mathcal{D}_k)} )} \sum_{r=0}^{T^\star_{kij}-t^\star_{kij}} \tilde{w}_{kr}
    \frac{\mathcal{B}( 1/2; a_{kr}^m, b_{kr}^m)}{\mathcal{B}\left( 1/2; a, b\right)}
\end{align}
and $a_{kr}^m=a+n_k+1-n_{ij}^{(k)} + t^\star_{kij}+r$, $b_{kr}^m=b+(n_k+1)(M-1) + n_{ij}^{(k)} - t^\star_{kij} - r$, with $t^\star_{kij}$, $T^\star_{kij}$ and $ \tilde{w}_{kr}$ defined in \Autoref{sec:clust_post_pred}, and $p({\mathcal{G}}^{(\mathcal{D}_k)} )$ is given in \eqref{eq:prop_cons_pred}.

\begin{figure}[h!]
\centering
\includegraphics[height=4.3cm]{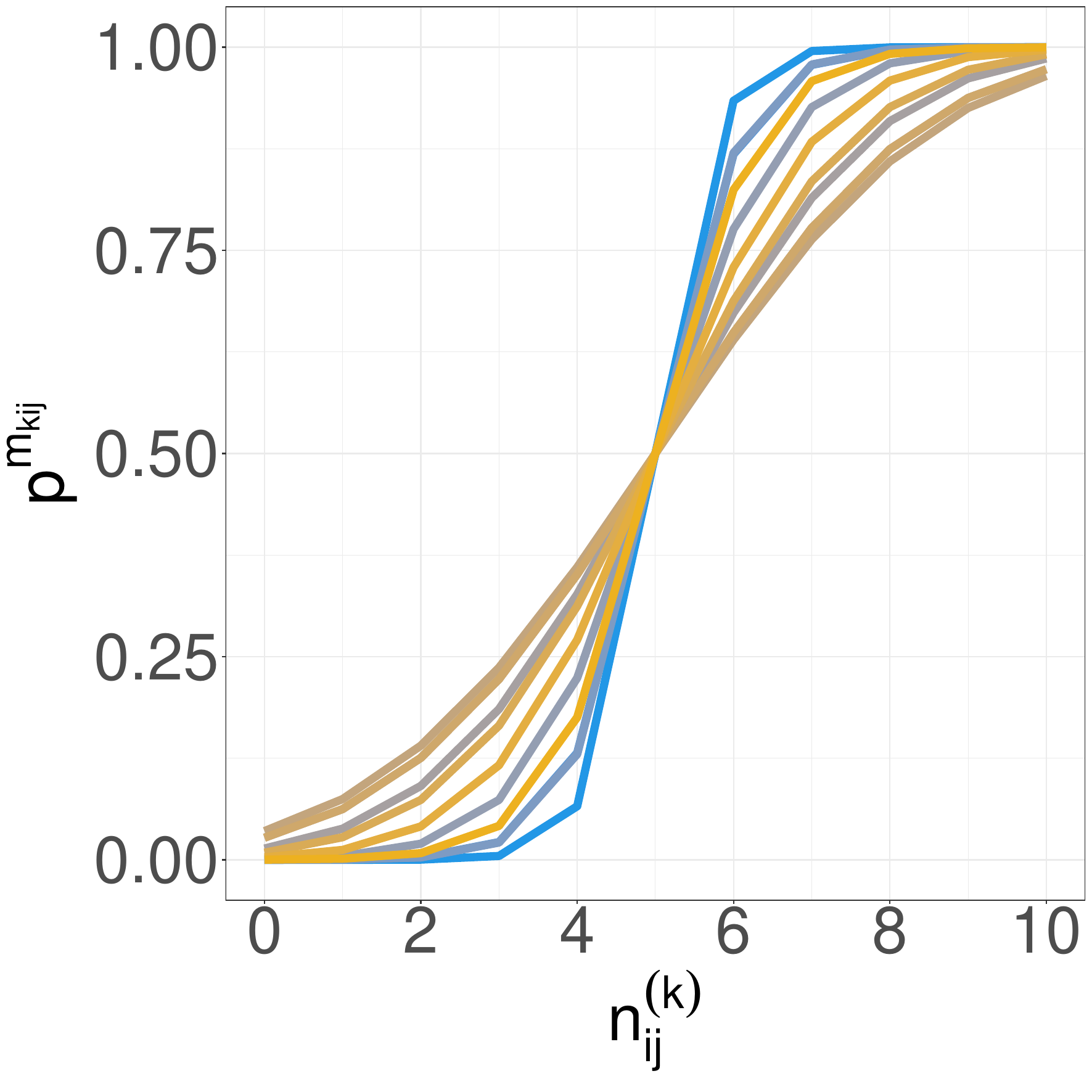}
\includegraphics[height=4.3cm]{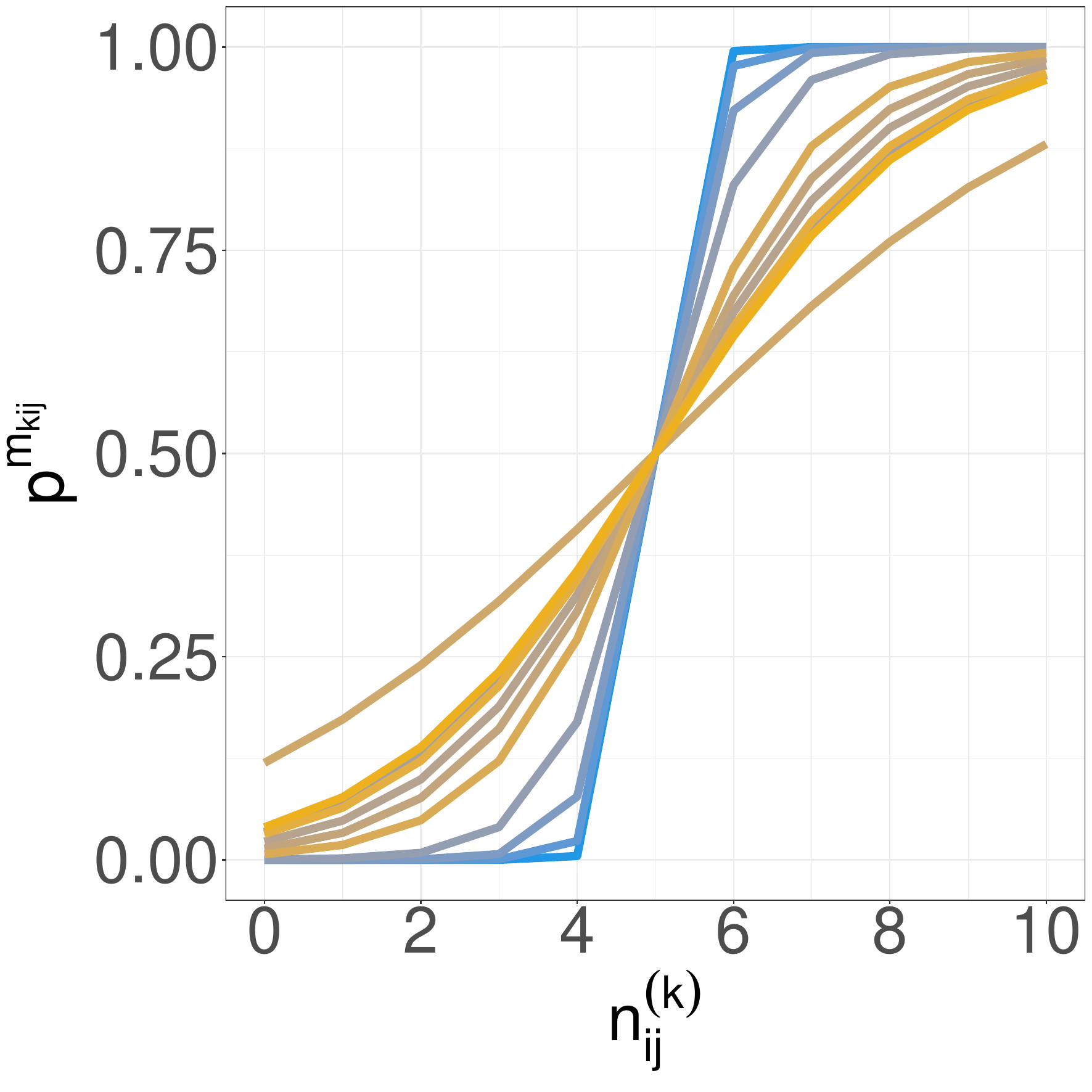}
\includegraphics[height=4.3cm]{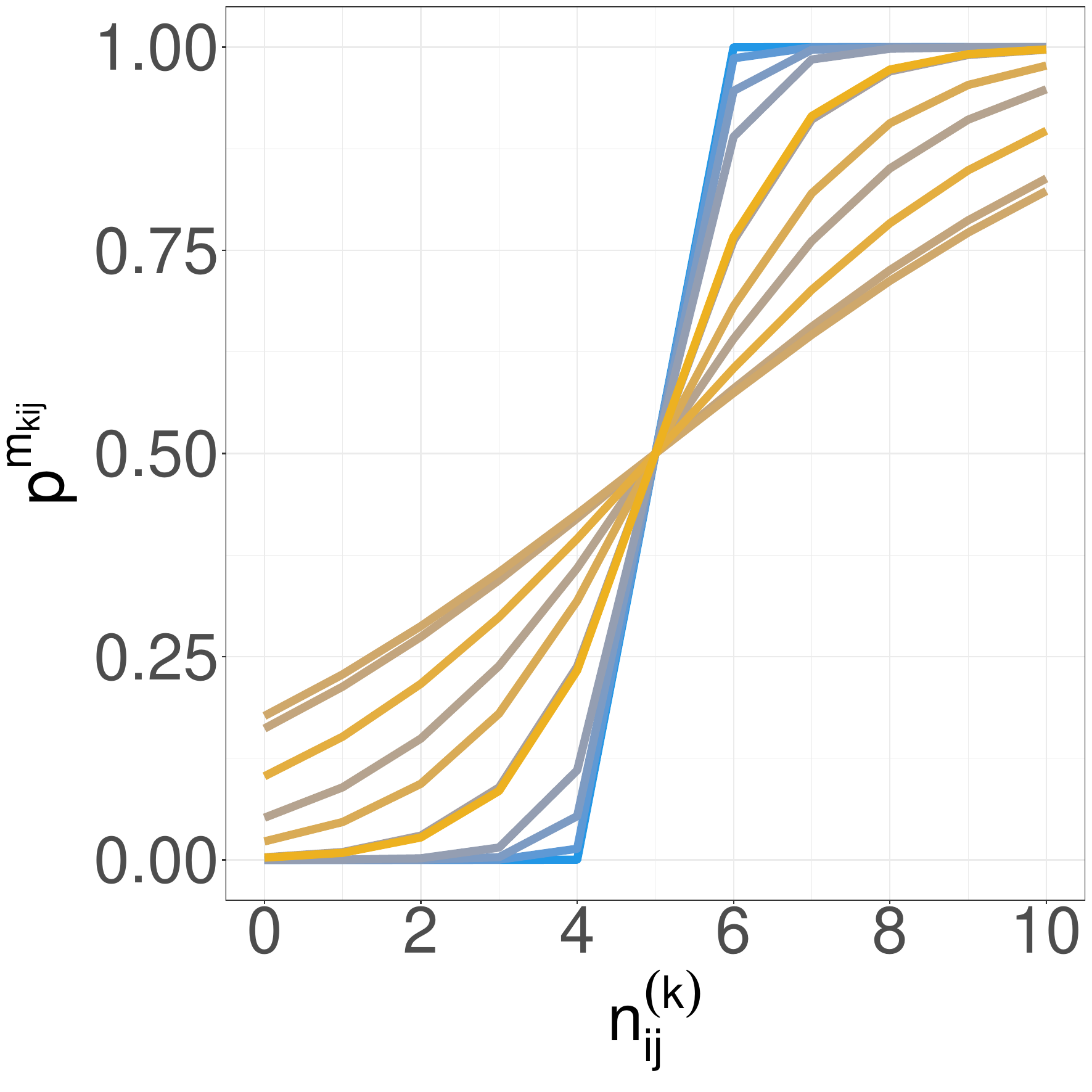}
\caption{\small  
Probability $p^m_{kij}$ in \eqref{eq:post_G_m_mar_prob}, with $\sum_{\{u,v\} \neq \{i,j\}} n_{uv}^{(k)} = z$, with $z$ taking $11$ equally-spaced values in $\{0,1, \ldots, (n_k+1)(M-1)\}$  (blue for low and yellow for high) and for $n_{ij}^{(k)} \in \{0,1, \ldots, n_k+1\}$, with $n_k+1=10$, and for $N \in \{ 3, 5, 10 \}$ (from left panel to right panel).}\label{fig:cls_post_prob_Gm}
\end{figure}

\Autoref{fig:cls_post_prob_Gm} shows how $p^m_{kij}$ varies as a function of $n_{ij}^{(k)}$, for $z=\sum_{\{u,v\} \neq \{i,j\}} n_{uv}^{(k)}$ ranging in 
$\{0,1, \ldots, (n_k+1)(M-1)\}$,
where $n_{uv}^{(k)}$ can change across $\{ u,v\}$ and can serve as comparison to the probability in \autoref{fig:cls_post_pred_prob_sparse}.

\section{Further details on the simulation study of Section 4}\label{app:simulation}
In \autoref{tab:table_centroids}, we report the specification of the parameters for the four data-generating processes used in Section 4 to generate the centroids $\mathcal{C}_{0k}$.

\begin{table}[h!]
\centering
\begin{tabular}{p{4cm} p{8cm}}
Graphical structure & Specification  \\
\hline \hline
\text{Scale-free} & We set the power law exponent of the degree distribution to $2$ and the sparsity to $0.2$. \\
\text{Small-world} & We set the degree of the lattice to $10$ and 
          the probability of rewiring to $0.2$.  \\
\text{Stochastic Block Model} & We set the number of blocks to $2$,
 with membership probabilities equal to $1/2$;
 the inclusion probabilities were set as $0.9$ and $0.1$ for
diagonal and nondiagonal blocks, respectively.\\
\text{Erdős--Rényi} & Probability of inclusion was set to $0.3$.\\
\end{tabular}
\caption{\small Graphical structures and the corresponding parameter specification
used to define the distribution centroids.}
\label{tab:table_centroids}
\end{table}

\subsection{Varying sample size with $\mathbb{L}^1$ distance}\label{sec:secD1}
\begin{figure}[t!]
\centering
\includegraphics[width = 0.94\textwidth]{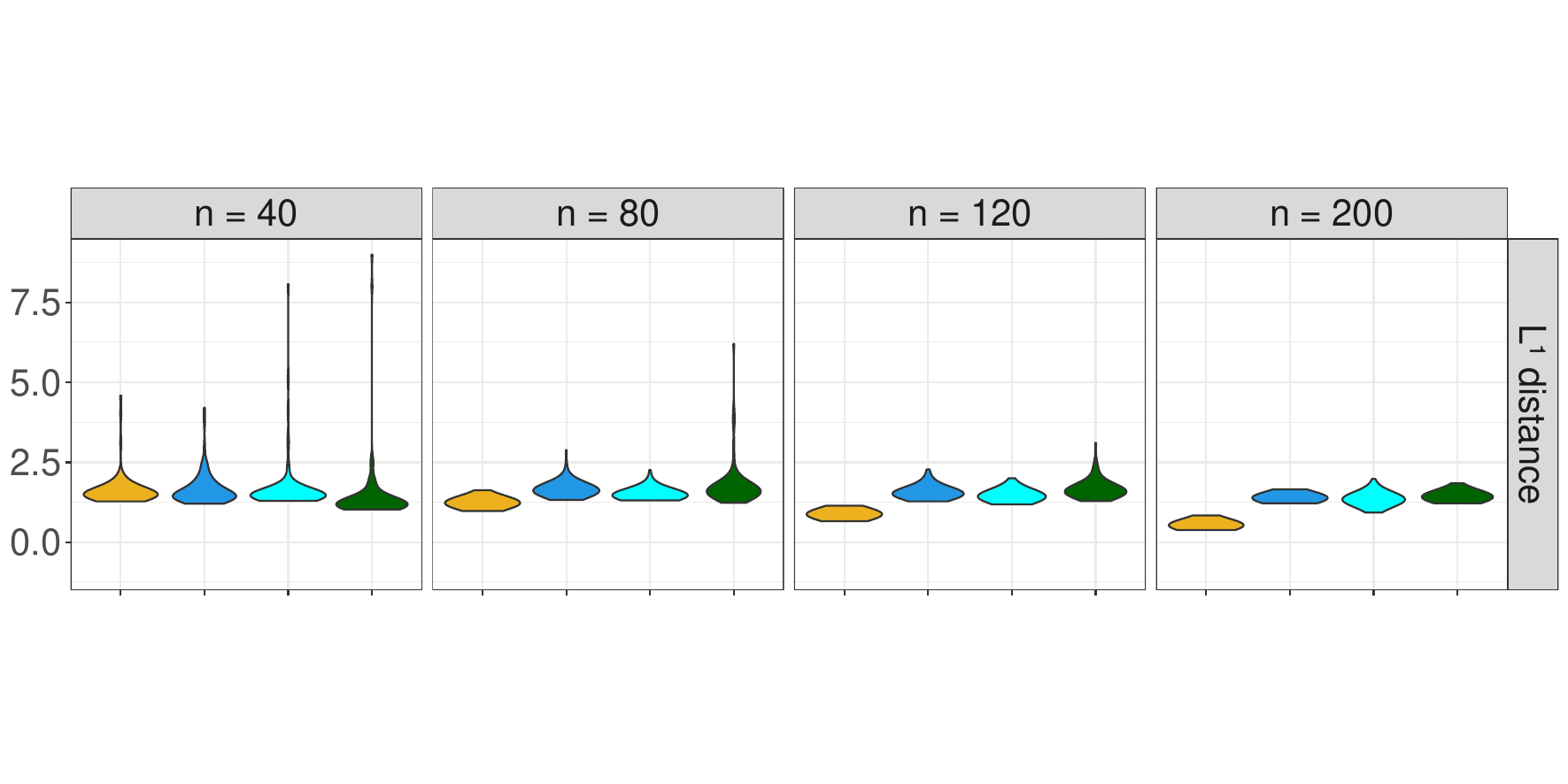}
\caption{\small Importance-sampling approximate distributions of $\mathbb{L}^{1}(p_*;\hat{f})$ distance for our method (yellow violins), and the methods of \citet{durante2017} (blue violins),
\citet{anastasia} (cyan violins) and
\citet{Signorelli} (green violins).
Distributions are estimated based on the analysis of 100 datasets. 
}\label{fig:fig_consistency_L1}
\end{figure}

We present additional simulation experiments investigating how the posterior mean 
 $$ \hat{f}(\cdot) = \mathbbm{E}[\tilde{f}(\cdot) \mid \mathcal{G}^{(1:n)}]=\frac{1}{c+n} \int_{\Theta} \psi\left(\cdot ; \vartheta \right)  \mathrm{d} P_{0}(\vartheta) + \frac{1}{T} \sum_{t=1}^T \sum_{k=1}^{K^{(t)}} \frac{n_k^{(t)}}{c+n} \; \psi(\cdot ; \vartheta_k^{*(t)})
 $$
concentrates around its true value 
$p_*(\cdot)=\sum_{k=1}^4 0.25p_{\text{CER}}(\cdot;\mathcal{C}_{0k},\alpha_{0k})$,
as a function of the sample size $n$, with $\hat{f}$ evaluated based on the posterior sample generated from Algorithm 1.
Unlike the study presented in Section 4.2, we focus here on the $\mathbb{L}^1$ distance as a metric on $\mathcal{P}_{\mathscr{G}_{\mathcal{V}}}$, and study the distribution of the distance between $p_*$ and $\hat{f}$ for finite samples of size $n \in \{40, 80, 120, 200\}$. The evaluation of $\mathbb{L}^1(p_*; \hat{f})$ requires summation over the graph space $\mathscr{G}_{\mathcal{V}}$, which is prohibitive even for moderate $N$. Thus, we propose an importance-sampling approximation of $\mathbb{L}^1(p_*; \hat{f})$. Namely,
\begin{align*}
\mathbb{L}^1(p_*; \hat{f}) &= \sum_{\mathcal{G} \in \mathscr{G}_{\mathcal{V}} } | p_*(\mathcal{G})-\hat{f}(\mathcal{G}) | =  \sum_{\mathcal{G} \in \mathscr{\mathcal{G}}_N } \frac{ |  p_*(\mathcal{G})-\hat{f}(\mathcal{G}) | }{p_*(\mathcal{G})}  p_*(\mathcal{G})  = \mathbbm{E}_{p_*}\left[  \frac{ |  p_*(\mathcal{G})-\hat{f}(\mathcal{G}) | }{p_*(\mathcal{G})}   \right]\\
& \approx \frac{1}{L} \sum_{l=1}^L  \frac{ |  p_*(\mathcal{G})-\hat{f}(\mathcal{G}) | }{p_*(\mathcal{G})}, 
\end{align*}
with $\mathcal{G}_l  \simiid  p_*$, for $l=1, \ldots, L$. The results are presented in \autoref{fig:fig_consistency_L1}, which shows that the posterior estimate $\hat{f}$ gets closer to $p_*$ as the sample size increases.
Our model appears to converge to $p_*$ faster than the models proposed by \cite{durante2017}, \citet{anastasia} and
\citet{Signorelli}. 
This additional study gives credibility to the robustness of our model with respect to the choice of the metric $d$ on $\mathcal{P}_{\mathscr{G}_{\mathcal{V}}}$.

\subsection{Additional simulation experiment}\label{sec:secD2}
\begin{figure}[h!]
\centering
\includegraphics[height=5cm]{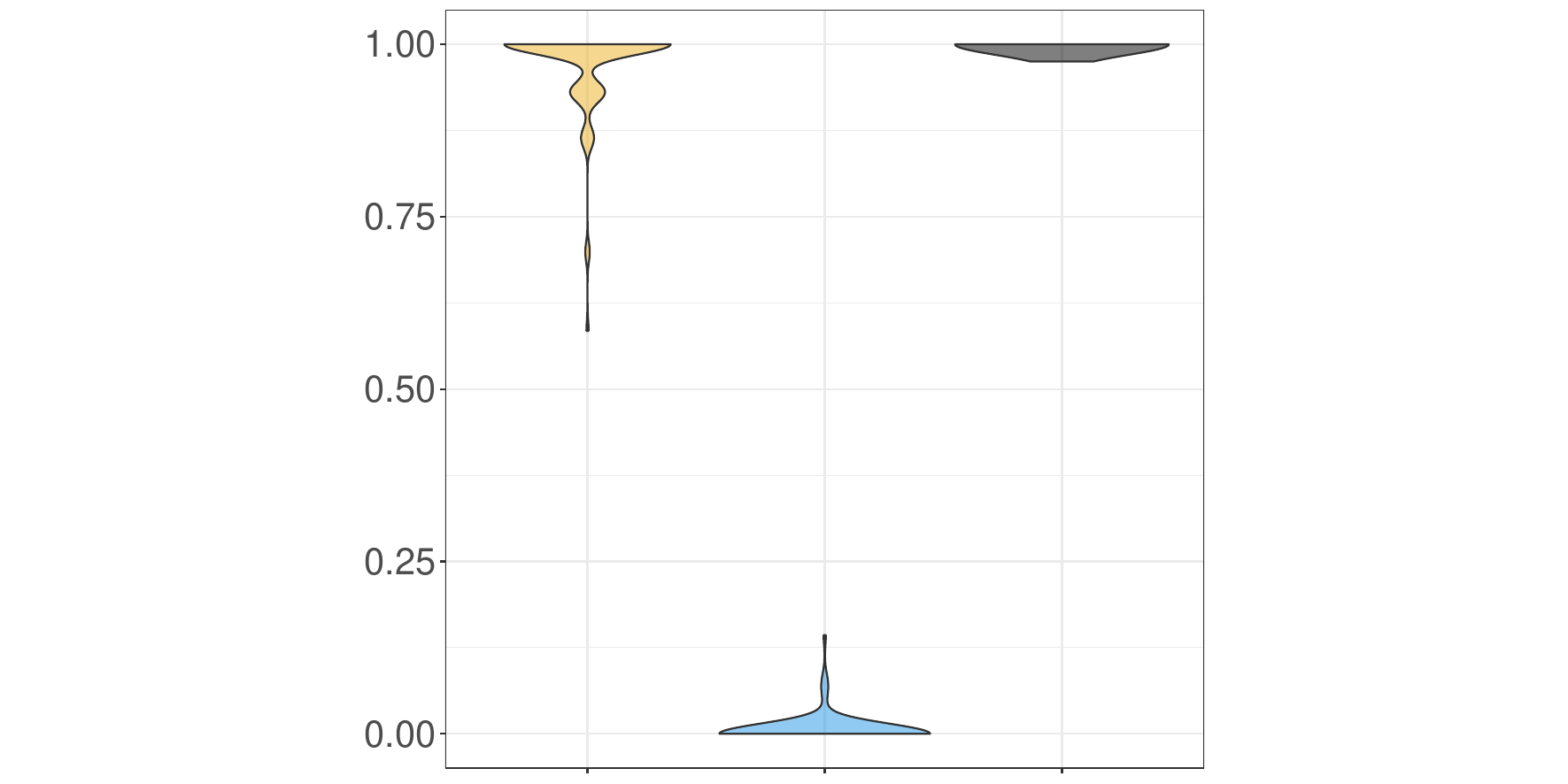}
\caption{\small Adjusted Rand index (left), clustering entropy (center), and clustering purity(right). Distributions are estimated based on the analysis of 100 datasets. See Section \ref{sec:secD2}.}\label{fig:cls_core_periphery}
\end{figure}

We present an additional experiment to assess the behavior of the DP mixture of CER kernels when more intricate connectivity patterns than those presented in Section 4 of the main paper are considered. Specifically, we consider the core-periphery structure, which may result, for example, from a non-assortative stochastic block model generative process,
in which nodes in the core are densely linked to each other and often to the periphery, and peripheral nodes are typically linked to the core but weakly connected with each other. As in Section 4 of the main paper, we focus on networks with $N = 20$ nodes. A set of $n=40$ observations are sampled from a two-component mixture of CER $p_*(\cdot)=0.5p_{\text{CER}}(\cdot;\mathcal{C}_{01},\alpha_{01})+0.5p_{\text{CER}}(\cdot;\mathcal{C}_{02},\alpha_{02})$, where 
the centroids $\mathcal{C}_{01}$ and $\mathcal{C}_{02}$ have a core-periphery \citep{core_periphery} and a Erdős--Rényi \citep{Erdos_Reny} structure, respectively, and 
the component-specific scales of variation are set equal to $\alpha_{01}=0.4$ and $\alpha_{02}=0.3$. \Autoref{fig:centroids_core_periphery} illustrates the generated centroids.  
\begin{figure}[h!]
\centering
\includegraphics[height=5cm]
{figures/plot_core_periphery_G_m_1.pdf}
\includegraphics[height=5cm]
{figures/plot_core_periphery_G_m_2.pdf}
\caption{\small Left: centroid \(\mathcal{C}_{01}\) with a core–periphery structure, core nodes shown in blue. Right: centroid \(\mathcal{C}_{02}\) with an Erdős–Rényi structure. See Section~\ref{sec:secD2}.} \label{fig:centroids_core_periphery}
\end{figure}
To assess the ability of our method to cluster multiple network data, we compare the
estimated partition to the true partition, which reflects the two-component mixture
structure of the data-generating model. We resort to three metrics: the adjusted Rand
index, clustering entropy and clustering purity. The results of our investigation are displayed in \Autoref{fig:cls_core_periphery}. The performance of our model appears robust to this more complex scenario.

\section{On the implementation of the competing methods}\label{app:data_analysis}
We detail here the specification of the hyperparameters for all models included in the comparison.\\
\Autoref{tab:hyperpar_durante} reports the hyperparameter settings adopted for the implementation of the method of \citet{durante2017} in the studies of Section~4, corresponding to the default options of the authors’ code. In the study of Section~5, the upper bound on the number of classes $H$ is set to twice the number of individuals, i.e.\ $H=60$.\\
For the method of \cite{josephs2025} in the simulation study of Section~4.1, we set the truncation levels for the number of classes $K$ and for the number of within-class node clusters $L$ to $15$, following the default configuration of the authors’ code. We employ the Incompatible Blocked Gibbs (IBG) sampler, which the authors report as yielding the best clustering performance.\\
For the method of \cite{anastasia}, uniform priors are assigned to all component-specific parameters, and representative networks with two node blocks are used, as in the default setting of the authors’ code. In the simulation studies of Section~4, the number of components is fixed to match the number of mixture components in the data-generating process, i.e.\ $K=4$. In the study of Section~5, we adopt the Sparse Finite Mixture extension of \cite{anastasia}, setting the upper bound on the number of clusters to $C_{\max}=60$ and placing a $\operatorname{Gamma}(a_e=1, b_e=400)$ hyperprior on the hyperparameter $e_0$ of the symmetric Dirichlet prior on the mixture weights $\bm{\tau}=(\tau_1,\ldots,\tau_{C_{\max}})$, to favour values of $e_0$ close to zero, as recommended by the authors.\\
For the method of \cite{Signorelli}, the number of components is set equal to the number of mixture components in the simulation studies of Section~4, i.e.\ $K=4$, and the unconstrained network model is adopted for the specification of the mixture components, where the number of parameters equals the number of edge pairs.
\begin{table}[h!]
\centering
\begin{tabular}{ccccccc}
	 Section & $R$ & $H$ & $a_1$ & $a_2$ & $\mu_l$ & $\sigma_l^2$  \\ 
		\hline\hline
		 4 & $10$ & $30$ & $2.5$ & $3.5$ & $0$ & $10$\\ 
	 5 &  $10$ & $60$ & $2.5$ & $3.5$ & $0$ & $10$
	\end{tabular} \caption{Hyperparameter specification for the method of \citet{durante2017}. 
The table reports: the upper bound on the latent space dimension $R$; 
the upper bound on the number of classes $H$; 
the hyperparameters $a_1$ and $a_2$ of the multiplicative inverse-Gamma prior; 
and the Gaussian prior mean $\mu_l$ and variance $\sigma_l^2$ for $Z$, 
for all $l = 1,\ldots, N(N-1)/2$.}\label{tab:hyperpar_durante}
\end{table}

\section{Further details on illustrations in Section 6}\label{app:largeN}
We present additional plots related to the application in Section 6.

\begin{figure}[b!]
\centering
\includegraphics[trim={0cm 0cm 0cm 0cm},clip,width=0.9\textwidth]
{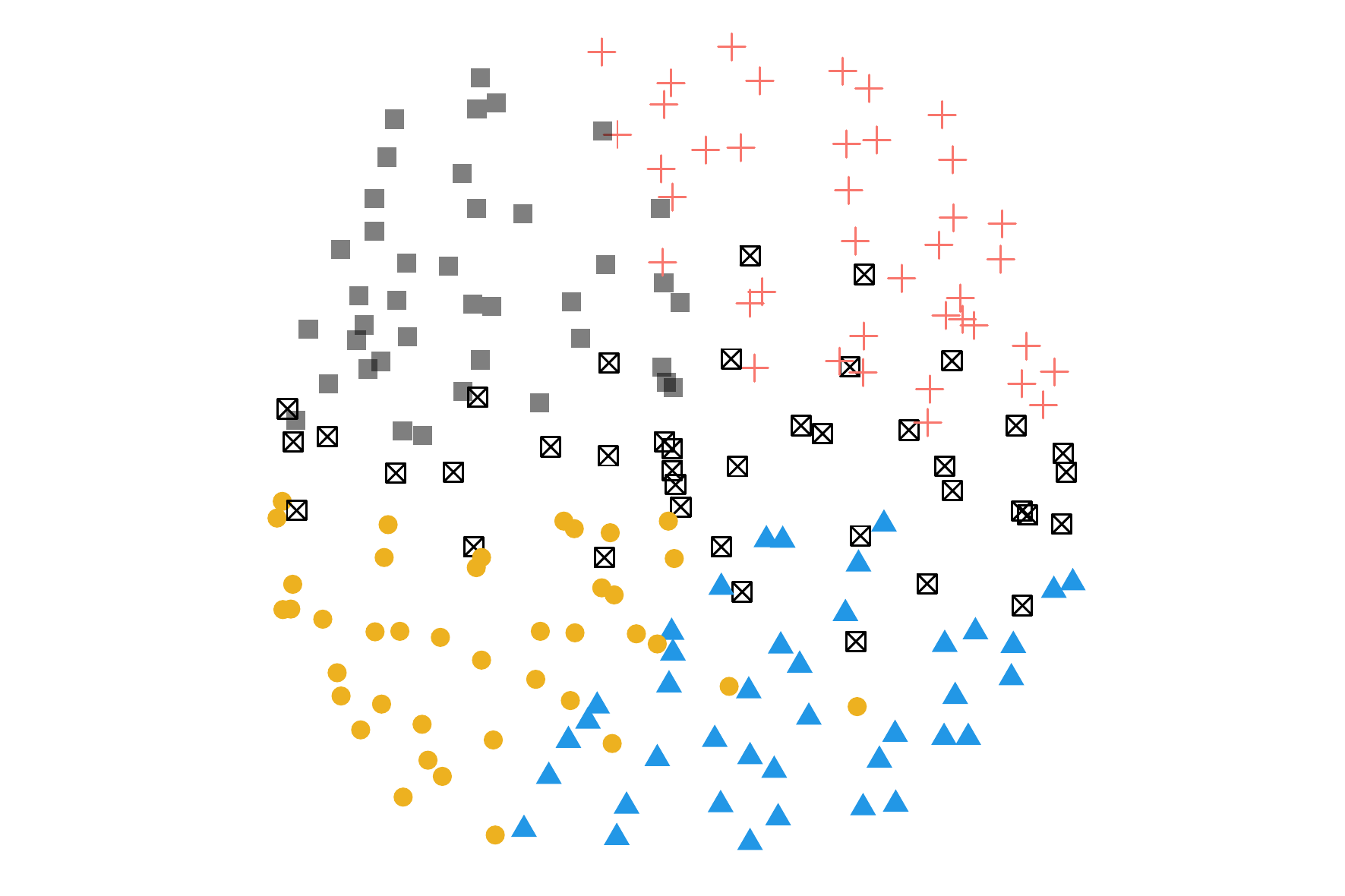}\caption{
A 2-D visualization (top-down projection) of the atlas with 200 ROIs, where colors and shapes represent the $m_{\text{sub}} = 5$ node cluster memberships identified through balanced clustering with $N_{\text{sub}} = 40$. 
}\label{fig:brain_node_cls_2}
\end{figure}

\begin{figure}[b!]
\centering
\includegraphics[trim={0.5cm 4cm 0cm 4cm},clip,width=1\textwidth]
{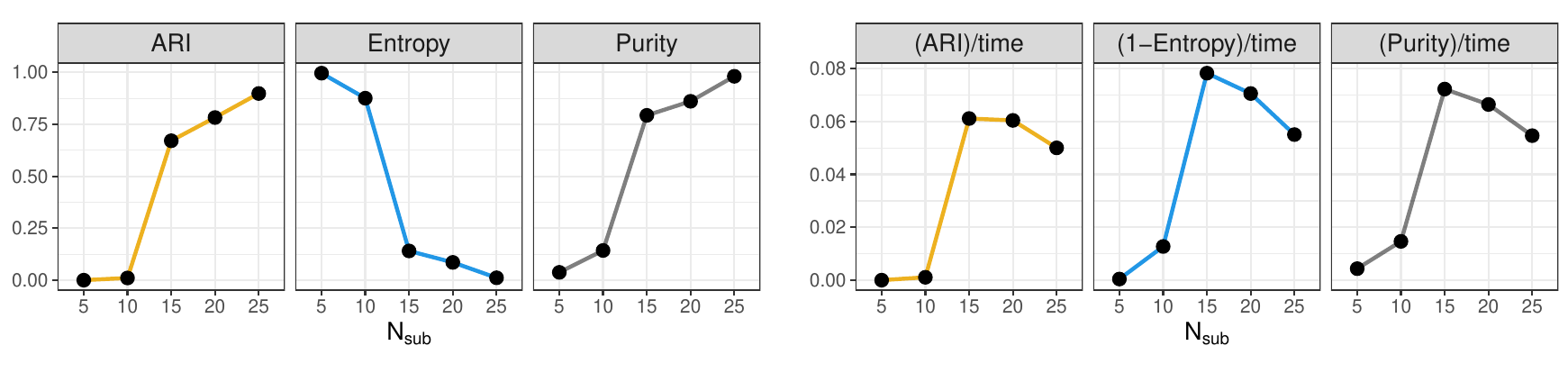}\caption{\small Clustering metrics comparing the partition estimated based on the consensus subgraph approach, with the one estimated with exact method on $48$ ROIs, for the Human Brain dataset based on $48$ ROIs, for $N_{sub}$ ranging in $\{5,10,\ldots,25\}$.
}\label{fig:large_N_48_wrt_exact_time_spatial_both}
\end{figure}

\clearpage
\subsection{Consensus subgraph clustering for large $N$ with nodes partitioned at random}\label{subsec:largeN_random}
\begin{figure}[h!]
\centering
\includegraphics
[trim={6.8cm 0cm 6.8cm 0cm},clip,width=0.31\textwidth]{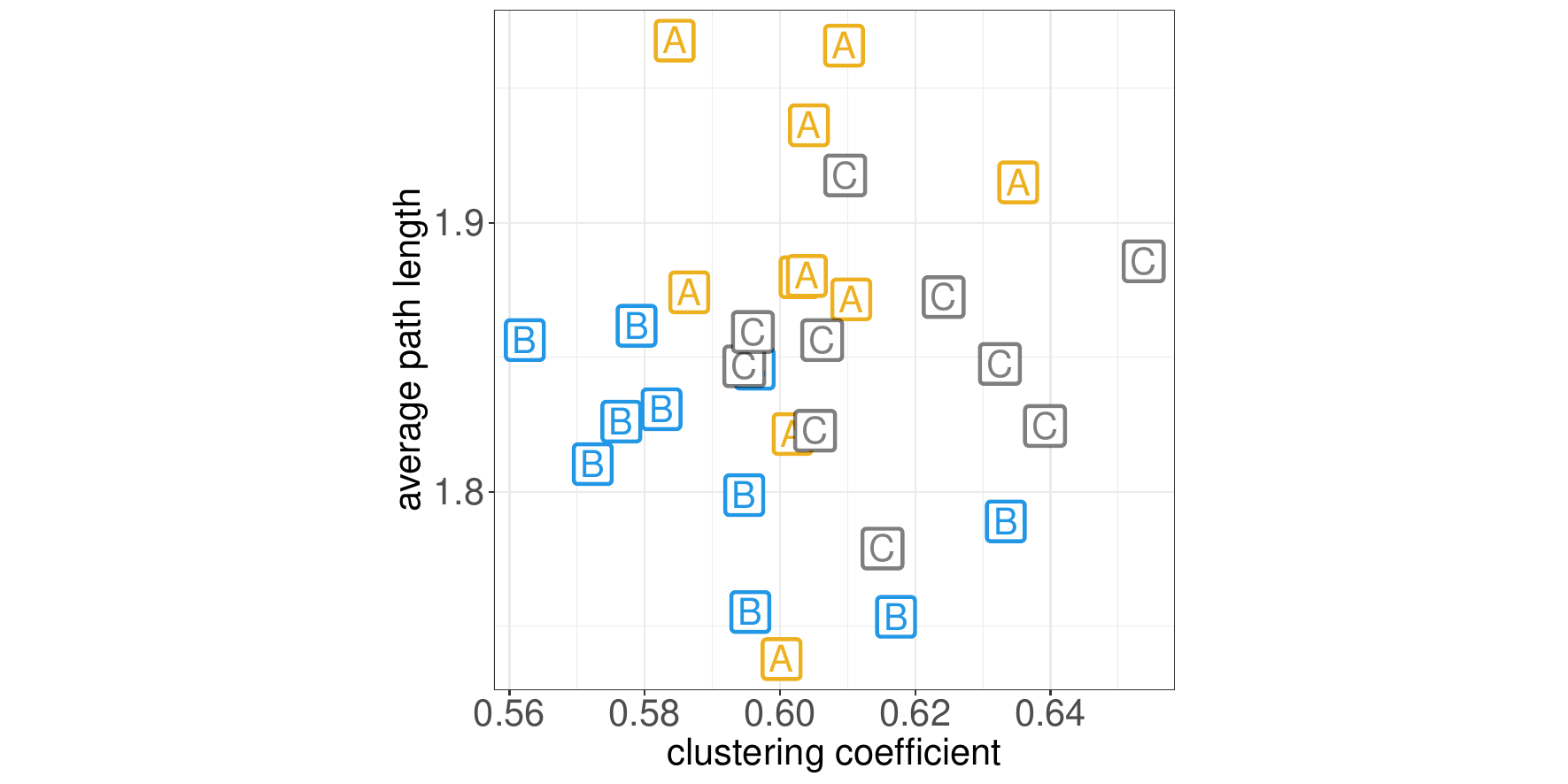}
\hspace{0.2cm}
\includegraphics[trim={6.8cm 0cm 6.8cm 0cm},clip,width=0.31\textwidth]{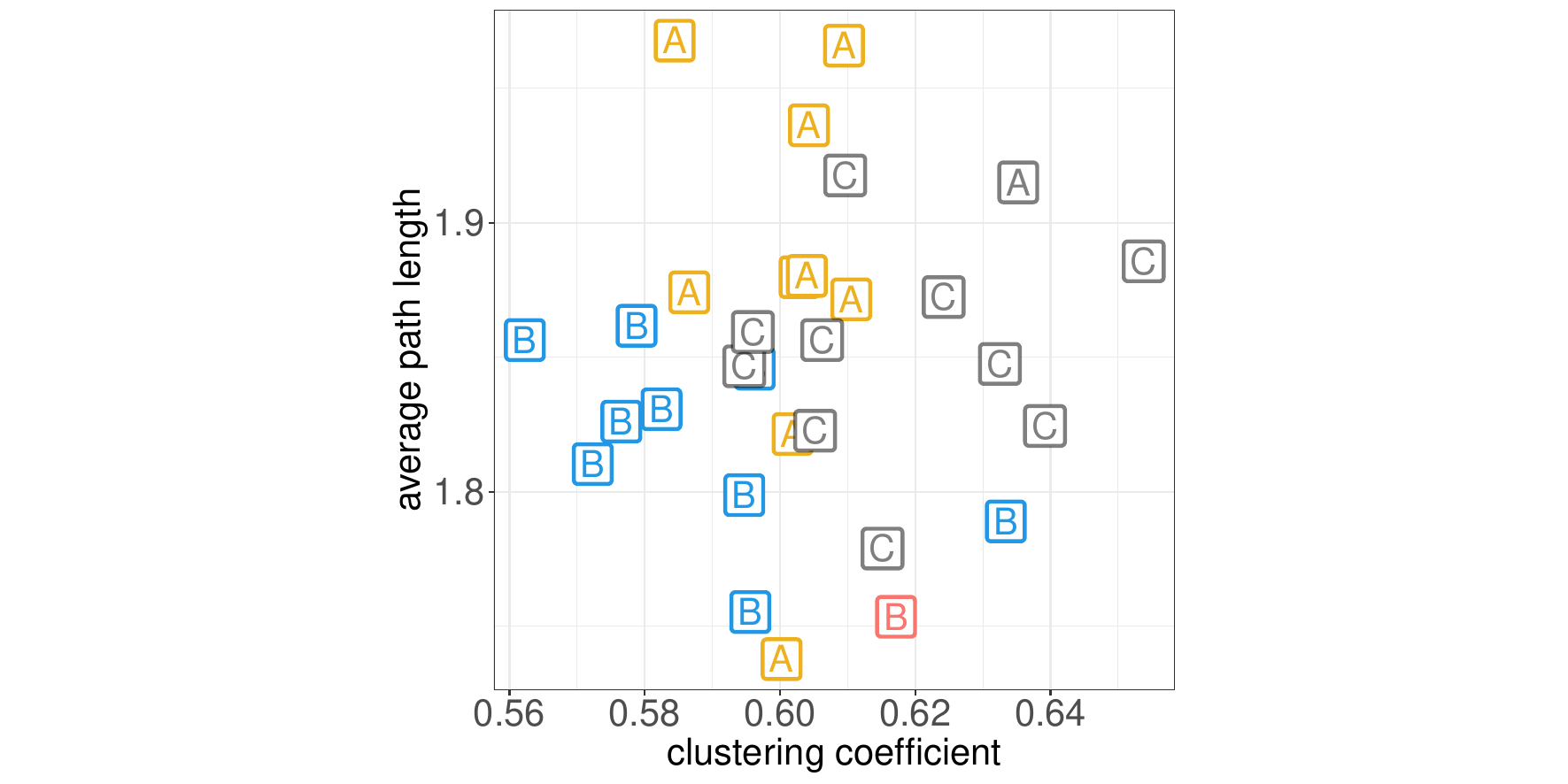}\hspace{0.2cm}
\includegraphics[trim={6.8cm 0cm 6.8cm 0cm},clip,width=0.31\textwidth]{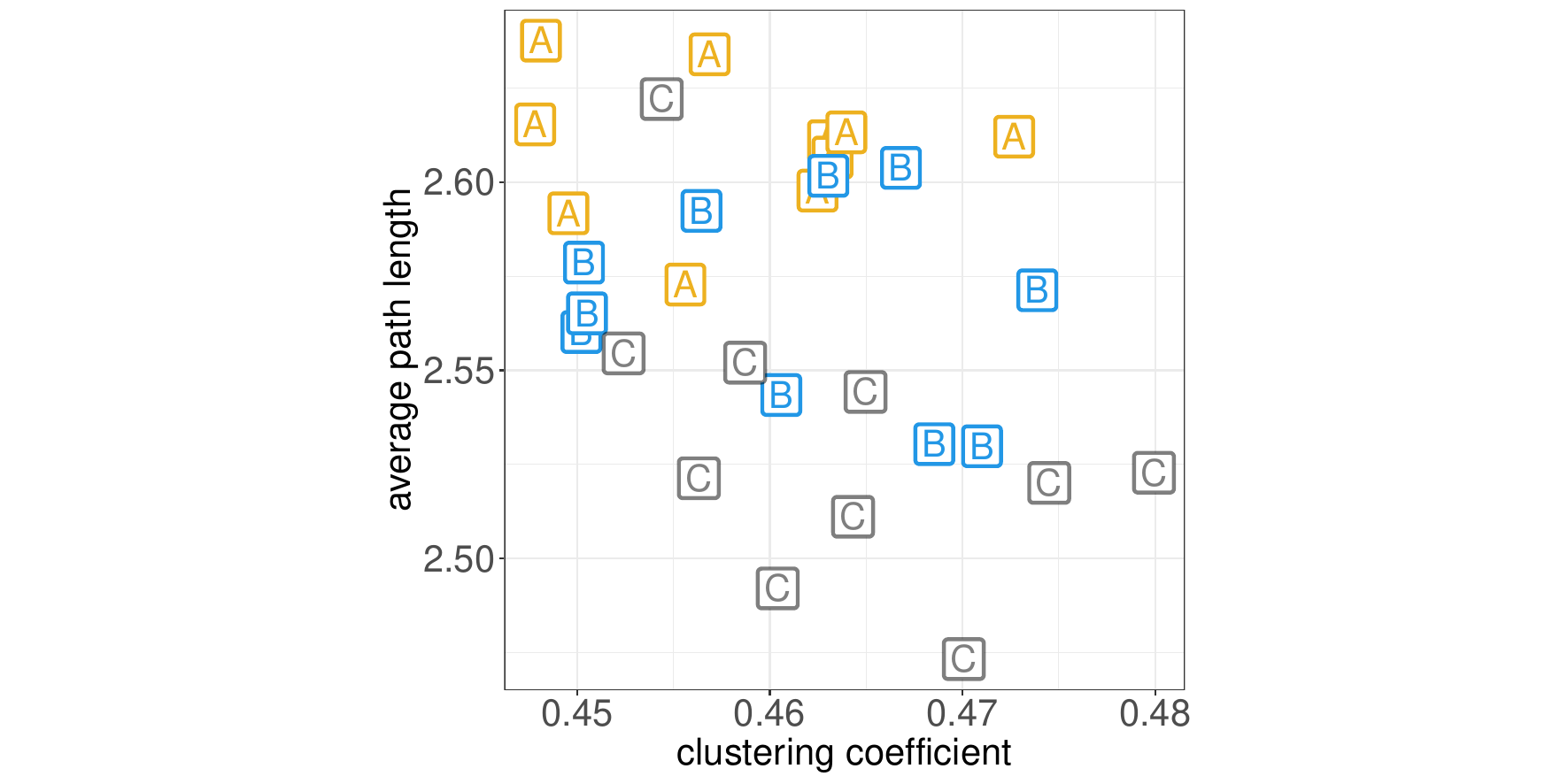}
\caption{\small  Scatter plots for the small-world properties of brain networks for three subjects in the dataset. 
Colors indicate the cluster membership, letters refer to the subject ID in the dataset, namely 0025443 (A), 0025445 (B) and 0025446 (C). Left panel: $N=48$, partition estimated via DP mixture of CER kernels. Central panel: $N=48$, partition estimated via consensus subgraph clustering with nodes partitioned at random. Right panel: $N=200$, partition estimated via consensus subgraph clustering with nodes partitioned at random. See Figure 7
for a comparison.}\label{fig:human_brain_random_split}
\end{figure}

For comparison with the analysis in Section 6, 
we performed consensus subgraph clustering on the human brain datasets, with 48 and 200 ROIs, by partitioning the nodes randomly, thus without utilizing the available spatial information on the nodes. This allows us to understand the impact of incorporating spatial information when partitioning the nodes. 
Interestingly, when nodes are partitioned at random, the clustering metrics computed on the estimated data clustering appear only slightly worse than those obtained in Section 6
based on the available spatial information. 
The results of our analysis are presented in \Autoref{fig:human_brain_random_split} and \Autoref{tab:table_cls_largeN_random}. $N_{\text{sub}}$ was set equal to 15 for both versions of the human brain datasets.

\begin{table}[h!]
\centering
\begin{tabular}{c c c c c }
$N$ ($N_{\text{sub}}$) & $\hat{K}$ & Adjusted Rand Index & Entropy & Purity \\
\hline\hline
$48\; (15)$ & 30 & 0.6642 & 0.1420 & 0.7970 \\
$200 \; (15)$ & 31 & 0.9490 & 0.0162 & 0.9699 \\
\end{tabular}
\caption{\small Human brain dataset. Estimated number of clusters and clustering metrics comparing the partition estimated based on the consensus subgraph clustering approach, with nodes partitioned at random, with the one implied by the presence of $30$ individuals in the study. 
See Table 3 for a comparison.
}
\label{tab:table_cls_largeN_random}
\end{table}


\end{bibunit}

\end{document}